\documentclass[twocolumn]{aastex631}
\usepackage{epsf}
\usepackage{psfig}
\usepackage{color}
\usepackage{float}
\usepackage [english]{babel}
\usepackage [autostyle, english = american]{csquotes}
\MakeOuterQuote{"}
\pdfoutput=1 
\usepackage{graphicx}
\usepackage[inline]{enumitem}
\usepackage{verbatim}
\usepackage{color}
\shorttitle{The Multi-phase IGM in Stephan's Quintet}
\shortauthors{Appleton et al.}

\newcommand{\Hi}{\ion{H}{1} }

\newcommand{\Hmol}{\mbox{H$_{\rm 2}$}}

\newcommand{\kms}{km~s$^{-1}$}

\newcommand{\nH}{$n_{\rm H}$}

\newcommand{\cc}{\mbox{cm$^{-3}$}}

\begin{document}

\title{Multi-phase gas interactions on subarcsec scales in the shocked IGM of Stephan's Quintet with JWST and ALMA}

\author{P. N. Appleton}
\affiliation{Caltech/IPAC, MC 314-6, 1200 E. California Blvd., Pasadena, CA 91125, USA. apple@ipac.caltech.edu}
\author{P. Guillard}
\affiliation{Sorbonne Universit\'{e}, CNRS, UMR 7095, Institut d'Astrophysique de Paris, 98bis bd Arago, 75014 Paris, France}
\affiliation{Institut Universitaire de France, Minist{\`e}re de l'Enseignement Sup{\'e}rieur et de la Recherche, 1 rue Descartes, 75231 Paris Cedex 05, France} 
\author{Bjorn Emonts}
\affiliation{National Radio Astronomy Observatory, 520 Edgemont Road, Charlottesville, VA 22903}
\author{Francois Boulanger}
\affiliation{UPMC Universite Paris 06, {\'E}cole Normale Sup{\'e}rieure, 75005 Paris, France}
\author{Aditya Togi}
\affiliation{Texas State University, 601 University Dr, San Marcos, TX 78666, USA}
\author{William T. Reach}
\affiliation{Universities Space Research Association, NASA Ames Research Center MS 232-11, Mountain View, CA 94035, USA}
\author{Kathleen Alatalo}
\affiliation{STScI, 3700 San Martin Drive,  Baltimore, MD 21218}
\author{M. Cluver}
\affiliation{Centre for Astrophysics and Supercomputing, Swinburne University of Technology, John Street, Hawthorn, 3122, Australia}
\affiliation{Department of Physics and Astronomy, University of the Western Cape, Robert Sobukwe Road, Bellville, 7535, South Africa}
\author{T. Diaz Santos}
\affiliation{Institute of Astrophysics, Foundation for Research and Technology-Hellas (FORTH), Heraklion, 70013, Greece.}
\affiliation{School of Sciences, European University Cyprus, Diogenes street, Engomi, 1516 Nicosia, Cyprus. }
\author{P-A Duc}
\affiliation{Université de Strasbourg, CNRS, Observatoire astronomique de Strasbourg (ObAS), UMR 7550, 67000 Strasbourg, France}
\author{S.Gallagher}
\affiliation{Institute for Earth and Space Exploration, Western University, 1151 Richmond St., London, ON N6A 3K7, Canada}
\author{P. Ogle}
\affiliation{STScI, 3700 San Martin Drive,  Baltimore, MD 21218}
\author{E. O'Sullivan}
\affiliation{Center for Astrophysics $|$ Harvard \& Smithsonian, 60 Garden Street, Cambridge, MA, 02138, USA}
\author{K. Voggel}
\affiliation{Université de Strasbourg, CNRS, Observatoire astronomique de Strasbourg (ObAS), UMR 7550, 67000 Strasbourg, France}
\author{C. K. Xu}
\affiliation{Chinese Academy of Sciences South America Center for Astronomy, National Astronomical Observatories, CAS, Beijing 100101, People’s Republic of China}
\affiliation{National Astronomical Observatories, Chinese Academy of Sciences (NAOC), 20A Datun Road, Chaoyang District, Beijing 100101, People’s Republic of China}
\begin{abstract}
We combine JWST and HST imaging with ALMA~CO(2-1) spectroscopy to study the highly turbulent multi-phase intergalactic medium (IGM) in Stephan's Quintet on 25-150 pc scales. Previous Spitzer observations revealed luminous H$_2$ line cooling across a 45 kpc-long filament, created by a giant shock-wave, following the collision with an intruder galaxy NGC~7318b. We demonstrate that the MIRI/F1000W/F770W filters are dominated by 0-0~S(3)~H$_2$ and a combination of PAH and
0-0~S(5)~H$_2$ emission. They reveal the dissipation of kinetic energy as massive clouds experience collisions, interactions and likely destruction/re-cycling within different phases of the IGM. In one kpc-scaled structure, warm H$_2$ formed a triangular-shaped head and tail of compressed and stripped gas behind a narrow shell of cold H$_2$. In another region, two cold molecular clumps with very different velocities are connected by an arrow-shaped stream of warm, probably shocked, H$_2$ suggesting a cloud-cloud collision is occurring. In both regions, a high warm-to-cold molecular gas fraction indicates that the cold clouds are being disrupted and converted into warm gas. We also map gas associated with an apparently forming dwarf galaxy. We suggest that the primary mechanism for exciting strong mid-IR H$_2$ lines throughout Stephan's Quintet is through a fog of warm gas created by the shattering of denser cold molecular clouds and mixing/recycling in the post-shocked gas. A full picture of the diverse kinematics and excitation of the warm H$_2$ will require future JWST mid-IR spectroscopy. The current observations reveal the rich variety of ways that different gas phases can interact with one another.

 \end{abstract}

\section{Introduction}

Since the discovery of a filament of radio continuum in the intergalactic medium (IGM) of the Stephan's Quintet \citep{Allen1972}, this compact galaxy group has been studied at many wavelengths to try to better understand that remarkable nature of its multi-phase intergalactic medium. As suspected in the early studies \citep{Moles1997,Sulentic2001,Xu1999}, the overall picture that has emerged is that one of the galaxies, NGC 7318b, is colliding with the diffuse intergalactic medium of the main group at a very high velocity creating a giant (45 kpc) filament of shocked gas seen from the X-rays \citep{Trinchieri2005,OSullivan2009}, in the UV \citep{Xu2005} and in ionized gas emissions  \citep{Xu2003,Iglesias-Paramo2012,Konstantopoulos2014,DuartePuertas2019,DuartePuertas2021}. Studies of the stellar populations along the giant shocked structure also suggested that star clusters are beginning to form there \citep{Gallagher2001,Fedotov2011}, although at a low rate,  perhaps because some of the  forming clusters lie inside bubbles of highly excited ionized gas \citep{Konstantopoulos2014}. 

\begin{figure*}
\includegraphics[width=0.96\textwidth]{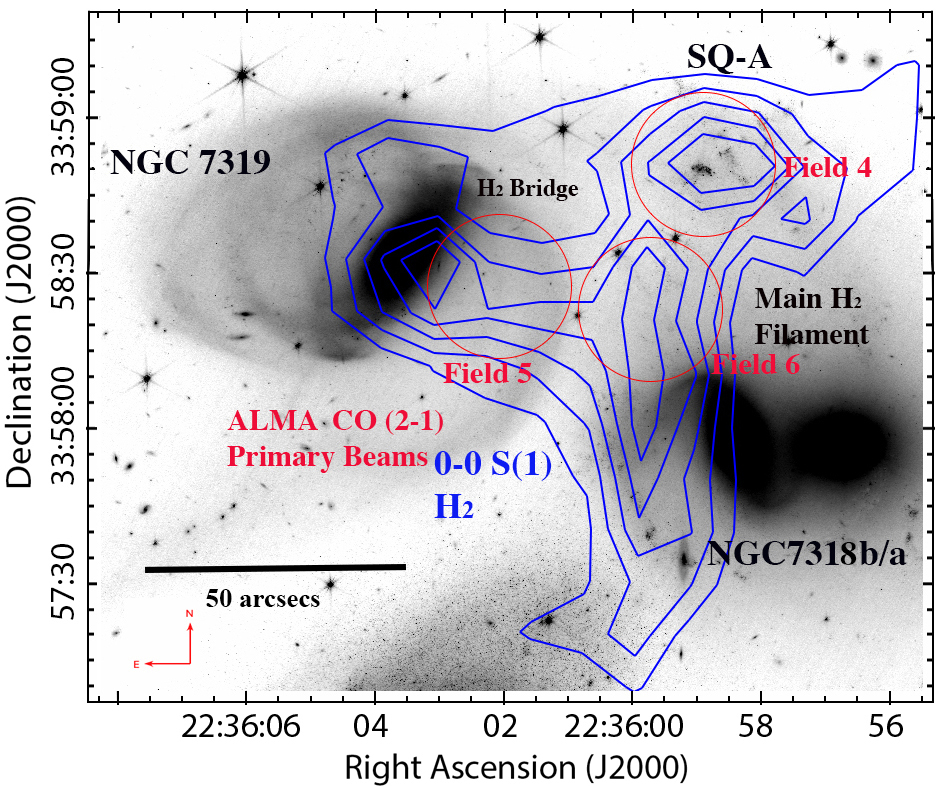}
\caption{Stephan's Quintet: A grey-scale representation of the NIRCam F150W ERO image of Stephan's Quintet with overlays of the {\it Spitzer} IRS image of the 0-0 S(1) H$_2$ line (blue contours) and the half-power primary beam size of the ALMA CO (2-1) observations (red circles) described in \S~\ref{sec:alma}. The faintest features seen in the F150W image are at a surface brightness level of 0.7-1 MJy sr$^{-1}$.  Contours of H$_2$ emission are in units of 0.3, 0.53, 0.75, 0.98, 1.2, 1.43, 1.65, 1.88 and 2.1 MJy sr$^{-1}$ from \citet{Cluver2010}.} \label{fig:intro}
\end{figure*}

The physical conditions within the gas along the main shock front  are far from simple and require more detailed study. Mid-IR spectroscopy of the main filament showed that the entire filament is radiating strongly in pure rotational lines of molecular hydrogen (\citealt{Appleton2006,Cluver2010}; See Figure~\ref{fig:intro}). An explanation of the excitation of such strong, $L(H_2) > 10^{41}$ erg\ s$^{-1}$,  molecular hydrogen emission from a region experiencing fast shocks from the intruder's high velocity was presented by \citet{Guillard2009}. The model assumes that the main intruder shock propagates into a clumpy pre-shock medium, heating low-density regions to X-ray temperatures, but causing mildly over-dense regions to collapse to form H$_2$ on grains on timescales shorter than the dynamical timescale of the collision. These molecular clouds experience low-velocity molecular shocks as they continue to dissipate mechanical energy from their surroundings. The interaction of these warm H$_2$ clouds with their surroundings is a key area that we will investigate with the James Web Space Telescope (JWST) and Atacama Large Millimeter Array (ALMA) in this paper. 

A more detailed analysis of the H$_2$ excitation properties of the warm H$_2$ in Stephan's Quintet was discussed in \citet{Appleton2017}, where it was shown that in order to heat so much H$_2$ ($10^9 M_{\odot}$) at low temperatures (150-400 K), a mix of low-velocity magnetic C-shocks ($\sim$5-10 \kms) and faster (15-25 \kms) J-shocks were needed. 

 Very broad spectral linewidths have been observed in the Stephan's Quintet filament. \citet{Guillard2012} used the IRAM 30m telescope to detect CO emission from several regions in the main filament and the bridge, and revealed line profiles of several hundred \kms~ in at least three separate kinematic components. In addition to gas detected at the systemic velocity of the intruder NGC 7318b (5774 \kms), and that of the remaining group members (6600 \kms), broad lines of CO emission were also detected at intermediate velocities, suggesting that molecular gas has formed out of diffuse shocked gas. Recent UV spectroscopic observations with the Hubble Space Telescope's (HST) Cosmic Origins Spectrograph (COS) targeted five regions along the main filament and in the bridge, and detected extremely broad Ly$\alpha$ emission (FWHM exceeding 1500 \kms) over small sampled regions of $\sim$1 kpc scale \citep{Guillard2022}. The lines were even broader in some cases than those seen in the [CII] line \citep{Appleton2013} using {\it Herschel}, suggesting some resonant scattering of UV photons. The detection of broad-line, powerful Ly$\alpha$ emission, broad [CII] neutral gas, and broad CO emission strongly supports the picture of a highly turbulent medium containing many gas phases resulting from a turbulent cascade of energy from the large to the small scales. 
 
Although the Early Release Observations (ERO) made by JWST of Stephan's Quintet cover the entire inner group (including NGC 7319, NGC 7318a/b, NGC 7317 and the foreground galaxy NGC 7320), this paper will concentrate on the main shocked filament and bridge that were observed by {\it Spitzer}~in spectral mapping mode \citep{Cluver2010,Appleton2017} using the InfraRed Spectrograph (IRS; \citealt{Houck2004}). As shown in Figure~\ref{fig:intro}, this included  the main north-south H$_2$ filament between NGC 7318b/a and NGC 7319, as well as the H$_2$ bridge connecting NGC 7319 to the main filament. It is known that much of the gas, including HI, ionized gas and molecular gas lies {\it outside} the main body of the galaxies. Although the previous IRS observations of the mid-IR lines covered much of this gas, it did not cover all.  The Long-low IRS coverage included all of the area shown in Figure~\ref{fig:intro}, but IRS Short-low coverage  was restricted to only the the main H$_2$ filament and part of the H$_2$ bridge (see \S~\ref{spitzer}).  

Data from the JWST and ALMA allow us to directly compare, for the first time at comparable resolution, the distribution of warm 0-0S(3) H$_2$ (as we will demonstrate via the F1000W MIRI Band) to cooler molecular hydrogen mapped through the low-J CO(2-1) line, as well as ionized gas emission (H$\alpha$ with  WFC3 HST, and Pa$\alpha$ with NIRCam F200W). Very hot gas, which forms an important IGM component in Stephan's Quintet is detected in X-rays throughout Stephan's Quintet, and especially in the main shock.  Its distribution cannot be compared in detail to our current observation, because even the deepest {\it Chandra} images \citep{OSullivan2009} do not detect enough photons at arcsecond scales to allow meaningful comparison. 

These observations will help us gain an understanding of how turbulent energy, driven mainly by the intruder galaxy, is dissipated from large driving scales to smaller dissipation scales, and how this affects the cooling of the gas through different gas phases and temperature regimes. In particular, how can we explain so much energy flowing out through low-velocity shocks needed to produce the high radiant line luminosity from warm molecular hydrogen discovered by {\it Spitzer}.  Although we do not  expect to probe down to the smallest dissipation scales, we will show that we already see differences in the distribution of different thermal phases in the IGM at $\sim$150~pc scales probed by ALMA and JWST.

The current paper will not try to address all aspects of the extended emission detected by JWST, but will primarily concentrate on two main objectives. Firstly, we compare the JWST MIRI images with the Spitzer IRS spectra and spectral images, to identify the dominant features present in the three MIRI filters (F770W, F1000W and F1500W). Secondly, we compare the distribution of warm molecular gas, cold molecular gas and ionized gas in three contrasting regions within the main IGM between NGC 7319 and NGC 7318b/a. The three regions we have chosen to emphasize in this paper were observed with ALMA in CO(2-1). The primary beams (fields-of-view) of the three ALMA pointings and their associated field numbers are also identified on Figure~\ref{fig:intro}.\footnote{Other fields were originally proposed in the ALMA proposal but only these three regions were actually observed}. We will the regions in more detail in \S 4.  Although we touch on the sparsity of recent star formation in two out of three of the studied fields, we leave a full discussion of the star formation properties and cluster formation to a later paper.

In \S~2 we will describe observations made with JWST, HST, the Atacama Large Millimeter Array (ALMA) and the Combined Array for Research in Millimeter-wave Astronomy (CARMA). 
Given that we do not yet have spectroscopy of the IGM in Stephan's Quintet, we present, in \S~3 a discussion of what each of the MIRI and NIRCam images is likely to contain by comparing the JWST images with {\it Spitzer} spectral maps of especially warm H$_2$ and PAH bands. In \S~4 we present the results of our zoomed-in study of three major emission complexes in the intergalactic medium observed at high resolution by ALMA in the CO (2-1) line. This allows us to synthesize from JWST, ALMA, and HST, the best possible information we have about the nature of the warm and cold molecular gas, the ionized gas and the formation of star clusters, all at subarcsec resolution. \S~4 also includes a discussion of the relative fraction of warm and cold molecular gas obtained for the regions using both the JWST and ALMA data. In \S~5 we discuss the results in the context of our understanding of turbulence and shocks which we believe largely dominate the gas dynamics of this multi-phase IGM. In \S~6 we present our conclusions.   

We assume in this paper a distance to Stephan's Quintet of 94 Mpc (for H$_0$ = 70 \kms Mpc$^{-1}$ and an assumed group heliocentric systemic velocity of 6600 \kms) for  consistency with previous work (e. g. \citealt{Appleton2017,Guillard2022}). At this distance 1 arcsec corresponds to a linear scale of 456 pc. The overall scale of the giant north-south intergalactic filament in Stephan's Quintet is 47 kpc.  

\section{Observations}\label{sec:observations}
\subsection{JWST images}

The first description of the NIRCam and MIRI ERO observations, and the choice of Stephan's Quintet as a target (PID~2732) is presented in \citet{Pontoppidan2022}. 
The NIRCam images were taken in 6 filters (F090W/F277W, F150W/F356W, F200W/F444W), with an integration time of approximately 20~min for each band. The FULLBOX 5-points "5TIGHT" dither pattern was used, resulting in a large rectangle mosaic of 3 pairs of dithered tiles, covering $6.3 \times 7.3$~arcmin$^2$.
The MIRI image covers a much smaller field of view than the NIRCam mosaic, i.e. the central galaxies, NGC7318a/b, NGC7319 (a Seyfert 2 galaxy), and NGC7320 (foreground galaxy), using 4 tiles in three MIRI bands, F770W, F1000W, and F1500W. The integration times were about 22~min per filter. The large cycling 8-points dither pattern was used to maximize the spatial coverage of a single tile.

We worked with the level 2b images retrieved from MAST\footnote{The data were obtained from the Mikulski Archive for Space Telescopes at the Space Telescope Science Institute, which is operated by the Association of Universities for Research in Astronomy, Inc., under NASA contract NAS 5-03127 for JWST.}.

\subsection{ALMA Observations}
\label{sec:alma}

Observations of CO(2-1) in the three fields of Stephan's Quintet were made with the ALMA 12m array in a mosaic observing mode on 21 July 2015 (ID: 2015.1.000241; PI: P.\,Guillard). The total integration time for the mosaic was 35 minutes. The spectral setup consisted of four spectral windows of 1.875 GHz with 3.9 MHz channels. One of the spectral windows was centred on the redshifted CO(2-1) ($\nu_{\rm rest}$\,=\,230.538\,GHz), while the remaining three spectral windows covered line-free continuum emission. The standard ALMA calibration plan included bandpass, phase, and flux calibration.

The data were calibrated with the scriptForPI.py calibration script that was included with the archival data, using the ALMA calibration pipeline version r32491 that is included with the Common Astronomy Software Applications (CASA) version 4.3.1 \citep{Casa2022}. After calibration, the continuum emission was subtracted in the ($u$,$v$)-domain by fitting a straight line to the line-free channels. The line-data where subsequently imaged using the CLEAN algorithm to produce data cubes with 20 km\,s$^{-1}$ channels and a resolution of 0.36$^{\prime\prime}$\,$\times$\,0.21$^{\prime\prime}$ (PA\,$\sim$\,10$^{\circ}$). 
The root-mean-square (rms) noise in these data cubes is 0.5 mJy\,beam$^{-1}$\,chan$^{-1}$. 

A primary beam correction was applied to produce accurate flux measurements. The half-power beam width of the primary beam is 22.8 arcsec (10.4 kpc). Total intensity maps of the CO(2-1) emission where made by integrating the signal across the channels where line emission was detected.

\subsection{CARMA CO Observations}
Observations of Stephan's Quintet were made with CARMA on 1 Aug 2010 (program c0593-PI. K. Alatalo) using 15 antennae in the CO (1-0) line. The original data was processed with MIRIAD \citep{Sault1995}. Data reduction and imaging were done in identical fashion to \citet{Alatalo2013}. The original data was cleaned \citep{Hogbom1974}, which resulted in a restored synthesized beam of 4.1 x 3.3 arcsec$^2$ (1.9 x 1.5 kpc). Velocities from extracted spectra have been converted to heliocentric velocities assuming a shift of -11~\kms~ from lsrk to heliocentric. Velocities are all quoted using the optical definition of velocity.

\subsection{HST Archival Images}
We de-archived calibrated images from the HST MAST archive for filters F665W and F336W  with the WFC3/UV instrument. For F665N, the image encompasses the H$\alpha$ line, and includes data based on five dithered observations, each taken with integration times of 5200s each in July and August 2009, and was obtained as part of a WFC3 Early Release Observations campaign (SM4/ERO). For F336W, the observations were obtained as part of proposal id 12301 (P. I. =  S. Gallagher) on 2011-10-27,  with a total integration time of 14474 s.
The resolution of the WFC3/UV observations is comparable with that achieved by NIRCam (0.05-0.06 arcsec). For F665W and F336W, the expected FWHM of a point source is 0.07 arcsec, and 0.075 arcsec respectively, corresponding to $\sim$30 pc at D = 98 Mpc.

\subsection{WCS Alignment of the JWST Images}

The observations were obtained from the JWST and HST MAST archive, and most of the images required small adjustments to the WCS coordinates for proper alignment. We initially used GAIA DR3 stars to align one image (the HST WFC3 F665N) to the DR3 system using the STScI software package WCSTweak. The NIRCam images were also similarly aligned. However, for MIRI,  the same method did not produce good results, probably because of the smaller number of stars  used for alignment. We used a refinement scheme to solve this problem as we need good alignment between all the images. This was especially important for one of the target fields that required continuum subtraction of the F1000W and F770W images with F1500W. First we created small subimages of the F1000W, F770W and F1500W images around the Field 6 area. The subimage contained at least 3 GAIA DR3 stars close to the target of interest. These subimages were then aligned more carefully by making small pixel-level shifts in each image to bring the subimages into close alignment with each other and the positions of the GAIA stars.  We applied the same method to Field 5 and 4 (see \S~\ref{sec:threereg}). The final results produced local MIRI images which aligned to better than 1 MIRI pixel (0.3 arcsec) over the small scale of the extracted regions. The results were validated where the CO ALMA images aligned with clearly related features.

Following \citet{Papadopoulos2008}, the uncertainty in the absolute astrometry of the ALMA data is $\delta\theta_{\rm bas}$\,=\,($\delta$\textit{{\textbf B}}\,$\cdot$\,$\Delta$\textit{{\textbf k}})/\textit{{\textbf B}}\,$\approx$\,($\delta\phi_{\rm bas}$/2$\pi$)$\langle\Theta_{\rm beam}\rangle$. The phase error $\delta\phi_{\rm bas}$\,$\sim$\,(2$\pi$/$\lambda$)($\delta$\textit{{\textbf B}}\,$\cdot$\,$\delta$\textit{{\textbf k}}), with $|\delta$\textit{{\textbf B}}$|$\,$\sim$\,0.1mm the assumed typical error in the baseline length, $|\delta$\textit{{\textbf k}}$|$\,=\,4.26$^{\circ}$ the distance to the phase calibrator J2216+3518, $\lambda$\,$\sim$\,1.33 mm the wavelength of the redshifted CO(2-1) emission, and $\Theta_{\rm beam}$\,$\sim$\,0.36$^{\prime\prime}$ the major axis of the synthesized beam. This results in $\delta\theta_{\rm bas}$\,$\sim$\,0.02$^{\prime\prime}$, which means that the uncertainty in the absolute astrometry of the ALMA data is small compared to that of the HST and JWST data\footnote{See also: https://help.almascience.org/kb/articles/what-is-the-absolute-astrometric-accuracy-of-alma}.

\section{Expectation of the emission observed through the MIRI and NIRCam filters in three study regions}

\begin{figure*}
\includegraphics[width=0.99\textwidth]{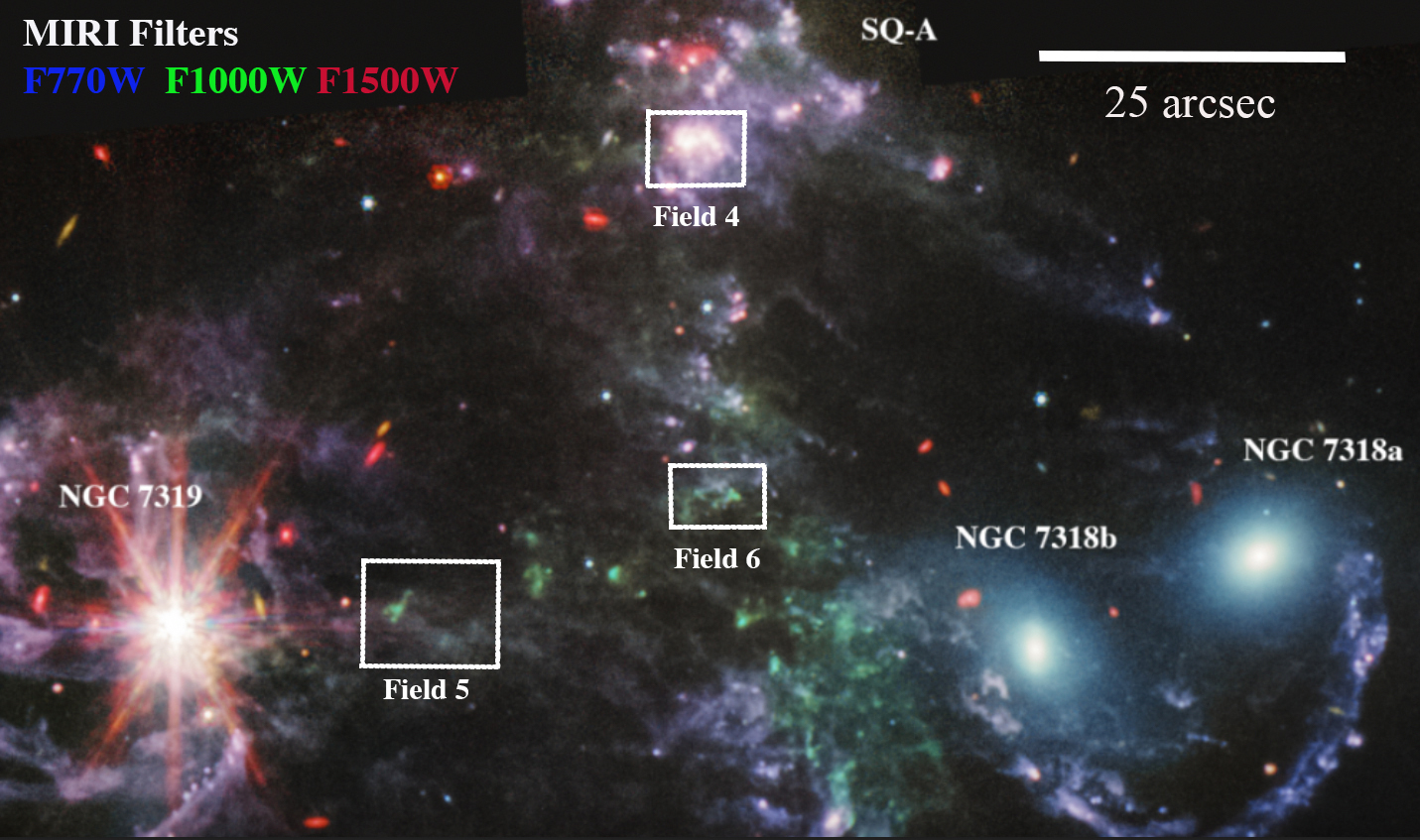}
\caption{Stephan's Quintet: A false-color representation of the MIRI F770W, F1000W and F1500W JWST image of the inner Stephan's Quintet, showing the three regions that are highlighted in the discussion that are detected near the primary beam centers in the ALMA CO (2-1) emission line observations. Field 6 is representative of the center of the main filament where previous observations with Spitzer show that the molecular hydrogen has the highest temperature. Field 5 lies in the bridge region between NGC 7319 and the main shock identified in \ref{fig:intro}. Finally, in Field 4, a bright region in the SQ-A star forming region is highlighted (see text). } 
\label{fig:fieldregions}
\end{figure*}

This section will explore our expectations for what emission processes are likely to contribute to the MIRI images based on previous {\it Spitzer} spectral mapping results. 
This paper will concentrates mainly on three regions which lie near the centers of the ALMA CO (2-1) observing fields. The regions are typical of the diversity of mid-IR spectral properties encountered in the IGM of SQ in general, and are the focus of this paper. They are highlighted in Figure~\ref{fig:fieldregions} as boxes (dotted lines) superimposed on one of the publicly released ERO images of the inner part of the group.  Field 6, which we will show is dominated by 10$\mu$m H$_2$ emission (green color in this figure), was targeted by HST COS \citep{Guillard2022} and shows extremely broad Ly$\alpha$ emission, broad [CII]158$\mu$m emission and contains some of the the warmest H$_2$ observed by {\it Spitzer} (e. g. \citealt{Cluver2010}). It is representative of a shock-heated region in the main N/S filament.  Field 5 is another H$_2$-dominated region which lies outside the main north/south H$_2$ filament, and observations at other wavelengths (e. g.  single disk CO observations by \citet{Guillard2012}, and {\it Herschel} [CII] observations of \citealt{Appleton2013}) show that this gas is also highly turbulent. It forms part of the bridge of H$_2$ emission seen by {\it Spitzer}. Field 4, by contrast, lies in a well-studied extragalactic star forming region, SQ-A (or the Northern Star Forming region) and has been studied extensively at optical wavelengths \citep{Gallagher2001,Xu2003}. Its star forming nature can be inferred from its bluish-white appearance in the false color map of Figure~\ref{fig:fieldregions} due to a combination of weaker H$_2$, strong PAH 7.7$\mu$m emission and dust continuum (See \citealt{Cluver2010,Appleton2017}.

\subsection {Comparison of JWST MIRI images with Spitzer IRS spectral mapping results}
\label{spitzer}

 Under conditions found in the disks of normal galaxies, strong 7.7 $\mu$m and 8.6 $\mu$m Polycyclic Aromatic Hydrocarbon (PAH) features are expected to dominate the MIRI F770W filter for low-redshift galaxies (e. g. \citealt{Smith2007}). In Stephan's Quintet  (\citealt{Appleton2006,Guillard2009,Cluver2010,Appleton2017}), much of the mid-IR spectrum of the IGM is dominated by strong pure rotational H$_2$ lines, and emission from [SiII]$\lambda$34.8$\mu$m. Unlike normal star formation regions, the majority of the main IGM filament in Stephan's Quintet exhibits very weak 7.7$\mu$m PAH complex emission and weak nebular lines like [SIII]$\lambda$18.7,33.5$\mu$m~\citep{Cluver2010}. This is the result of low star formation rates in the main filament (measured to be between 0.05 and 0.08 M$_{\odot}$ yr$^{-1}$; \citealt{Cluver2010}), and emission lines more typical of fast shocks than HII regions (e. g. \citealt{Konstantopoulos2014}). Examples of these H$_2$ dominated spectra have been presented in Figures 8, 13 and 14 of \citet{Cluver2010}, and Figure 15 and 16 of \citet{Appleton2017}. The  {\it Spitzer} IRS spectra showed that only a small minority of places  along the main filament (including parts of the SQ-A region) are spectra found with line flux ratios more typical of star forming regions. 

\begin{figure*}
\includegraphics[width=0.96\textwidth]{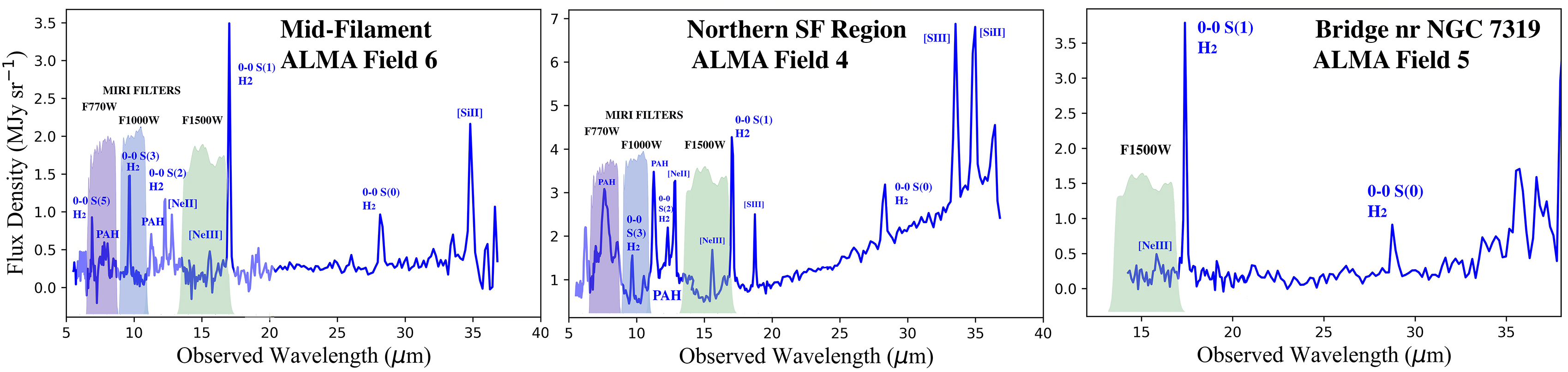}
\caption{Spitzer IRS spectra of three 5.5 x 5.5 arcsec$^2$ (2.5 x 2.5 kpc$^2$) regions centered on the three fields studied at high resolution in CO (2-1) emission with ALMA and at 7.7 and 10~$\mu$m  with JWST MIRI imaging (see Figure~\ref{fig:Field6}). The MIRI broad band filter transmissions are superimposed to emphasize that the F1000W MIRI filter detects almost exclusively emission from the S(3) line and that the 7.7~$\mu$m band captures a combination of the weak 7.7$mu$m PAH complex and emission from S(5) H$_2$. The 15~$\mu$m MIRI bandpass detects mainly faint dust continuum and some (weak) [NeIII] emission. ALMA Field 5 was not covered by the Short-Lo IRS module. In regions of strong star formation the F15000W filter can also be contaminated with the wings of the 17$\mu$m PAH complex \citep{Smith2007}}. 
\label{fig:IRSSpec}
\end{figure*}

 In Figure~\ref{fig:IRSSpec}a-c, we compare the filter transmission band-passes for the MIRI imaging bands (F770W, F1000W and F1500W) directly with the Spitzer IRS spectra taken within Fields 4, 5 and 6 (Figure~\ref{fig:intro}).  The MIRI filter F770W is likely to contain emission primarily from the 7.7$\mu$m PAH complex in regions where star formation dominates, but in other regions, especially those with strong H$_2$ emission, the 0-0~S(5) can also potentially contribute, as in Figure~\ref{fig:IRSSpec}a.  The F1000W filter, on the other hand, almost exclusively contains line emission from the S(3) mid-IR pure rotational H$_2$ line (See also Figures 15e and h of \citealt{Appleton2017}).  Based on the presented spectra, we expect the MIRI F1500W filter to be dominated by faint dust continuum from the general IGM, as well as from warm dust from embedded star clusters, and sometimes the extended wing of the broad 17$\mu$m PAH complex \citep{Smith2007}--see center panel of Figure~\ref{fig:IRSSpec}.  

\begin{figure*}
\includegraphics[width=0.98\textwidth]{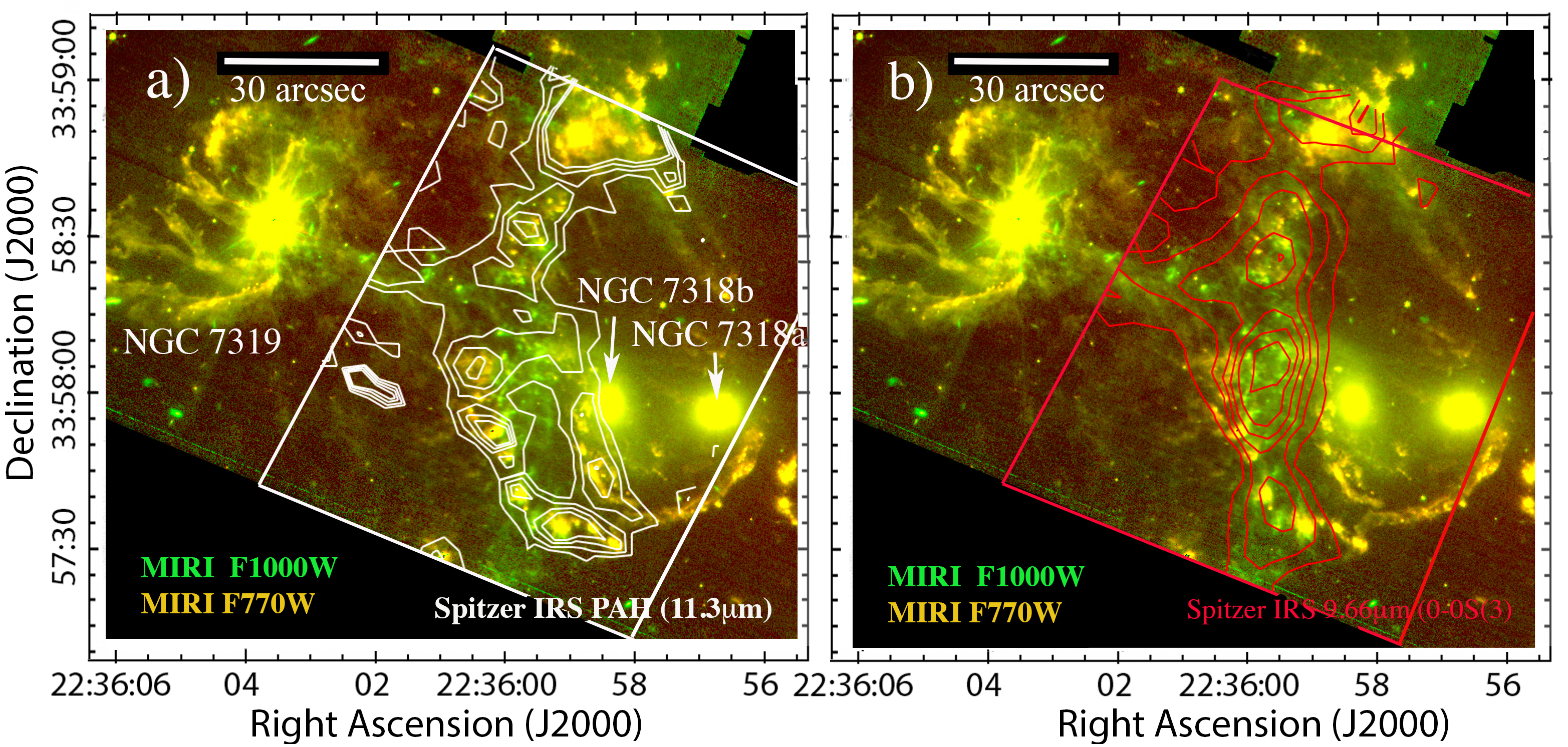}
\caption{Stephan's Quintet: A false-color representation of the MIRI F770W and F1000W JWST images with contours of rest-frame (a) 11.3$\mu$m PAH and (b) 9.66$\mu$m emission derived from spectral cubes obtained from the Spitzer IRS Short-low mapping of Stephan's Quintet. The white and red solid lines show the extent of the IRS spectral mapping (See \citealt{Cluver2010}). The figure demonstrates how the S(3) line isolated in the Spitzer cube follows closely the 10$\mu$m MIRI emission (green), and the PAH features follow the MIRI 7.7$\mu$m image which contains contributions from the 7.7$\mu$m PAH and H$_2$ emission (See also Appleton et al. 2017, and Figure~\ref{fig:IRSSpec} of this paper).} \label{fig:IRScomp}
\end{figure*}

A clean separation between the mainly PAH dominated emission in the F770W MIRI filter and the 0-0S(3)H$_2$-dominated F1000W band can be seen in many regions along the main IGM filament and in the bridge region in Figure~\ref{fig:IRScomp}. Here we overlay the contour maps obtained by isolating a PAH band\footnote{We used the IRS 11.3$\mu$m PAH rather than the 7.7$\mu$m PAH, because the signal to noise ratio was much better in this PAH feature in the {\it Spitzer} data.} from the IRS spectral map from {\it Spitzer} in Figure~\ref{fig:IRScomp}a (white contours) with a two-color image of the JWST MIRI F1000W (green) and the MIRI F770W (orange).  In Figure~\ref{fig:IRScomp}b we overlay the 0-0 S(3) H$_2$ contours (red) over the same image. The overlays show that the white contours of PAH emission from Spitzer follow quite closely the brighter emission in the F770W dominated regions, whereas the red contours of H$_2$ follow closely the green emission originating mainly from the F1000W filter. This figure demonstrates how the 10$\mu$m MIRI filter, as we expected from the individual IRS specta, detects preferentially  H$_2$ emission, while the F770W filter detects those regions with dominated by PAH emission. As shown in \citet{Cluver2010} the PAH emission is relatively faint in the main shock, but this figure emphasise those regions with stronger PAH emission.   This near-isolation of the H$_2$ emission in many regions of the IGM is a result of the relatively narrow width of the F1000W filter, and the lack of other lines or bands that can seriously contaminate it for the redshift of Stephan's Quintet.

\subsection{Dust continuum}

Observations using the F1500W MIRI filter are expected to detect mainly warm dust emission, similar to that mapped by {\it Spitzer} at 24$\mu$m with MIPS \citep[see][]{Guillard2010}. However, in the SQ-A region (Figure~\ref{fig:IRSSpec}b), the dust continuum is also contaminated by [NeIII] emission. As noted above the very strong 0-0S(1) line is, fortuitously, redshifted out of the F1500W band. 

Before discussing results for each of the zoomed-in regions in the next section, we mention the removal of faint dust emission from F1000W and F770W at the center of Field 6 (near the center of the main IGM filament in SQ).   In this field, faint dust continuum in the F1500W filter was seen in the vicinity of the structures we are interested in (see \S~4). We therefore removed the faint dust continuum from both the F1000W and F770W images, using the {\it Spitzer} IRS spectrum of the region as a guide. First we created small sub-images of the field in all three bands centered on the main structure of interest and corrected them for small offsets in the WCS as discussed in \S \ref{sec:observations}. After removing a median sky background from each subimage, we then subtracted the F1500W image from both the F770W and F1000W images, scaling the continuum by a factor of 0.9 to account from a slight rise in continuum seen between 7.7, 10 and 15 $\mu$m in the IRS spectrum of the region.  This allowed us to better define the distribution the H$_2$ and PAH features. The effect was quite small (the dust emission was quite weak compared with the emission seen in the 7.7 and 10$\mu$m bands) on the resulting fluxes of the F1000W and F770W band. Without spectroscopy of the region on the same spatial scale as the observed features, this method is necessarily imperfect, and where we estimate the fluxes for the features observed in Field 6 in the next section, we include larger uncertainties to take this into account.  For the sub-region near the center of Field 5, no obvious dust structure was seen in the F1500W band near the structures of interest, and so we did not attempt to continuum subtract the H$_2$ and PAH-dominated emission from the F1000W and F770W images. For Field 4 --- centered in the SQ-A star forming region --- our attempts at subtracting a scaled version of the dust dust continuum led to a severe over-subtraction of the emission in the F770W and F1000W bands for most reasonable dust scale factors. This is likely due to a combination of factors.  Firstly at that position (see Figure~\ref{fig:IRSSpec}b), we have shown that the 15$\mu$m band shows emission from not only dust continuum, but also relatively strong [NeIII], as well as broad PAH-band emission from the 17$\mu$m "plateau" complex (see \citealt{Smith2007}). These additional contaminants make it difficult to isolate the warm dust component in the 15$\mu$m image. Therefore, for this star formation dominated field, we conclude that the F1000W and F770W MIRI filters are hopelessly contaminated by dust emission in way that cannot easily be corrected without high resolution spectroscopic data from the MIRI MRS. This significantly limits our discussion of the molecular hydrogen properties in Field 4. For that region we limit our discussion of the warm H$_2$ properties determined from the {\it Spitzer} IRS observations.

 \subsection{Near-IR images}
 We also present images in this paper obtained through the NIRCam F200W filter. F150W provides a near-IR continuum band relatively clear of emission lines at the redshift of Stephan's Quintet (0.022 $<$ z $<$ 0.025), with the well-known near-infrared [\ion{Fe}{2}] lines falling just blueward or redward of the filter transmission.  The bandpass of the F200W filter is sensitive to emission from the hydrogen recombination line Pa$\alpha$, and starlight.  Diffuse emission from this line becomes obvious when we compare some of those images with those obtained in the F150W NIRCam filter, and by comparison with the HST H$\alpha$ images. 

JWST/MIRI, which is able to detect warm H$_2$ through the pure rotational transitions, can significantly improve our knowledge of how energy is dissipated in the IGM down to the 0.3 arcsec scales of giant molecular cloud complexes ($\sim$140 pc). NIRCam, with its ability to probe down to scales of 0.05-0.06 arcsec (23-27 pc at 1.5 and 2$\mu$m respectively) can probe ionized gas and stellar associations and clusters at scales at which sporadic star formation along the main filament is observed.

\section{Results}
\label{sec:threereg}

\begin{figure*}
\includegraphics[width=0.96\textwidth]{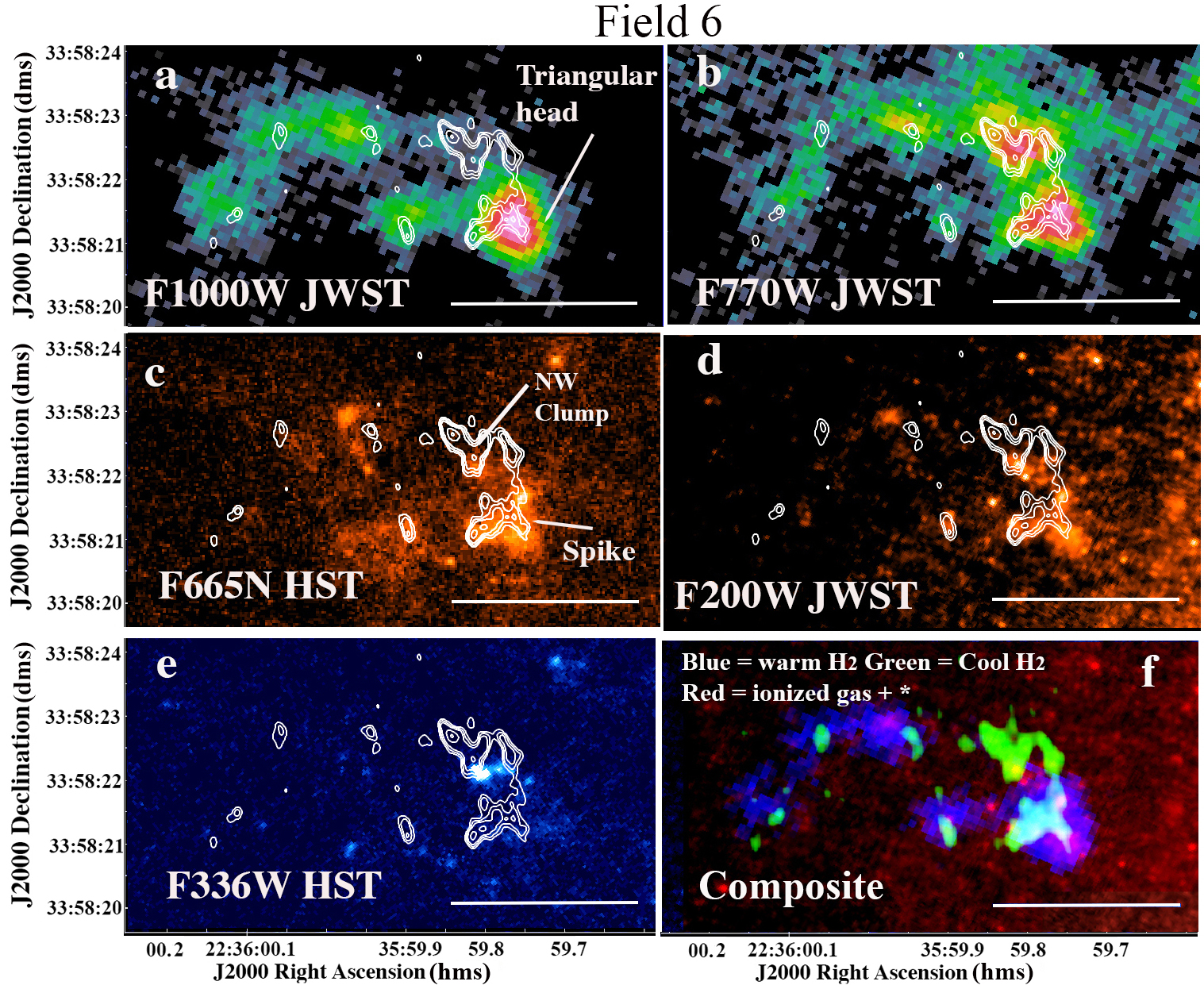}
\caption{A zoom-in on the center of the ALMA Field 6 (see Figure~\ref{fig:intro}) comparing (white) contours of CO (2-1) emission to (a) continuum-subtracted pure warm 0-0S(3) H$_2$, (b) continuum-subtracted warm 0-0 S(5) H$_2$ plus 7.7$\mu$m PAH complex emission, (c) H$\alpha$ emission from F665N WFC3 HST, (d) mix of IR (red compact) stellar sources and Pa$\alpha$ emission in the NIRCam f200w filter, (e) blue star clusters from F336W WFC3/UV HST \citep{Gallagher2001}. In (f) we show a composite RGB color representation of the F1000W (mainly warm H$_2$), ALMA CO (2-1) emission (cool H$_2$), and the F200W (Pa$\alpha$ plus star clusters) images. For clarity, this sub-figure does not overlay the CO contours. The MIRI identifications of the contributing lines/PAH bands are clear from the Spitzer IRS spectra of a region 5.5 x 5.5 arcsec$^2$ (2.5 x 2.5 kpc$^2$) in Figure~\ref{fig:IRSSpec}a, which spatially covers this entire region. All images were carefully aligned using GAIA DR3 stars close to the region shown, leading to relative uncertainties in astrometry of $\sim$ 0.15 arcsec. The F770W and F1000W JWST images were smoothed to the same resolution as the F1500W image before subtracting the carefully-aligned dust continuum. The white scale bar is 3 arcsec (1.4 kpc) in length. A more detailed description of the CO emission and its kinematics is given in Figure~\ref{fig:FigureA1}.} \label{fig:Field6}
\end{figure*}

\begin{figure*}
\includegraphics[width=0.99\textwidth]{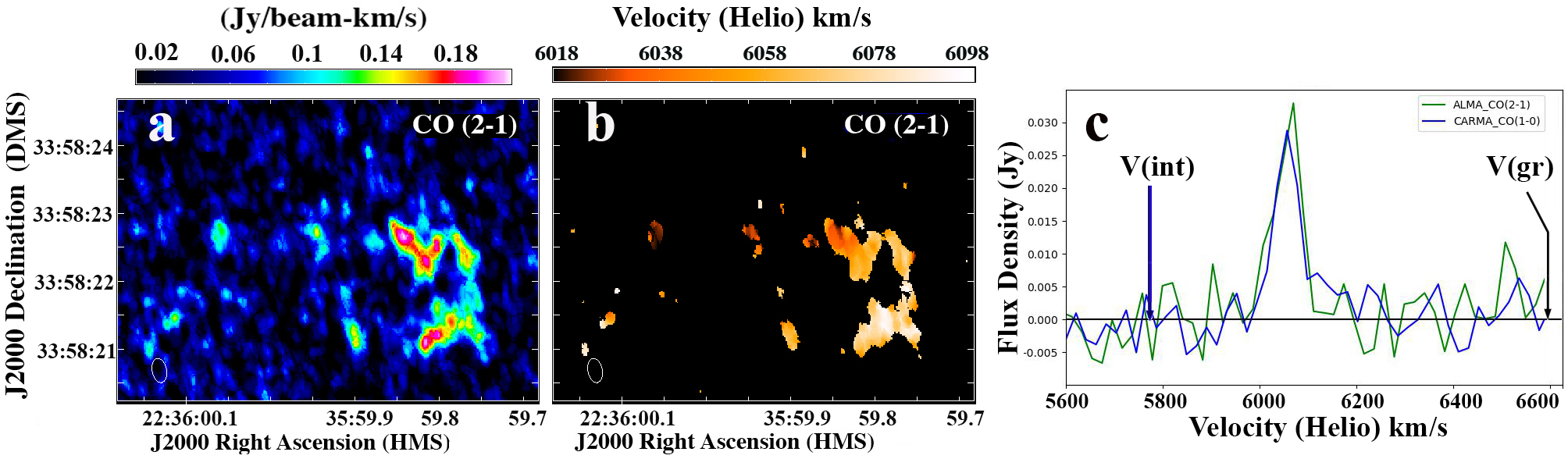}
\caption{ALMA Field 6 region (labeled in Figure~\ref{fig:fieldregions}): (a) CO (2-1) integrated emission, (b) the intensity-weight mean velocity map of the same region, and (c) the ALMA CO (2-1) and CARMA CO (1-0) spectrum of the region 5 x 5 arcsec (2.5 x 2.5 kpc$^2$) centered on same region. Effective synthesized beam-shapes (0.36 x 0.21 arcsec$^2$ = 164 x 96 pc$^2$ FWHM) are shown graphically in the left-hand corner of each figure. Arrows indicate the radial velocity of the intruder V(int) and the barycentric velocity of the main group (V(gr). } \label{fig:Field6_moments}
\end{figure*}


\begin{figure*}
\includegraphics[width=0.98\textwidth]{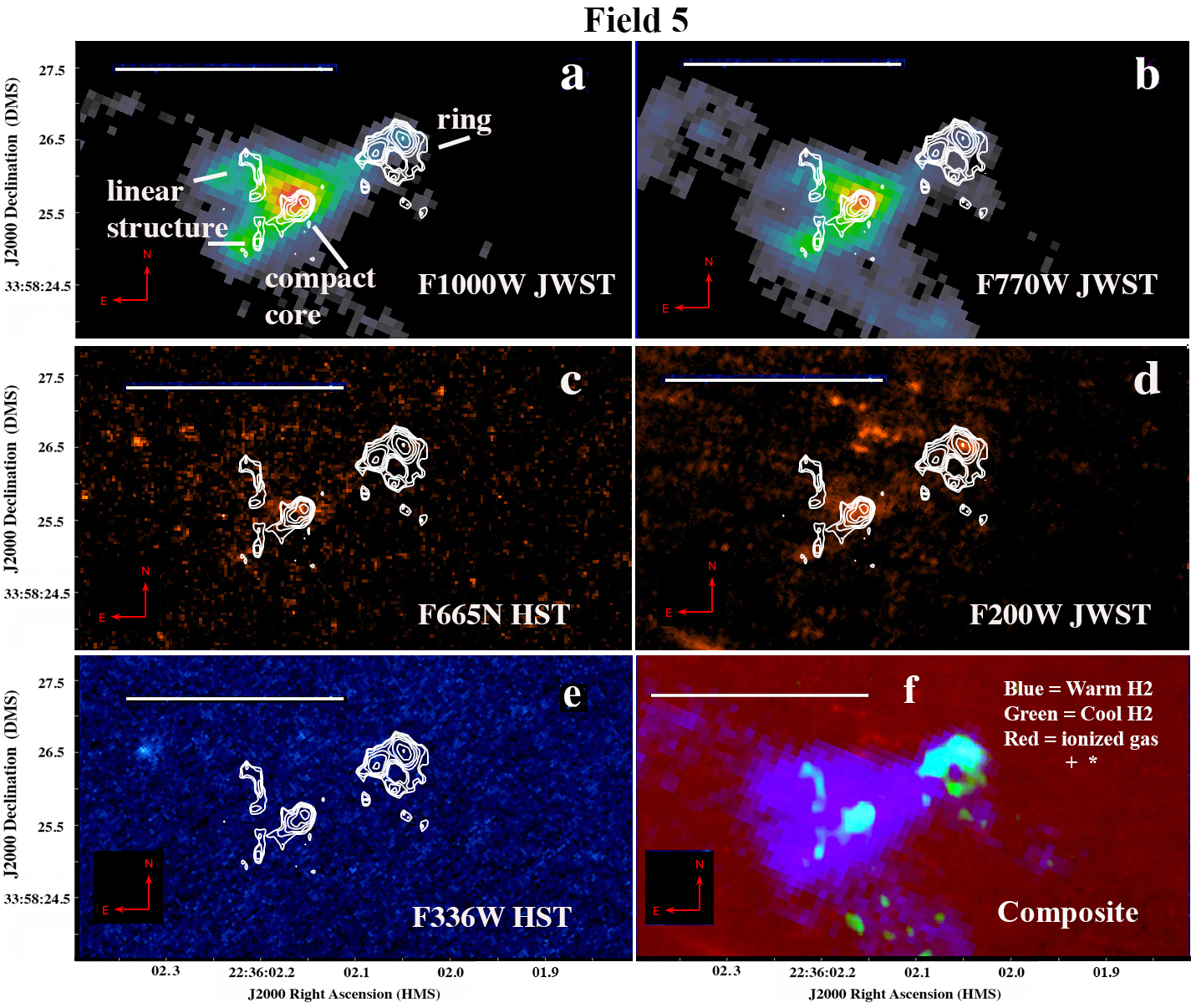}
\caption{A zoom-in on the center of the ALMA Field 5 (see Figure~\ref{fig:intro}) comparing (white) contours of CO (2-1) emission to (a) pure warm 0-0S(3) H$_2$,  (b) warm 0-0 S(5) H$_2$ plus 7.7$\mu$m PAH complex emission, (c) H$\alpha$ emission from F665N WFC3 HST, (d) mix of IR (red compact) stellar sources and Pa$\alpha$ emission in the NIRCam F200W filter, (e) blue star clusters from F336W WFC3/UV HST. In (f) we show a composite RGB color representation of the F1000W (mainly warm H$_2$), ALMA CO (2-1) emission (cool H$_2$), and the F200W (Pa$\alpha$ plus star clusters) images. For clarity, this sub-figure does not overlay the CO contours. The MIRI identifications of the contributing lines/PAH bands are likely from the Spitzer IRS spectra of a region 5.5 x 5.5 arcsec$^2$ in Figure~\ref{fig:IRSSpec}, which in this case only includes the longer IRS LL wavelengths. All images were carefully aligned using GAIA DR3 stars close to the region shown, leading to relative uncertainties in astrometry of $\sim$ 0.15 arcsec. The F770W and F1000W JWST images were smoothed to the same resolution as the F1500W image before subtracting the dust continuum. The white scale bar is 3 arcsec (1.4 kpc) in length.} \label{fig:Field5}
\end{figure*}

\begin{figure*}
\includegraphics[width=0.96\textwidth]{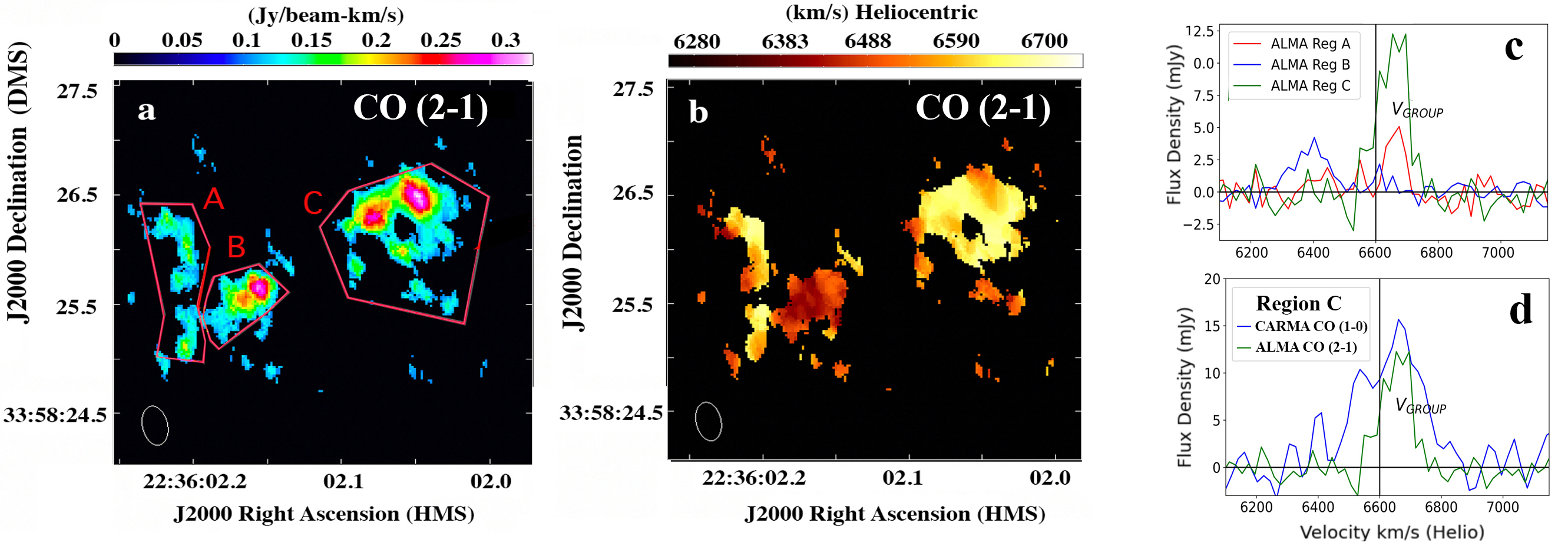}
\caption{ALMA Field 5 region: (a) the CO (2-1) integrated emission at a resolution of 0.36 x 0.21 arcsec$^2$, and (b) the intensity-weight mean velocity maps of the same region, (c) the ALMA CO (2-1) spectra of the three different extracted regions shown as a red polygons in (a). Note the very different systemic velocity of the compact structure B.  In (d) we overplot the CARMA CO (1-0) over a 5 x 5 arcsec$^2$ aperture (blue line), compared with the velocity of the ring C.  The effective ALMA beam shapes (FWHM ellipses) are shown graphically in the bottom left hand corner of the images. The black vertical line on the spectra show the barycentric systemic velocity of the V(group). } \label{fig:Field5_moments}
\end{figure*}

\begin{figure*}
\includegraphics[width=0.99\textwidth]{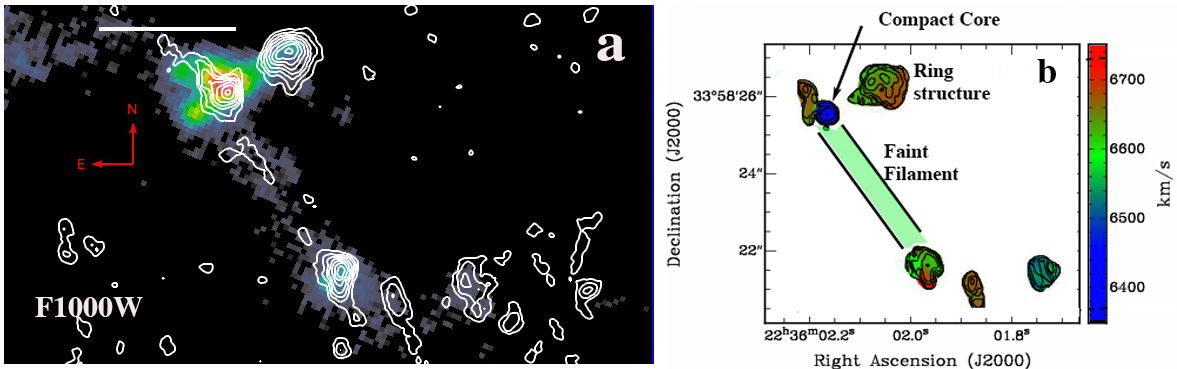}
\caption{Extended CO (2-1) emission for ALMA Field 5 over a larger area than Figure~\ref{fig:Field5}. In (a), the CO emission (white contours) have been smoothed to a resolution 1 x 1 arcsec$^2$ (450 x 450 pc$^2$) and superimposed on the warm 0-0 S(3) H$_2$ detected within the F1000W filter. In (b) the smoothed version of the velocity field is shown for the same region. This emphasizes the unusually low radial-velocity for the compact core compared with the other surrounding regions. The long filament connecting the brighter CO structures is seen in the warm H$_2$. The filament's radial velocity is shown schematically because it is weak, but has the same velocity as the majority of the other structures except the compact core.  The white scale bar in (a) is 3 arcsec (1.4 kpc) in length. } \label{fig:Field5extended}
\end{figure*}

\begin{figure*}
\includegraphics[width=0.99\textwidth]{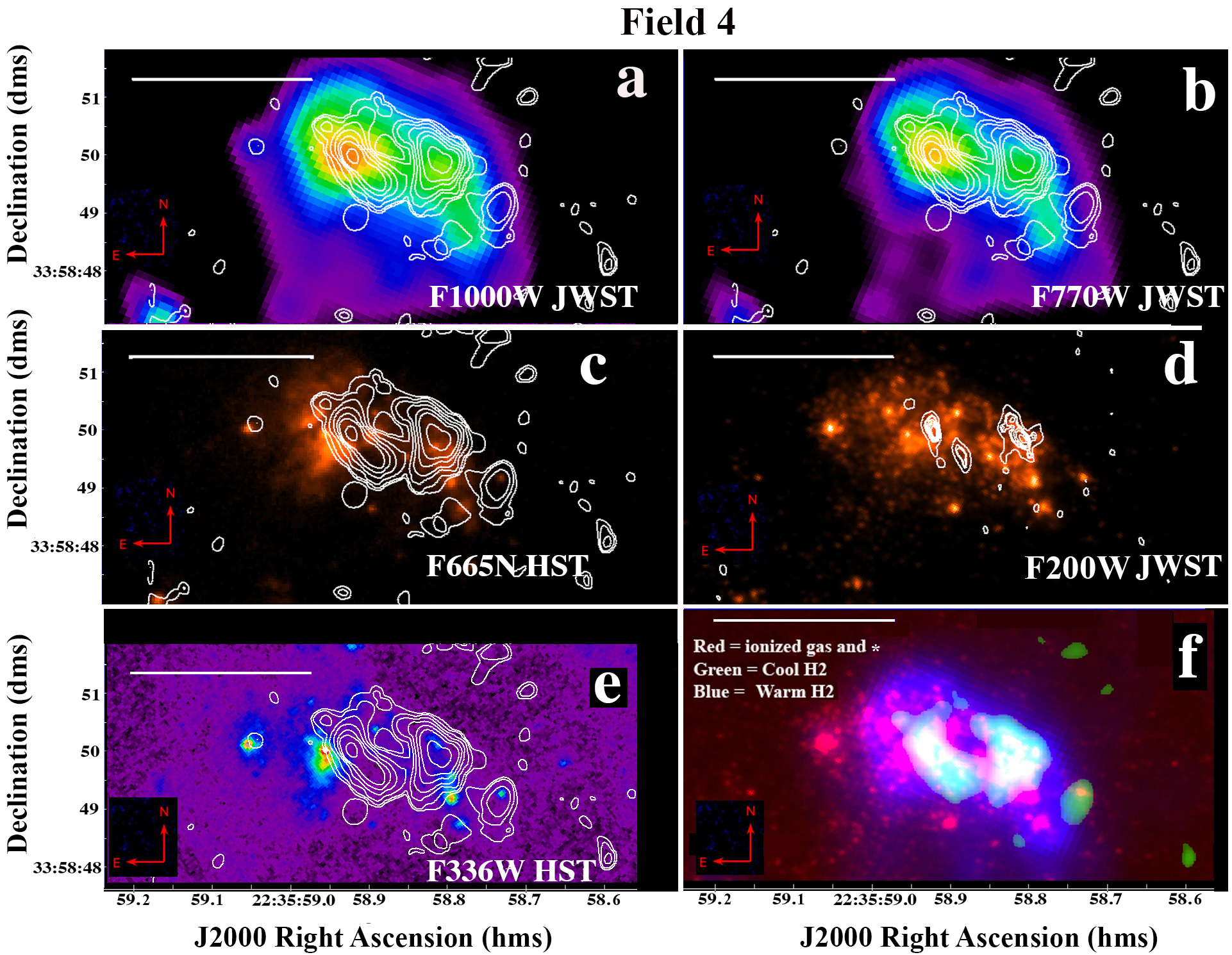}
\caption{CO (2-1) emission (white contours) at the center of ALMA Field 4 (the SQ-A northern star forming region; see Figure~\ref{fig:intro}) at 2 different spatial resolutions. The CO emission in (a), (b), (c), (e) and (f)  have been smoothed to a resolution 1 x 1 arcsec$^2$ (456 x 456 pc$^2$), whereas in (d) the full resolution (0.36 x 0.21 arcsec$^2$ (164 x 96 pc)) CO emission is shown. CO contours are superimposed on (a) the band containing the warm 0-0S(3) H$_2$ line, (b) the band dominated by 7.7$\mu$m PAH complex emission for these northern star forming regions, (c) H$\alpha$ emission from F665N WFC3 HST, (d) mix of IR (red compact) stellar sources and Pa$\alpha$ emission in the NIRCam f200w filter, (e) the blue star clusters from F336W WFC3/UV HST. In (f) we show a composite RGB color representation of the F1000W (mainly warm H$_2$), ALMA CO (2-1) emission (cool H$_2$),  and the F200W (Pa$\alpha$ plus star clusters) images. For clarity, this sub-figure does not overlay the CO contours. Unlike Figure~\ref{fig:Field6} and Figure~\ref{fig:Field5}, because of strong contamination of the F1500W continuum band by [NeIII] emission (see Figure~\ref{fig:IRSSpec}(b)) no continuum subtraction was made for the MIRI images for (a) and (b).  All images were carefully aligned using GAIA DR3 stars close to the region shown, leading to relative uncertainties in astrometry of $\sim$ 0.15 arcsec. } \label{fig:Field4}
\end{figure*}
\begin{figure*}
\includegraphics[width=0.99\textwidth]{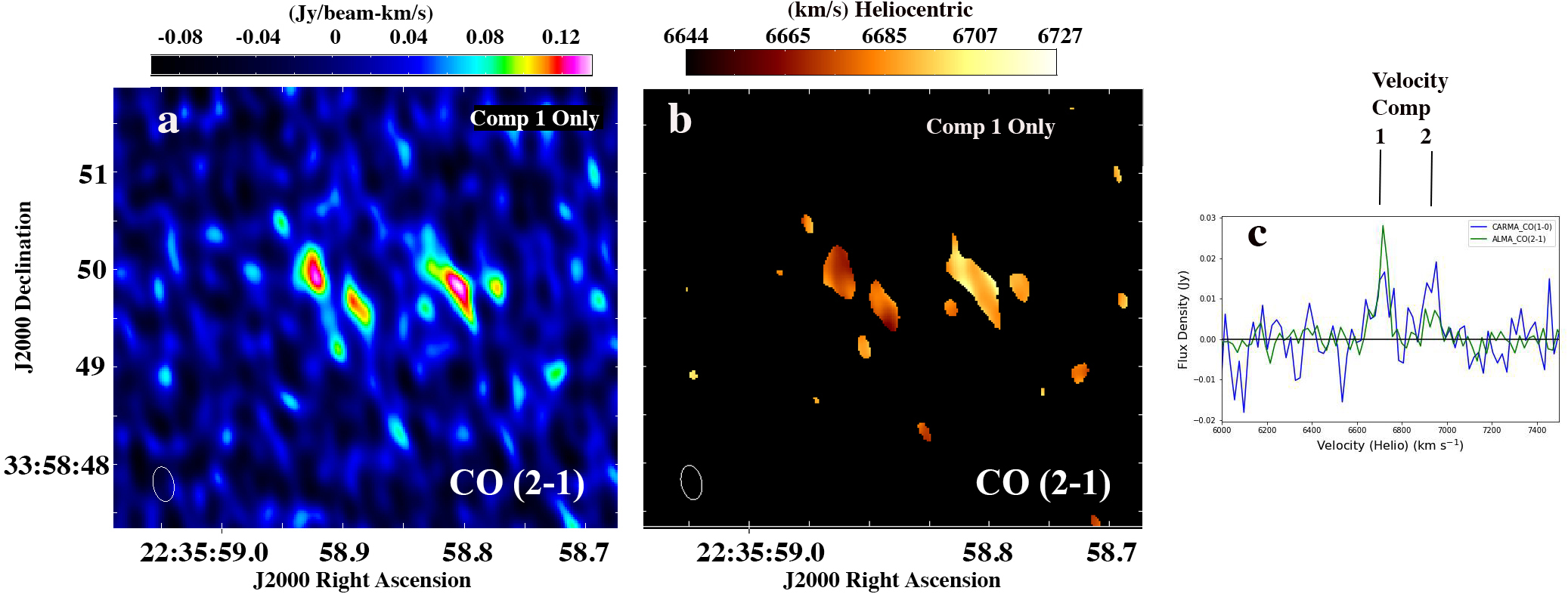}
\caption{ALMA Field 4 in the SQ-A region: (a) CO (2-1) integrated emission associated with the brighter lower-velocity Component 1) in the center of ALMA Field 4 associated with the star clusters shown in Fig.\ref{fig:Field4}, (b) the intensity-weight mean velocity map of the same region, and (c) the ALMA CO (2-1) and CARMA CO (1-0) spectrum of the region 5 x 5 arcsec centered on same field. Note that both ALMA and CARMA show a higher velocity CO component (Comp 2) associated with the region which is faint and diffuse in the ALMA observations. (a) and (b) only show the structure for Comp 1. Effective synthesized beam-shapes (0.36 x 0.21 arcsec$^2$ FWHM) are shown graphically in the left-hand corner or each figure.} \label{fig:Field4_moments}
\end{figure*}

\subsection{Analysis of the three regions}

In this section we explore the comparison of archival HST and JWST MIRI data with the ALMA CO (3-2).

\subsubsection{Field 6: Within the main shock-dominated North/South filament}

In Field 6 (Figure~\ref{fig:Field6}a) we show the integrated CO (2-1) emission from cold molecular gas superimposed on the warm H$_2$ emission. The warm H$_2$ gas (assumed to dominate the F1000W filter from the 0-0 S(3) H$_2$ line) is distributed in an elongated structure extending over 4-5 arcsecs ($\sim$ 2 kpc at D = 94 Mpc) with a triangular-shaped hot-spot to the south-west, and a series of fainter clumps  extending to the east. The overall distribution of warm H$_2$ has the appearance of head-tail structure. The CO emission (white contours) displays a clumpy narrow shell-like structure associated with the warm H$_2$ head, and a scattering of CO clumps along the tail, extending over the same overall extent, but not correlating in detail with the hotspots in the warm H$_2$. 
 
 Based on the published IRS spectra, Figure~\ref{fig:Field6}b is interpreted as a mix of H$_2$ (0-0S(5)) and weak PAH emission. The dominant triangular-shaped head structure seen in the 10$\mu$m image is also strongly represented here.  A second major clump of 7.7  $\mu$m emission is seen to the NW, which is weak at 10$\mu$m. It is plausible that this feature is PAH dominated, whereas the emission from the compact head (which is seen also at 10$\mu$m) may be dominated by warmer than normal H$_2$ which would emit strongly at in the 0-0 S(5) line. Without mid-IR spectroscopy, this cannot be verified.  
 
 This distribution of ionized gas can be inferred from the narrow-band HST F665N filter images (dominated by H$\alpha$ emission)  
 in Figure~\ref{fig:Field6}c) and the F200W NIRCam image (dominated by Pa$\alpha$ emission and near-IR starlight) in Figure~\ref{fig:Field6}d. 

 Both images show significant extended (presumed ionized gas) emission associated with the head of the warm H$_2$ including a protrusion, or "spike" labeled in  
 Figure~\ref{fig:Field6}c). The feature is also seen in CO emission. Extended(ionized) gas also follows more closely the distribution of the warm H$_2$ in the tail, than it does the CO emission. This is well demonstrated in the composite image Figure~\ref{fig:Field6}f,  which shows an RGB representation of the 10$\mu$m (warm H$_2$ , CO emission (cold H$_2$) and F200W (ionized gas and stars). Gemini optical spectroscopy \citep{Konstantopoulos2014} which covered this region at a spatial resolution of 1 arcsec (458 pc) strongly suggested that the ionized gas is shock-excited, which would explain why the ionized gas follows the warm H$_2$ emission in the head and the clumpy tail, and is less correlated with the CO emission.   
 
The spatial resolution of the F200W image is far superior to the HST image, and it is clear that there are several compact sources on scales of  $<$ 0.05-0.6 arcsec (23-27 pc) embedded in the ionized gas component. These sources are likely stellar associations, or unresolved super star clusters \citep{Gallagher2001}. They are generally rather sparsely distributed and poorly correlated with the warm or cool molecular hydrogen. 

One exception is a group of compact sources and extended F200W emission associated with the southern tip of the NW clump of CO emission identified in Figure~\ref{fig:Field6}c. These sources may be evidence of recent star formation associated with the head-tail structure. Evidence of this comes from the bright u-band sources associated with the same region shown in Figure~\ref{fig:Field6}e.  Although a full analysis of the NIRCamcolors of the point sources in Stephan's Quintet will be discussed in a separate paper, we know these blue regions have the optical colors of young clusters with ages estimated to be between 3-5 Myrs based on previous HST UBV$_m$VI photometry (\citealt{Fedotov2014}; see Table A4 for more details). The existence of young star formation near the NW CO clump may also explain why that region exhibits PAH emission, since PAHs are believed to be excited by UV radiation from young stellar associations.    

The kinematics of the diffuse ionized gas centered on the head from optical spectroscopy shows extremely broad line widths ($>$800-1200 \kms) in several key emission lines (e. g. [OIII]5007, [OI]6300, and H$\alpha$) \citep{Konstantopoulos2014,Guillard2022}. The broad line-widths have been attributed to a combination of turbulence and large-scale motions associated with gas along the line-of-sight.  Even broader wings in the Ly$\alpha$ profile were also seen at this position.  We will argue in \S~5 that the warm H$_2$ emission detected by JWST at this position may be responsible for UV scattering of Ly$\alpha$ emission within the turbulent regions, and if so, should exhibit broad line-widths.    

While we cannot yet measure the kinematics of the warm gas, the CO observations provide kinematic information about the cold molecular phase.   Figure~\ref{fig:Field6_moments}a and b shows the integrated CO surface density (moment 0) map and the mean velocity field (moment 1) map of the CO emission respectively.  The systemic velocity of the CO emission falls between the velocity of the intruder galaxy NGC 7318b (V(int) = 5770 \kms) and that of the velocity barycenter of the main group (V(group) = 6600 \kms). This is consistent with gas which has been decelerated in the collision of the intruder with the group-wide gas.  However, the linewidth of the entire region in CO is less that 120 \kms, as shown in the integrated ALMA CO (2-1) spectrum of the entire region and the CARMA CO (1-0) spectrum shown in Figure~\ref{fig:Field6_moments}c. This width is significantly smaller than that seen in the ionized gas, implying that the cold molecular gas does not take part in the turbulent motions seen in the ionized gas phase. More extensive kinematic information for the CO emission is presented in the Appendix-A, where we show individual spectra extracted from many regions (Figure~\ref{fig:FigureA1}), along with tabulated properties of the CO emission on the scale of the ALMA beam (Table~\ref{tab:COobs}). The tip of the head structure has the highest radial velocity, and the clumps show a trend to lower velocities as one moves east into the tail, reaching the lowest velocity in one of the most easterly clumps. We will argue later that this suggests material is being stripped from the head into the tail through ram-pressure stripping with a hot medium.   

The velocity dispersion\footnote{Hereby measured as the FWHM of the emission line profile, or 2.35 $\times$ the sigma of these mainly Gaussian lines.} of the individual clumps around the structure show a mix of unusually broad line widths (up to 100 \kms~FWHM) and narrow lines ($<$ 40 \kms). In Table~\ref{tab:COobs}, we show that the region near the "spike" feature seen in the ionized gas, is unusually broad (FWHM = 102 \kms). Away from the head, several regions in the tail of the structure show line widths ranging from 60-90 \kms (FWHM) (see Figure~\ref{fig:FigureA1}) to values lower than 40 \kms. Given the masses of the individual molecular clumps of $\sim$few$\times$ 10$^6 M_{\odot}$ (Table~\ref{tab:COobs}), it is unlikely that the clumps with high velocity dispersion are gravitational bound\footnote{For example, for a clouds of diameter 100 pc (just less than the resolution of ALMA) and given a typical gas surface density of $\Sigma_{gas}$ = 170 M$_{\odot} pc^{-2}$), the velocity dispersion for a cloud in gravitational (virial) equilibrium would be  $\sigma^2 =(3/5) \pi G R_{cloud} \Sigma_{gas}$. To be in equilibrium $\sigma$ = 8.3 \kms, or a FWHM $\sim$ 20 \kms.}.   The broader lines are consistent with  turbulent motions on the sub-arcsec ($\sim$ 100-150 pc) scale in the colder gas component, while others have much narrower lines.  

One such region occurs in the CO emission clump closest to the bright blue star clusters described earlier. Here the CO emission is essentially unresolved (40 \kms or less) perhaps suggesting turbulent motions are calmer there.
Table~\ref{tab:COobs} presents information about the line fluxes and estimated H$_2$ masses for all the regions measured in detail with ALMA. 

\vspace{0.04 in}
\subsubsection{Field 5: A shocked structure in the bridge region}
\vspace{-0.02 in}
Field 5 lies outside the main molecular filament in Stephan's Quintet, and is closer to large face-on galaxy NGC 7319. It forms part of an  apparent bridge of H$_2$ emission discussed by \citet{Cluver2010}. The JWST images (e. g. Figure 4) show that it is the most easterly of  series of irregular shock-dominated (appearing green in that image) clouds that run approximately East/West across the field and may not be
physically connected. 
In contrast to Field 6 which contained a single coherent CO structure,   Field 5 contains three separate bright structures visible in the CO maps (labeled in Figure~\ref{fig:Field5}a). These consist of a clumpy CO ring about 1 arcsec across (458 pc), a compact elongated core 1.5 arcsec to the SE of the ring, and a broken linear filament of gas running north-south 0.5 arcsec further east of the compact core. The direction of the elongation of the core points towards the center of the ring. Both Figure~\ref{fig:Field5}(a) and (b) show that the compact CO core lies near the brightest emission at the center of the arrow-shaped 10$\mu$m (warm H$_2$) structure.  As the arrow-shaped emission narrows down towards the NW, it connects to the northern part of the CO emission from the ring.  Figure~\ref{fig:Field5}(c) shows diffuse H$\alpha$ emission from the compact CO core but little emission from the ring.  The higher spatial resolution of the NIRCam F200W image shows that the corresponding Pa$\alpha$ in the core is also elongated Figure~\ref{fig:Field5}(d).  Faint F200W emission is also seen in the brightest CO ring clump.  The U-band image, Figure~\ref{fig:Field5}e,  shows no obvious emission near any of the CO molecular structures.  Although not shown here, the F150W NIRCam image shows point-like sources in the northern part of the CO ring, suggesting that the F200W image is revealing a mix of HII regions and compact sources within the ring. The composite image, showing the three gas phases in one figure emphasizes the warm H$_2$ connection between the ring and the core  (Figure~\ref{fig:Field5}f). 

A string of clumpy sources at 2$\mu$m is clearly seen approximately 1 arcsec to the east and north of the ring (F200W image). It is not clear if these sources have any real connection to the CO structures, as some could be background galaxies. One of the clumps has a faint U-band association (see Appendix-B and Table A3) suggesting recent star formation. Near-IR spectroscopy may help to determine if they are are physically associated with Stephan's Quintet by providing evidence of HII region-like emission lines at the Quintet's redshift, such as Pa$\alpha$). 

The kinematics of CO gas in the compact core is strikingly different from that of the ring and the linear structure.  Figure~\ref{fig:Field5_moments}a and b, shows the integrated and radial velocity map of the CO (2-1) emission, and spectra of the three different CO emitting structures are presented in Figure~\ref{fig:Field5_moments}c. The velocities within the partial ring increase clockwise around the ring from 6631 \kms~ in the north, reaching the highest value in the south and east quadrant of 6688 \kms.  In contrast, the compact core  (which lies near the peak in the warm H$_2$ emission) has a much lower (negative 250 \kms) radial velocity than the ring ($\sim$ 6400 \kms). The core also shows a velocity shear along its length of about 30 \kms over 0.5 arcsec (200 pc), as shown in the Appendix-A. Finally the third linear structure to the east of the compact core, shares velocities similar velocity to that of the ring, with velocities ranging from 6620~\kms in the south, to 6680~\kms in the north. 

An in-depth view of the kinematics of Field 5 in given in the Appendix-A, including individual spectra, Figure \ref{fig:FigureA2}, and 
 a table of CO properties, Table~\ref{tab:COobs2}. The spectra show that individual CO clumps in all of the three CO structures exhibit broad CO line-widths at the scale of the ALMA beam (~$\sim$150 pc). The compact core itself has a velocity dispersion of 145 \kms (some of this may be due to the gradient seen along the core), and all components of the ring have line-widths which lie in the range 60-120 \kms (FWHM). The linear structure shows narrower line profiles in the range 48-87 \kms. Appendix-A also shows that although the majority of the gas in the ring has much higher radial velocities than the core structure, one small segment of the ring (at the point nearest the core) has a velocity similar to the core. This is supported by the CARMA spectrum, Figure~\ref{fig:Field5_moments}d, which shows a blueward component associated with the average spectrum of the ring. This strengthens the idea that the two CO structures are somehow related despite their discrepant radial velocities.  We will discuss in the next section the possibility that the warm bridge connecting the CO ring and core  may be splashed warm molecular gas caused by the collision of a dense molecular cloud moving at high (blueward) velocity with respect to more diffuse gas associated with material in the group. 
 
Some of the CO emission in Field 5 may be associated with a larger filament and other CO clumps, as seen in a larger-scale (10-15 arcsec = 4.5-7 kpc) image,   Figure~\ref{fig:Field5extended}a, where we present the ALMA CO emission map smoothed (to 1 x 1 arcsec$^2$) to bring out fainter features. This reveals a very faint filament of CO gas and several fainter southern CO concentrations.  Both the filament and the southern CO structures have faint warm H$_2$ counterparts.  Filaments like this one, in this case extending over 6-10 arcseconds in scale (3kpc-5kpc), seem to be common in the MIRI maps of SQ (e. g. Figure~\ref{fig:fieldregions}). Figure~\ref{fig:Field5extended}b shows again the velocity field of the ring, compact core and linear structure, this time in relation to the larger-scale structure in the region. This further emphasises how the compact core has such a different velocity from all the other structure in the region. The velocity of the filament is poorly determined in CO because it is so faint, but seems to have a velocity similar to that of the average velocity of the group.  Mid-IR spectroscopy would be needed to provide more information about the filament since it is well detected in the MIRI 10$\mu$m image (presumed H$_2$ emission).

\subsubsection{Field 4: In the northern star-forming region, SQ-A}

Finally we present the distribution of gas phases and star clusters 
in the Field 4 SQ-A region. The region has been discussed previously in terms of a possible tidal dwarf galaxy \citep{Moles1997,mendes2001}, although this is disputed by \cite{Xu1999} who point out that the kinematics do not show simple rotation.  Figure~\ref{fig:Field4} shows a closer correspondence between the various gas phases, compared with Fields 5 and 6. Panel (a) and (b) of Figure~\ref{fig:Field4} show a close correspondence between the CO distribution, the 10$\mu$m H$_2$ image and the 7.7$\mu$m PAH-dominated image. In some of the panels we have smoothed the CO distribution to an effective resolution of 1 arcsec (458 pc).  The close correspondence between these three ISM indicators is not unexpected since previous IRS spectroscopy of the region showed that the total H$_2$ to PAH ratios in that region are more typical of normal Photo Dissociation Region (PDR) regions associated with HII regions. Also ground-based spectroscopy shows that the optical emission line ratios from the ionized gas are consistent with HII-region like spectra \citep{Konstantopoulos2014}. Extended patchy ionized gas is seen in  panels (c) and (d) of Figure~\ref{fig:Field4}. A sharp boundary in the H$\alpha$ distribution near the point where the CO emission starts to rise on the eastern edge, suggesting significant extinction of H$\alpha$ light due to a dust lane associated with the denser molecular gas. This is supported by the fact that the same "edge" is not seen in the F200W NIRCam filter (Figure~\ref{fig:Field4}d)  where the Pa$\alpha$ is likely to be much less extinguished that the H$\alpha$. \citet{Konstantopoulos2014} measured high optical extinction in this region based on ground-based optical spectra.  The emission seen in the F200W filter, Figure~\ref{fig:Field4}d,  show many slightly extended clump sources scattered across an elongated "disk". The overall scale of this possible starburst dwarf system, is 5 $\times$ 1.5 arcsec$^2$ (2.3 kpc $\times$ 700pc). The U-band image supports the idea of strong extinction in the region of the peak CO emission. In Table A4 we also present UBV$_m$VI photometry of some of the brighter well defined sources \citep{Fedotov2014}. This study shows that the brighter clumps have ages based on the optical photometry ranging from 6-36 $\times$ 10$^{7}$ yrs, much longer than that of the blue clumps we investigated in Field 6. This may suggest a more mature stellar population, although because of higher extinction here, NIRCam photometry would provide a more definitive result.  

The structure of the CO is that of two interleaved half-shells (very well seen in the smoothed version of the CO contours which are shown in all the panels of Figure~\ref{fig:Field4} except (b) and (d)).  As with the previous descriptions, Figure~\ref{fig:Field4}f is an RGB composite of the F1000W (red, mainly H$_2$ dominated), the smoothed ALMA CO (2-1) (green, cool H$_2$) and F200W (red, Pa$\alpha$ plus starlight) images to show the spatial relationship between the different gas phases and star clusters.  

The kinematics of the system led \citet{mendes2001} to identify the region as a potential tidal dwarf because of its clear rotation using Fabry-Perot observations with the Canada-France- Hawaii Telescope (CFHT).  They derived a systemic velocity of 6680 \kms, an asymmetric rotation curve, and a high velocity dispersion of 115 \kms, which may have been the result of a blend of two different optical emission line components. 

Our CO spectra do not show obvious rotation.  We show the integrated CO (2-1) emission map (moment 0) and velocity field map (moment1) for Field 4 in  Figure~\ref{fig:Field4_moments}. Although the overall kinematics of the CO gas in Field 4 shows lower systemic velocity on the eastern side (6700 \kms), and a higher value to the west (6740 \kms), the velocity range is narrow (FWHM = 47 \kms) centered on the barycentric velocity of the main SQ group. The overall systemic velocity is in reasonable agreement with the optical emission line velocities, although there is little real hint of rotation. This may argue against the formation of the object within a tidal stream, where the rotation is expected as a remnant of tidal shear.  As we show in Table~\ref{tab:COobs3} and Figure~\ref{fig:FigureA3} the velocity dispersion of the clumps at the highest resolved scale is narrow, which is very different from the more turbulent emission in Field 5 and 6. 

Although not shown in Figure~\ref{fig:Field4}, there is a second very faint ALMA CO velocity component (6900 \kms) which is seen at a higher amplitude in CO (1-0) observed with the larger beam of CARMA data. It is very clear as a double-horned profile in the CARMA spectrum of Figure~ \ref{fig:Field4_moments}c.  This emission must be quite diffuse and is only just detected with ALMA. A similar high-velocity CO component was seen associated with SQ-A by \citet{Guillard2009} with the 30-m IRAM telescope. There is a hint of this higher-velocity component in the kinematic maps obtained in  H$\alpha$ by \citet{DuartePuertas2019} where the ionized emission is quite localized to SQ-A. So, like Field 5, we find two CO components associated with a major CO concentration in SQ with quite discrepant velocities. In this case, the two components are roughly coincident and so its difficult to know whether the warm H$_2$ gas, which dominates the 10$\mu$m JWST image, originates in two clouds in physical contact, or are simply chance associations. Only mid-IR spectroscopy will allow us to fully explore this possibility.

\subsection{Mid-IR Line luminosities, warm and cold H$_2$ masses}
\begin{figure*}
\includegraphics[width=0.99\textwidth]{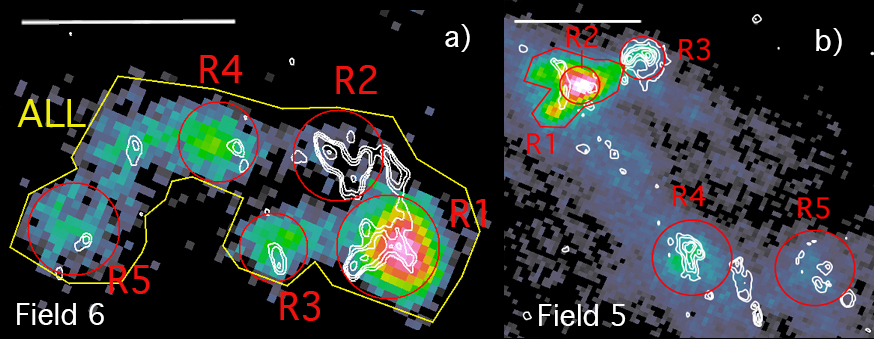}
\caption{Regions used to determine the brighter warm H$_2$ flux hotspots and CO spectral extractions areas for (a)  Field 6, including 5 regions (red circles, and yellow full region, and b) Field 5 showing the 5 extraction regions. Larger regions (not shown) covering an area 5.5 x 5.5 arcsec$^2$ were also extracted, corresponding to the the area sampled by {\it Spitzer} IRS. The properties of the extracted fluxes and CO spectral properties are presented in Table 1 and Table 2. Contours are of the integrated CO (2-1) emission. Smaller extraction regions concentrating on the properties of the small-scale CO emission for all three regions are presented in the Appendix-A. The white scale bar is 3 arcsec in length.} 
\label{fig:regionsdef}
\end{figure*}
\begin{deluxetable*}{llllllllll}
\tablecolumns{10}
\tablewidth{0pc}
\tablecaption{Measured and Derived Properties of Warm H$_2$ in Field 5 and 6} \label{tab:table1} 
\tablehead{
\colhead{Region\tablenotemark{a}} &
\colhead{Filter/Band} & 
\colhead{Assumed} &
\colhead{Flux Density\tablenotemark{b}} &
\colhead{Area} &
\colhead{Bandwidth} &
\colhead{H$_2$~Line Flux} &
\colhead{L(0-0S(3))} &
\colhead{n\tablenotemark{c}} &
\colhead{$M_{H2}$\tablenotemark{d}} 
\\
\colhead{} & 
\colhead{} & 
\colhead{Transition} &
\colhead{($\mu$Jy)} &
\colhead{(arcsec$^2$)} &
\colhead{($\mu$m})&
\colhead{ ($\times 10^{-18} W m^{-2}$}) &
\colhead{ ($\times 10^{32}$ W)} &
\colhead{} &
\colhead{ ($\times 10^6~M_{\odot}$)} 
\\
\colhead{[1]} & 
\colhead{[2]} & 
\colhead{[3]} &
\colhead{[4]} &
\colhead{[5]} &
\colhead{[6]} &
\colhead{[7]} &
\colhead{[8]} &
\colhead{[9]} &
\colhead{[10]} 
}
\startdata
F5-R1     & F1000W & 0-0S(3)           &  44~($\pm$7)   &  2.8  &  2.0 &  2.6~($\pm$0.4) & 2.8~($\pm$0.4) & [4.2]\tablenotemark{c} &  5.6~($\pm$1.0) \\              
F5-R2     & F1000W & 0-0S(3)           &  17~($\pm$2)   &  0.6  &  2.0 &  1.0~($\pm$0.2) & 1.0~($\pm$0.2) & [4.2]\tablenotemark{c} &  2.1~($\pm$0.3) \\                   
F5-R3    &  F1000W & 0-0S(3)           &  7~($\pm$1)    &  0.9  &  2.0 &  0.4~($\pm$0.1) & 0.4~($\pm$0.1) & [4.2]\tablenotemark{c} &  0.9~($\pm$0.2) \\
F5-R4    & F1000W  & 0-0S(3)           &  20~($\pm$3)   &  2.7  &  2.0 &  1.2~($\pm$0.2) & 1.3~($\pm$0.2) & [4.2]\tablenotemark{c} &  2.5~($\pm$0.4)\\
F5-R5    & F1000W  & 0-0S(3)           &  14~($\pm$2)   &  2.6  &  2.0 &  0.8~($\pm$0.1) & 0.9~($\pm$0.1) & [4.2]\tablenotemark{c} &  1.8~($\pm$0.3) \\
F5-IRSarea & F1000W & 0-0S(3)           & 87~($\pm$13)   &  30.3  &  2.0 &  5.2~($\pm$0.8) & 5.5~($\pm$0.8) & 4.2($\pm$0.02) &   11.0~($\pm$1.9) \\
F5-IRS    & LL\tablenotemark{e}    &  0-0S(1)           & -      &  30.3  &  -   &  4.01\tablenotemark{e} & 4.2 &   -   &  - \\
F5-IRS    & LL\tablenotemark{e}    &  0.0S(0)           & -      &  30.3  &  -   &  0.50\tablenotemark{e} & 0.5 &   -   &  - \\
\\
F6-All   &  F1000W &   0-0S(3)        &  58~($\pm$9)   &   11.0 & 2.0 &  4.3~($\pm$0.6) & 4.5~($\pm$0.7) & [4.5]\tablenotemark{c}   & 14.4~($\pm$2.5)\\           
F6-R1  & F1000W &  0-0S(3)             &  20~($\pm$3)   &    1.6 & 2.0 &  1.2~($\pm$0.2) & 1.2~($\pm$0.2) & [4.5]\tablenotemark{c}   &  4.0~($\pm$0.7) \\
F6-R2  & F1000W &  0-0S(3)             &  6.0~($\pm$0.9)    &    1.3 & 2.0 &  0.4~($\pm$0.1) & 0.4~($\pm$0.1) & [4.5]\tablenotemark{c}   &  1.2~($\pm$0.2) \\   
F6-R3  & F1000W &  0-0S(3)             &  5.4~($\pm$0.8)   &    0.7 & 2.0 &  0.3~($\pm$0.1) & 0.3~($\pm$0.1) & [4.5]\tablenotemark{c}   &  1.1~($\pm$0.2)\\
F6-R4  & F1000W &  0-0S(3)             &  7.3~($\pm$1.1)   &    1.0 & 2.0 &  0.4~($\pm$0.1) & 0.5~($\pm$0.1) & [4.5]\tablenotemark{c}   &  1.5~($\pm$0.3) \\
F6-R5  &  F1000W & 0-0S(3)             &  7.7~($\pm$1.2)   &    1.3 & 2.0 &  0.5~($\pm$0.1) & 0.5~($\pm$0.1) & [4.5]\tablenotemark{c}   &  1.5~($\pm$0.3) \\
F6-IRSbox &  F1000W &    0-0S(3)       &  76~($\pm$11)   &   30.3 & 2.0 &  4.6~($\pm$0.7) & 4.8~($\pm$0.7) &  4.50($\pm$0.02)    & 15.3~($\pm$2.5) \\
\\
F4-IRSonly\tablenotemark{f} & SL-LL & 0-0S(0)-0-0S(5) & - & 30.3 & - & 2.1 & 2.2 & 5.00~($\pm$0.03) & 14.8~($\pm$2.2) 
\enddata
\tablenotetext{a}{The regions for F5 (Field 5) and F6 (Field 6) are shown graphically in Figure~\ref{fig:regionsdef}.}
\tablenotetext{b}{The flux density is measured from the surface brightness in the JWST MIRI image over the apertures shown in Figure~\ref{fig:regionsdef}, assuming a pixel size of 0.11 x 0.11 arcsec$^2$.}
\tablenotetext{c}{Power law index n, assuming the warm H$_2$ column density follows a power law with temperature of the form $\delta$$N~\propto T^{-n}$$\delta$T, following the work of \citet{Togi2016}, and as discussed for Stephan's Quintet by \citet{Appleton2017}. We assume that each of the regions extracted from the MIRI image have the same power law as the average for the whole regions, which is a significant limitation of the current analysis.}
\tablenotetext{d}{Estimated mass in warm H$_2$ for T$>$ 100~K assuming a value for the powerlaw index, n, for the resolved regions that is the same as that measured by the IRS over a larger 5.5 x 5.5 arcsec$^2$ region. These warm H$_2$ mass estimates for T$>$~100~K employ the method of \citet{Togi2016}. For Field 6, n was derived from the full mapping with both LL and SL IRS modules described in \citep{Appleton2017}. For Field 5, was measured using {\it Spitzer} IRS Long-low data for 0-0S(0) and 0-0S(1), combined with the flux measured for 0-0S(3) from the MIRI data. Uncertainties in the final masses represent propagation of the formal uncertainties, and do not include the unknown effect of possible variations from region to region in the actual power law index, which we have assumed here to be constant.}
\tablenotetext{e}{Line fluxes measure directly from extracted IRS spectrum from Long-Low (LL) IRS spectrograph shown in Figure~\ref{fig:IRSSpec}b.}
\tablenotetext{f}{Unambiguous H$_2$ emission was not easily extracted from the F1000W data for this region (see text).  This entry is taken from the IRS data from \citet{Appleton2017} for comparison with the other regions. Note the large value for the power law index (n) indicative of cooler H$_2$ temperatures near the northern tip of the SQ filament in the SQ-A region. }
\end{deluxetable*}
\begin{deluxetable*}{lllllll}
\tablecolumns{7}
\tablewidth{0pc}
\tablecaption{Integrated properties of CO emission in Field 5 and 6 over same areas as Table 1 \label{tab:table2}}
\tablehead{
\colhead{Region} &
\colhead{V$_{helio}$} & 
\colhead{FWHM} &
\colhead{$\Sigma(S_v\Delta V$)} &
\colhead{$\Sigma(S_v\Delta V$)\tablenotemark{a}} &
\colhead{M(H$_2$)\tablenotemark{a}} &
\colhead{M$_{gas}$\tablenotemark{b}} 
\\
\colhead{} & 
\colhead{} & 
\colhead{} &
\colhead{CO(2-1)} &
\colhead{CO(1-0)} &
\colhead{$\times 10^6$} &
\colhead{$\times 10^6$}
\\
\colhead{} & 
\colhead{(\kms)} & 
\colhead{(\kms)} &
\colhead{(Jy-\kms)} &
\colhead{(Jy-\kms)} &
\colhead{(M$_{\odot}$)} &
\colhead{(M$_{\odot}$)}
\\
\colhead{[1]} & 
\colhead{[2]} & 
\colhead{[3]} &
\colhead{[4]} &
\colhead{[5]} &
\colhead{[6]} &
\colhead{[7]}
}
\startdata
F5R1A &  6401 ($\pm$6) &   101 ($\pm$15) &      0.9 ($\pm$0.1) &     0.28 ($\pm$0.05) &   18.8 ($\pm$3.0) &   25.6 ($\pm$4.1) \\
F5R1B &  6646 ($\pm$4) &    73 ($\pm$10) &      0.6 ($\pm$0.1) &     0.19 ($\pm$0.03) &   13.0 ($\pm$2.0) &   17.7 ($\pm$2.7) \\
F5R2 &  6388 ($\pm$4) &   119 ($\pm$10) &      0.8 ($\pm$0.1) &     0.25 ($\pm$0.02) &   16.7 ($\pm$1.5) &   22.7 ($\pm$2.0)  \\
F5R3 &  6656 ($\pm$2) &   105 ($\pm$ 5) &      1.5 ($\pm$0.1) &     0.47 ($\pm$0.02) &   31.5 ($\pm$1.5) &   42.8 ($\pm$2.0)  \\
F5R4 &  6618 ($\pm$4) &    84 ($\pm$ 8) &      1.1 ($\pm$0.1) &     0.33 ($\pm$0.04) &   22.4 ($\pm$2.4) &   30.4 ($\pm$3.3)  \\ 
F5R5 &  6537 ($\pm$5) &    67 ($\pm$12) &      0.6 ($\pm$0.1) &     0.20 ($\pm$0.04) &   13.5 ($\pm$2.7) &   18.3 ($\pm$3.6)  \\
F6-ALL &  6054 ($\pm$2) &    91 ($\pm$4) &      4.7 ($\pm$0.2) &      1.5 ($\pm$0.1) &   99.1 ($\pm$4.4) &  134.8 ($\pm$6.0)  \\
F6R1 &  6073 ($\pm$1) &    47 ($\pm$2) &      1.1 ($\pm$0.1) &      0.3 ($\pm$0.0) &   23.0 ($\pm$1.1) &   31.3 ($\pm$1.4)  \\
F6R2 &  6040 ($\pm$2) &    93 ($\pm$4) &      1.2 ($\pm$0.1) &      0.4 ($\pm$0.0) &   26.1 ($\pm$1.3) &   35.4 ($\pm$1.7)  \\
F6R3 &  6051 ($\pm$2) &    41 ($\pm$5) &      0.2 ($\pm$0.0) &      0.1 ($\pm$0.0) &    4.3 ($\pm$0.6) &    5.9 ($\pm$0.8)  \\
F6R4 &  6009 ($\pm$2) &    37 ($\pm$4) &      0.3 ($\pm$0.0) &      0.1 ($\pm$0.0) &    5.9 ($\pm$0.7) &    8.1 ($\pm$1.0)  \\
F6R5 &  6103 ($\pm$4) &    48 ($\pm$10) &      0.2 ($\pm$0.1) &      0.1 ($\pm$0.0) &    4.9 ($\pm$1.1) &    6.7 ($\pm$1.5) 
\\
\enddata
\tablenotetext{a}{We assume a conversion between a line flux at CO~(1-0) to that of CO~(2-1) of
S$_{CO(1-0)} /S_{CO(2-1)}$ = $(\nu_{1-0}/\nu_{2-1})^2 (r_{21})^{-1}$, where $\nu_{1-0}$ and $\nu_{2-1}$ are the rest frequencies of the transitions, and $r_{21}$ is assumed to be 0.8, similar to that measured in the Taffy galaxy bridge \citep{Zhu2007}, which shares many similarities with the gas in Stephan's Quintet (see \citealt{Appleton2022}). The H$_2$ masses presented here are for an X$_{CO}$ value =  X$_{CO,20}$ = N(H$_2$)/I$_{CO}$ = 2 $\times$ $10^{20}$ cm$^{-2}$ ($K-$\kms)$^{-1}$, the standard value assumed for our Galaxy.  In the text we discuss how this may not be applicable in such a turbulent region. $M_{H2}$ = $7.72 \times 10^3 D^2 \Sigma(S_v\Delta$V)(1+z)$^{-1}$, where D = 94 Mpc, and $\Sigma(S_v \Delta V$) is the CO$_{(1-0)}$ line integral with $S_{v}$ in Jy and the velocity V of the gas in \kms. We assume z = 0.02.}
\tablenotetext{b}{Total gas mass M$_{gas}$ = M$_{H2}$ $\times$ 1.36, includes a 36$\%$ correction for Helium \citep{Bolatto2013}.}
\end{deluxetable*}

\vspace{-0.75 in}
In Table~\ref{tab:table1}, we present measured and derived properties of the warm H$_2$ emission derived from both the {\it Spitzer} IRS data of \citet{Appleton2017} and from measured fluxes from the F1000W image for Field 5 and Field 6, where we are confident that the emission is primarily from the 0-0S(3) H$_2$ line. Unlike the tables in the Appendix-A, which are concerned with illustrating the detailed kinematics and cold gas properties at the scale of the CO clumps, here we are concerned with the masses of the warm H$_2$ and the cold molecular mass integrated over the same larger areas, presented in Table~\ref{tab:table2}. For Field 4, where we were unable to properly correct for underlying dust emission across the map, we only quote the properties of the H$_2$ from the IRS data on a large scale for comparison with Field 5 and 6. 

In Table~\ref{tab:table1}, Column 1 gives the name of the regions defined in Figure~\ref{fig:regionsdef} for Field 5 and 6. The regions were chosen to cover the main warm H$_2$ features. 
Column 2 gives the MIRI filter used, or the {\it Spitzer} module used (see later). For most of the regions we extract the equivalent 0-0S(3) flux from the JWST F1000W image. The pure-rotational molecular hydrogen transition is listed in Column 3. Column 4 gives the measured H$_2$ flux density measured from the JWST F1000W over an area in arcsecs given in Column 5. We convert these flux densities to equivalent line fluxes, Column 7, using the filter bandwidth listed in Column 6. The line luminosity of the 0-0S(3) line is given in Column 8 assuming a distance to SQ of 94 Mpc. We also present properties of warm H$_2$ taken from {\it Spitzer} for the regions, and the appropriate IRS modules and H$_2$ rotational transition is listed. These are used below in the calculation of H$_2$ masses. Uncertainties are provided in parentheses for each measured quantity.

To calculate the total warm H$_2$ mass in the regions, we make the following assumptions. First, since in most cases we only have the line flux of the 0-0S(3) line derived from the MIRI image, we must assume something about the temperature distribution of the gas in order to estimate the total warm H$_2$ mass.  Since the regions we are interested in were observed with {\it Spitzer} over a larger area, but over many of the rotational H$_2$ lines, we can assume, as an approximation, that the temperature distribution measured by the {\it Spitzer} IRS \citep{Appleton2017} can be applied to all the smaller regions resolved by MIRI imaging. In that paper we showed that the excitation diagrams for many regions extracted over SQ were almost always consistent with a single power law relationship between the warm H$_2$ column density and the H$_2$ temperature of the form $\delta$N(H$_2$)~$\propto T^{-n}\delta$$T$, following the work of \citet{Togi2016}. For Field 6 the {\it Spitzer} data provided a  power law index $n$ = 4.5, which is quite shallow compared with normal HII regions, and consistent with warm gas that is shock heated \citep{Appleton2017}. 

Because Field 4 (the region in SQ-A) was ambiguous regarding isolating the H$_2$ line in the F1000W filter because of contamination from neighboring broad wings of PAH features (Figure~\ref{fig:IRSSpec}), we did not attempt to measure the masses of smaller regions in this field. Rather we provide, for comparison with Field 5 and 6, the IRS-measured values. This analysis yielded a much steeper value of $n$ = 5. This is consistent with our observation that Field 4 is dominated by star formation. 

Field 5 is a little more complicated. Unlike the other cases, this region lies outside the full spectral mapping region performed by {\it Spitzer}, and only the IRS Long-low module was used to cover this area. This provided measurements of the 0-0 S(0)28$\mu$m and 0-0S(1)17$\mu$m transitions, but not the shorter wavelength transitions. However the JWST observations from MIRI can provide the 0-0 S(3) flux, if we extract over the same extraction region (referred to as F5-IRSarea in Table~\ref{tab:table1}) observed by {\it Spitzer}. With three points on the excitation diagram given in the table, we were able to measure a power-law index of 4.2. The flat value for $n$ is comparable with some of the warmest regions of Stephan's Quintet~\citep{Appleton2017}. 

The above measurement assumes that the MIRI and {\it Spitzer} data can be compared reasonably accurately--given that one is obtained from an image, and the other from a spectrometer. To check the validity of this comparison, we compared the measured 0-0S(3) line flux from IRS with the flux we derived from the extraction of the same area for Field 6. As Table 1 shows, we measured a line flux of 4.6 ($\pm$ 0.7) $\times 10^{-18} ~W m^{-2}$ for the 0-0 S(3) line from the JWST image. With the IRS, we measured (region 142 of \citet{Appleton2017}) a lineflux of 4.7 ($\pm$ 0.3) $\times 10^{-18}~ W m^{-2}$. The agreement provides some confidence that we are measuring the fluxes correctly using the  MIRI F1000W image.  

Having established an average power-law index, we now apply that same value of $n$ appropriate to the smaller extracted regions for Field 5 and 6 (Column 9 of the table). Without spectroscopic observations on the sub-arcsec scale, this is obviously an approximation. Using the 0-0S(3) flux measured by JWST, we estimate the total H$_2$ mass in these regions integrated over the power law down to a temperature of 100K. These masses (and uncertainties) are given in Column 10 of Table~\ref{tab:table1}. We caution that these uncertainties do not include the unknown effect of possible variations, from region to region, in the actual power-law index, which we have assumed is equal to the average value derived over a much larger area.

From this analysis, we see that the warm H$_2$ mass of the entire arrow-head like structure in Field 5 (F5R1) is $5.6~(\pm 1.0)\times10^6 M_{\odot}$ and its core (F6R2) contains almost half of the mass, $2.1~(\pm 0.3) \times10^6 M_{\odot}$. The partial ring structure (F5R3) has a much smaller warm H$_2$ mass of $0.9~(\pm0.2)\times10^6 M_{\odot}$. The two more distant regions of warm H$_2$ gas at the end of the filament in Field 5 have larger warm gas masses of (F5R4) $2.5~(\pm 0.4)$ and (F5R5) $1.8~(\pm0.3)\times10^6 M_{\odot}$ respectively. 

In the case of the head-tail structure of Field 6, the triangular hotspot  (F6R1) has a warm H$_2$ mass of $4~(\pm0.7)\times10^6 M_{\odot}$ and most of the clumps in the tail have masses of 1-2$\times10^6 M_{\odot}$. The total H$_2$ mass integrated over the whole structure is $14\pm3)\times10^6 M_{\odot}$ suggesting that there is considerable extended emission in the tail. 

Field 4, 5 and 6 have similar total warm H$_2$ masses when integrated over the larger 5.5 x 5.5 arcsec$^2$ (2.5 x 2.5 kpc) area measured by {\it Spitzer} IRS.  

How do these warm H$_2$ masses compare with the mass of molecular gas measured by ALMA?  In Table\ref{tab:table2} we present the measured and derived properties of the gas from the ALMA observations. We use the integrated CO (2-1) line profile for the same regions as those used in Table~\ref{tab:table1} for the warm H$_2$. Here we are concerned with measuring the total CO line profile associated with each of the regions for which we measure warm H$_2$ emission. Column 1 gives the region description. Column 2 and 3 provide the mean heliocentric velocity and full-width half-maximum (FWHM)  of the emission after fitting each region with a Gaussian line profile.  We note that for F5R1, there we two detected CO emission features referred to as F5R1A and B. This is because R1 contains emission from both the compact core feature and the linear filament. The inclusion of two distinctly different structures (with different kinematics) within the one region is clearly seen in Figure\ref{fig:regionsdef}b. These structures will be discussed in more detail in Appendix-A. Column 4 provides the line integral of the CO (2-1) emission in Jy-\kms\ and Column 5 provides an estimate of the equivalent CO (1-0) emission, based on an assumed ratio of the integrated intensity of the main-beam temperature in the two lines ($r_{21} = I[12CO(2-1)]/I[12CO(1-0)$)
Values of r$_{21}$ range from 0.76 to 0.9 in normal galaxies at low and intermediate redshift, with a slightly larger range for perturbed galaxies \citep{Braine1992,Daddi2015}. We assumed a value of 0.8 to be similar to that measured in the highly turbulent Taffy Bridge \citep{Zhu2007}. Column 6 provides an estimate of the  H$_2$ mass for an assumed value of the X-factor, $X_{CO}$.  For the purposes on the discussion, we present the H$_2$ masses in Column 6, and the total gas mass in Column 7 assuming $X_{CO}$is the average Galactic value. This value is likely to be significantly overestimated in regions of strong shocks and turbulence. The footnotes to the table provide more details about the definitions and assumptions used in the calculations.  

Bearing in mind the uncertainties in both the warm and cold molecular masses, we can estimate the ratio of M(H$_2$)$_{warm}$/ M(H$_2$)$_{cold}$ by comparing the warm gas masses estimated by JWST with those from ALMA. For the F5R1 regions (the arrow-head structure in Field 5) the ratio of the warm to cold H$_2$ masses is 5.6/(25.6+17.7) = 0.13. This assumes that both the CO linear filament and the CO core both contribute to the total mass. Its not clear that the linear structure is actually detected in warm CO, so the ratio could be much higher if that object was excluded (0.22). However, in reality, since $X_{CO}$ is likely to be much smaller than Galactic (say 1/5 $X_{CO,20}$), then the fraction of warm gas with T $>$ 100 K is likely to be very large (almost equal to the cold H$_2$). These values are significantly greater than that seen in normal galaxies. F5R3, which includes the partial ring of CO emission has a warm/cold H$_2$ ratio of 0.02, which is much more like a normal galaxy. Regions F5R4 and R5 show warm to cold ratios somewhat in between the two extremes (0.08 and 0.1). These regions lie at the opposite end of a warm filament (only weakly detected in CO) that connects them to the other Field 5 structures.  The warm triangular hotspot in Field 6 shows an average ratio of warm to cold H$_2$ = 0.12. Again, if we assume $X_{CO}$ is 1/5 of the Galactic value, this would imply a significant warm component. F6R2, has a much lower ratio of warm to cold gas (0.03), a number which is closer to that of normal star forming galaxies. It is interesting that this is the region which emits strongly in the F770W filter, which is consistent with this containing a PDR. The other clumps in the tail have unusually large values of warm to cold CO (in the range 0.18-0.22), reflecting the relative faintness of the CO in the clumps in the tail of Field 6. The overall conclusion is that except for the regions that show obvious star formation (PAH emission and compact star clusters embedded in the emission) the warm to cold H$_2$ ratios are unusually large for both Field 5 and Field 6. We will discuss this further in the discussion in the context of shocks and turbulence. 

We note that our ALMA radio continuum observations at 230GHz might, in principal, provide useful information about the mass of cool dust associated with the target regions. Inspection of the continuum maps did not detect dust emission from any of the regions of interest down to an rms upper-limit of 0.03~mJy/beam. We assume a dust mass 
$M_{dust}= S_{\nu} D^2 / (\kappa_{\nu} B(\nu,T_{dust}))$, where S$_{\nu}$ is the flux density at frequency $\nu$, D is the distance to the galaxy, and B($\nu,T_{dust}$) is the black-body function at frequency $\nu$ and temperature T$_{dust}$. Assuming a reasonable average value for the dust absorption coefficient $\kappa_{1.3mm}$ of 0.03 m$^2$ kg$^{-1}$ (extrapolating from 500 $\mu$m from \citealt{Clark2019}), T$_{dust}$ = 20K, we estimate a 3$\sigma$ upper-limit to the dust mass of $M_{dust}~<~5.6~\times~10^5~M_{\odot}~beam^{-1}$. A dust temperature of 20K is consistent with previous {\it Herschel} observations of dust emission on much larger scales \citep{Appleton2013}. If the typical gas to dust ratio for galaxies \citep{Sandstrom2013} is 80-100 (this number may be higher if dust is destroyed in the shock waves), the corresponding upper limit to the gas mass is at least a factor of ten larger than the CO-derived gas masses within the same beam. This suggests that the continuum observations are not deep enough to provide useful physical constraints on the cold dust properties of the observed regions. Observations at higher frequencies would be desirable to potentially detect cold dust from these regions.

\section{Discussion}

SQ is a fantastic example to study the dynamical interaction between gas phases in a group environment. The big leap forward in spatial resolution offered by the JWST images compared to Spitzer sheds light on the morphological structures of the multi-phase gas, and the physical processes at the origin of the formation and excitation of the molecular gas in the IGM. We discuss what we are learning about those two aspects in the following sections.

\subsection{Morphology of the molecular gas emission and large-scale kinematics}
\label{subsec:discussion_1}


\begin{figure*}
\includegraphics[width=0.96\textwidth]{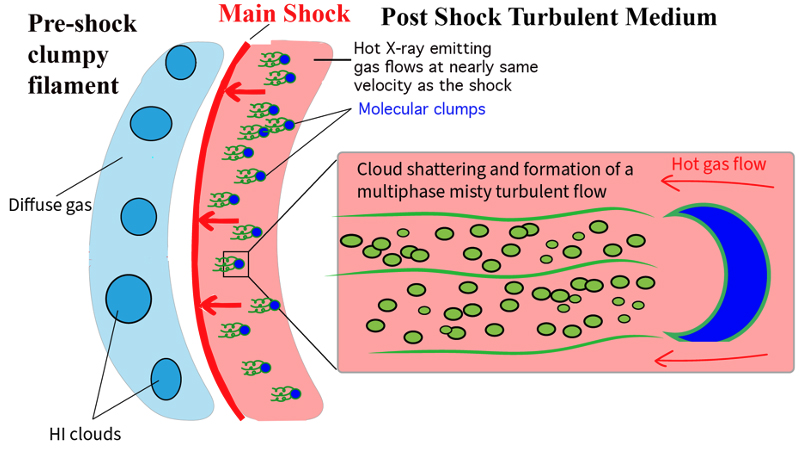}
\caption{Sketch of the scenario leading to the formation of multiphase gas in the post-shock region. The relative motion between the intruder, NGC~7318b, and the tidally-stripped filament drives a large-scale shock (40 kpc-long, symbolised with the red line) in the low-density gas (the light blue filament). The pre-shock medium being multiphase, shock velocities are slower in the denser (atomic) regions (darker blue clouds), allowing molecular gas (dark blue clumps) to form in less than the crushing timescale of the \Hi\ clouds \citep[see Sect.~\ref{subsec:discussion_1} and][]{Guillard2009}. We argue in  Sect.~\ref{subsec:discussion_2} and~\ref{subsec:discussion_3} that the small-scale (a few 100~pc) \Hmol\ and CO structures  observed JWST and ALMA may be the result of the fragmentation of the clouds and subsequent reformation of a "foggy" molecular gas (symbolised as tiny green droplets). At small scales, as sketched in the zoom-in panel on the right, the dynamical interaction between the flow of hot gas (red arrow) and the resulting cloudlets  drives a cycling across ISM phases, leading to the formation of extended and turbulent \Hmol\ tails. }
\label{fig:schema}
\end{figure*}

In order to explain the formation of large masses of molecular gas and high luminosity in the pure rotational ground-state transitions of \Hmol, \citet{Guillard2009} presented a model in which a large-scale driving shock wave (travelling at 600-800~\kms) from the intruder passes into a multi-phase medium of gas in a filament of tidal debris caused by a previous tidal interactions between group members (see Figure~\ref{fig:schema} for a schematic view). We summarize here that scenario, which was based on observations with limited spatial coverage and  resolution with \textit{Spitzer}, and then discuss what we are learning from the new JWST plus ALMA data presented here about the link between the large-scale dynamics in the group, and the formation and excitation of the molecular gas in the IGM. 

The large-scale shock wave driven by the intruder, NGC7318b, passes over the tidal, multiphase filament, heating low density gas (\nH$~<2 \times 10^{-2}$~\cc) to X-ray emitting temperatures (see \citealt{OSullivan2009}), and destroying any dust grains present there (illustrated in Fig.~\ref{fig:schema} as the light orange medium).  Regions in the pre-shock region with over-densities greater than 10 (typically \Hi\ gas with \nH$~>~0.1$~\cc) compared to the volume-filling gas are heated to lower ($~<~10^6$~K) temperatures, can cool and form molecules on dust grains as they survive the slower shocks (sketched as the dark blue clouds in Fig.~\ref{fig:schema}). Dust grains are expected to survive in shocks with velocities lower than 200\kms\ \citep{Jones2011}. \citet{Guillard2010} have indeed shown that dust survives in the denser regions of the IGM of the galaxy collision, and that the PAH and continuum infrared emission is consistent with the dust associated with the warm ($> 150$ K) \Hmol\ gas, heated by a UV radiation field of intensity comparable to that of the solar neighborhood, $G_0 = 1.4$. \citet{Natale2010} also showed that a diffuse IR dust emission component is associated with the collisional cooling of the hot halo gas, provided that dust is injected into the hot plasma by tidal stripping or stellar feedback. In the next section, we discuss that the dynamical interaction between the cold and hot gas can lead to the same physical conditions down stream from the shock, as thermally unstable gas is produced within the turbulent mixing layers \citep{Gronke2022}, leading to a cycle of molecular clouds breaking up and reforming in the flow. This is sketched as the turbulent tail in the zoom-in panel on the right-hand side of Fig.~\ref{fig:schema}. 

An efficient transfer of the bulk kinetic energy of the galaxy collision to turbulent energy within the molecular gas, associated with a continuous cold gas destruction-reformation cycle, is required to make \Hmol\ a dominant coolant of the postshock gas \citep{Guillard2009}. In the next section we discuss how the spatial decoupling between cold and warm gas at GMC scales, revealed by the comparison between ALMA and JWST images, is shedding light on the physical processes that drive this energy transfer mediated by thermal instability and a dynamical interaction between gas phases.

%
%



\subsection{Cycling and redistribution of matter across gas phases: shattering and growth of cold clouds in the IGM}
\label{subsec:discussion_2}

The ALMA and JWST data presented in this paper reveals complex and turbulent structures, with large velocity dispersions (40 to 100~\kms) at GMC scales, and a spatial decoupling of cold and warm molecular gas. Although a quantitative interpretation of these data is difficult and beyond the scope of this paper, we qualitatively discuss a scenario that relate the cold and warm \Hmol\ distributions to theoretical work of cold clouds shattering in their dynamical interaction with the hot volume-filling hot plasma and  subsequent gas cooling \citep{Gronke2020}. 

When the shock wave hits the multiphase IGM filament, the clouds are compressed on a crushing timescale $t_{crush} = L_c/V_c$, where $V_c$ is the shock velocity in the cloud and $L_c$ the size of the preshock cloud. The shocked clouds experience dynamical instabilities and mixing at the interface with the hot gas flow, as well as thermal instabilities induced by gas cooling, which provide an additional fragmentation mechanism occurring over the gas cooling timescale, $t_{cool}$, as the shock moves into the cloud. 
The gas stripped from the cloud may cool and condense as a result of thermal instability. This occurs if the gas cooling time $t_{cool}$ is smaller than $t_{crush}$. 
There is a wide range of parameter space (typically \nH$\approx 0.1-100$~\cc, $L_c=1-100$~pc) where $t_{crush} > t_{cool}$, i.e. molecular gas can form before the shock has crossed the cloud and thus before dynamical instabilities fully develop \citep[see Fig.~4 in][]{Guillard2009}. The cooling time in the warm molecular cloudlets is mostly driven by \Hmol\ cooling, therefore the dust-to-gas mass ratio could, in principle, indirectly affect the \Hmol\ re-formation in the postshock gas, and therefore the cooling time. However, modelling of the dust-to-gas ratio, and the the PAHs to very small grains abundance ratio, based on Spitzer and Herschel spectroscopy and photometry, show that these are comparable to the Galactic value in the molecular gas \citep{Guillard2010}. This again suggests that dust survives the slower shocks driven by the dynamical instabilities and mixing into the molecular cloudlets.

Because of thermal instability, the post-shock clouds shatter in many clumplets, which then are seeds for further cooling-induced condensation, but also mixing \citep{Farber2022a}. In this parameter space where the cooling time is shorter than the cloud destruction timescale, this shattering process causes the gas to cycle through the ISM phases with the reformation of cold clumpy gas. Cloud growth depends on density, as well as on local heating processes \citep{Jennings2022}. Given the large masses of diffuse molecular gas observed in the IGM where atomic \Hi\ gas is undetected, qualitatively, the post-shock physical conditions must be in the density and irradiation regime where the cloud evolution is dominated by shattering and perhaps cooling-driven coagulation.

Many of the structures detected with JWST along the main north-south molecular filament  are shaped like cometary tails, with Field 6 being a prime example (see Fig.~\ref{fig:regionsdef}). They could be signatures of a misty \Hmol\ outflow driven by the dynamical interaction between cold and hot ISM phases, e.g. ram pressure stripping of CO clouds as the hot X-ray emitting gas flows past them. This is illustrated in the zoom-in panel of Fig;~\ref{fig:schema}. The high abundance of warm H$_2$ relative to the CO derived cold gas mass discussed in \S~4.1 for both Field 6 and Field 5 lends support to this idea, despite their different spatial positions in the SQ group. We will return to these differences later.    
The pressure-driven flow velocity is larger than the turbulent gas velocity because the droplets are tiny. At the scale of an \Hmol\ droplet, the difference between the turbulent velocity dispersion and the flow velocity is even larger. \citet{Farber2022a} suggest that in typical CGM or galaxy cluster environments, cloud shattering during the cooling process could be responsible for in-situ formation of a fog of warm (1000~K) gas embedded in hot ($10^{7-8}$~K) plasma, with molecular shattering length scales of the order of 0.1--100~pc. This would imply that fragments formed in such manner are very small, with a low volume filling fraction. A detailed investigation of the physical conditions, which could be achieved with JWST \Hmol\ spectroscopy, would allow us to determine whether the cloudlets can coagulate, or whether the molecular streams we observe remain misty.

\subsection{Multi-scale kinematical structures and survival of the cold gas}
\label{subsec:discussion_3}

The very complex, multi-scale kinematical and morphological  structures observed at high resolution with JWST and ALMA points towards a highly clumpy (misty) structure of the molecular gas in the IGM of SQ. This could explain why the  Ly$\alpha$ line emission is so strong and the line profile so broad \citep{Guillard2022}. Indeed, the cloudlets must have a small volume-filling fraction but their surface filling factor must be close to unity so Ly$\alpha$ photons can escape through multiple scatterings off the clump surfaces without being absorbed in the inter-clump, dust-free plasma.

The ALMA and JWST observations also shed light on a puzzle raised by the H$_2$ observations by Spitzer \citep{Appleton2006}, as well as single-dish CO observations, which showed extreme velocity dispersion ($\approx 1000$~\kms) of the molecular gas on kpc-scales. Those large scale velocity gradients, also observed in other gas phases, could partly originate from coherent bulk flows \citep{Guillard2022}. 
The H$_2$ molecules can only survive at shock velocities $\lesssim 20$~\kms, which is much smaller than those velocities at kpc scales, but also smaller than the linewidths  of the CO clouds. The ALMA observations presented in this paper indeed show that the CO clumps have velocity dispersions in the range of 40--100~\kms. Those velocities are also higher than the shock velocities needed to account for the  excitation of the pure rotational lines of \Hmol\, of the order of 5-20~\kms\ \citep{Guillard2009, Lesaffre2013, Appleton2017}. 

The tiny \Hmol\ droplets are unlikely to survive over the length of the tails, which supports the idea that they are continuously destroyed and reformed out of the more diffuse gas. This could explain why some of the CO clumps (in the tail of field 6 for instance) have high velocity dispersions (see Fig.~\ref{fig:FigureA1}), as they are reformed out of pressure-driven outflow from the parent cloud.
This will need to be further investigated with future \Hmol\ spectroscopy with JWST to study the velocity gradients of the warm \Hmol\ gas at the scales of the ALMA observations. Spectroscopy would also
allow us to see whether this fog of small warm clouds are being accelerated in the bulk flow of the hot X-ray emitting gas, and would likely differ from the motion of the massive CO clouds. The temperature of the warm droplets may also increase along the tail as the foggy clouds progressively mix and become heated in the faster flow. This might be evident in the excitation diagram of H$_2$ when a full complement of H$_2$ rotational lines can be measured. Those observations will motivate a direct comparison with future numerical simulations. 

The comparison between the warm and cold molecular gas spatial distributions in the IGM of SQ show that the powerful mid-IR emission originally discovered with Spitzer spectroscopy does not primarily have a one to one correlation with the cold CO(1-0) emitting gas. This supports the view of the model put forward in \citet{Guillard2009}, which involves a turbulent energy cascade and dissipation of mechanical energy within the warm \Hmol\ phase, which are both driven by the dynamical interaction between the hot and cold gas phases. The complex morphology and kinematics of the warm and cold \Hmol\ structures shows that this dynamical interaction occurs over a wide range of scales. This is expected from an energy cascade over many orders of magnitudes in spatial scales, from tens of kpc at the injection scale of mechanical energy from the galaxy collision to sub-pc dissipation scale through line emission \citep{Guillard2009}.

\subsection{Does the proposed molecular cloud outflow model apply across the whole SQ system?}
\label{subsec:discussion_4}

As we have stressed earlier, the high luminosity in the warm molecular hydrogen lines over a large part of  Stephan's Quintet system requires a common mechanism to convert a fraction of the kinetic energy resulting from the large-scale shock wave into rotational molecular hydrogen via low-velocity shocks \citep{Guillard2009,Appleton2017}. We have seen that, of the three regions we studied in detail here, the one that most closely fits the proposed cloudy outflow model is the head-tail structure in Field 6, where we seem to  be witnessing a possible break-up of a large cold molecular cloud into warmer H$_2$ in both the head and tail. 

Although modeling of the actual structure of a cold molecular cloud in a hot wind will require further simulation, its likely that cometary structures should be common along the main shock wave, and the diversity of structures should be similar to those seen in the cloud-survival sequences of models by \citet{Farber2022b}.  Those model sequences, where clouds are entrained in a fast wind, show that in some cases the cooler head and tail structures can survive for long periods, whereas in other cases, depending on the properties of the clouds, they can dissolve fairly quickly. Indeed, one of the models has a morphology remarkably similar to the shell-like CO and warm H$_2$ structure in the head of Field 6, including the spike feature at the head of the clump.  Many of the shapes of the bright H$_2$ emitting clumps seen along the main shock front in Stephan's Quintet (Figure~\ref{fig:fieldregions}) show suggestions of head/tail structures, and have filamentary structure very reminiscent of the modeled shapes seen when the head/tail structures survive. Others have shapes similar to the models where the clouds have lost their heads, and are in the process of dissolving. Therefore it is plausible that our picture of the foggy outflowing cloud picture in applicable to a large number of cloud structures seen in the main shock. 

In addition,  Figure~\ref{fig:fieldregions} shows head-tail structures far to the south and west of the main SQ-A region. One of those tails extends over 30-40 arcsecs (10-20 kpc!). These structures may represent the interaction of dense molecular clumps with the outer hot halo of the intruder far downstream of the main shock. More observations of the cold and warm molecular gas, and their kinematics will be needed to test this idea. 

In the case of our study of the Field 4 region in SQ-A, it is less clear that the cloud outflow model is applicable, although we note that a faint irregular tail is seen extending to the east of the main concentration of gas and stars. \citet{Xu1999} has previously suggested that SQ-A is too young to have formed as a tidal dwarf galaxy, but is more likely a starburst region created as part of the main shocked filament.  The H$_2$ gas temperature measured by {\it Spitzer} in SQ-A is much lower than other regions in the main filament \citep{Appleton2017}, and the gas may have cooled to the point that stars can form in large quantities there. Although understanding how those stars form is beyond the scope of this paper, we have shown in \citet{Guillard2009}, that the large ($>$ 50 pc), dense (n$_H$~$>$~100 cm$^{-3}$)  pre-shock clouds are expected to be gravitationally unstable in the hot gas at this pressure (see their Fig.~4). Therefore, their evolution will be different than the more diffuse gas, and should trigger star formation. This may explain star formation in SQ-A, and other star forming regions scattered in the IGM. Those large pre-existing clouds have crushing times longer than the SQ collision age (believed to be a few tens of millions of years). However, such clouds must be rare, otherwise we would see major bursts of star formation all over the main filament, which is not observed.

The case of Field 5 (in the bridge between NGC 7319 and the main filament) is interesting. As we have shown in \S~4, the morphology of the warm and cold gas is not so obviously a head-tail structure, although there is a warm gas filament running through the region. This region, may owe its overall morphology and kinematics to another process--for example the collision of a dense molecular concentration (the compact core) with a diffuse region near NGC 7319, driving a ring into the target material. The elongation of the compact CO emitting core along the direction which points towards the center of the CO emitting ring, and the existence of a clump of CO in the ring at the same velocity as the core may support a dynamical connection. Future warm H$_2$ spectroscopy will determine whether the warm gas connecting the two CO structures with disparate radial velocities are related, or merely chance alignments of two different cloud systems along the line of sight.  Such events, if they occur, are likely to be relatively rare, and unlikely to provide a generic mechanism for the dissipation of collisional kinetic energy into widespread warm H$_2$ emission.

Finally we come to the complex network of filaments around NGC 7319, which, for the most part, appear pink in the false color representation of Figure~\ref{fig:fieldregions}. This implies that the 7.7$\mu$m and 15$\mu$m emission is much stronger here than in the regions which are warm H$_2$ dominated. 

The irregular radial structure of the filaments might suggest a nuclear origin connected with the AGN in NGC 7319. The JWST MRS data in the central region of the AGN indeed reveal a spatially-resolved ionized and molecular outflow, with kinematical evidence of an interaction between the AGN jet and the host ISM   \citep{Pereira-Santaella2022}. Perhaps previous jet activity from the AGN has somehow formed these strange filaments. 

\section{Conclusions}
We have made a detailed comparison, at sub-arcsec resolution, between previous HST, recent ERO JWST NIRCam and MIRI imaging and ALMA CO (2-1) imaging spectroscopy of three targeted regions approximately 3 kpc in size in the large-scale (40-50 kpc) intergalactic gas in Stephan's Quintet. Based on a comparison with a full spectral map by {\it Spitzer} and observations made by JWST, we demonstrated that much of the gas along the main molecular filament detected in the F1000W/MIRI filter is dominated by strong emission from the 0-0 S(3) line originating in warm molecular hydrogen. This allowed us to reveal differences in spatial distribution of the warm H$_2$ gas phase, measured by JWST, and the colder CO-emitting gas, at comparable subarcsec resolution in three regions targeted by ALMA. Our study has come to the following conclusions:  

\begin{itemize}

\item{The three zoomed-in regions we studied show a great variety of structure in the warm H$_2$ when compared with the cold molecular gas on scales from 150 pc to 3 kpc. In many cases diffuse ionized gas (observed by HST and F200W/NIRCam) follows the warm H$_2$ distribution in the two structures believed to be shock-dominated (ALMA Field 5 and 6), whereas the cold molecular gas was more clumpy, and not always strongly correlated with the warm H$_2$. 
Both structures show an overabundance of warm H$_2$ over the colder gas with reasonable assumption about the conversion from CO line intensity to total H$_2$ mass. 
In the star formation dominated region (at the center of ALMA Field 4), the 10$\mu$m and 7.7$\mu$m bands are well correlated, as expected in regions dominated by PDRs.}

\item{In the (ALMA Field 6) region observed near the center of the main shocked filament, the warm H$_2$ emission resembles a  head-tail structure. Strong CO emission forms a partial shell of cold H$_2$ which is embedded in a bright extended hotspot of warm H$_2$ (F1000W emission), containing approximately $4 \times 10^6$~ M$_{\odot}$ of warm gas if we assume the excitation of the gas is similar to that inferred on a larger scale by {\it Spitzer}. This triangular-shaped head also shows extended ionized gas. The tail is composed of clumps of warm and cold H$_2$ emission scattered in two streams eastwards from the head. Near the apex of the triangular warm H$_2$ head and in some regions along the tail, CO linewidths of 60 $<$ FWHM (\kms) $<$ 100 \kms~ are observed suggesting quite turbulent (probably gravitational unbound) gas. We suggest that we are witnessing the shattering of a dense cold molecular gas cloud in a fast hot wind, leading a fog of warm H$_2$ (see below). We believe that this process is a commonly occurring phenomenon in Stephan's Quintet, judging by the existence of other head-tail strcutures along the main shock and elsewhere in the group.}

\item{In another shock-dominated warm H$_2$ region (ALMA Field 5) which is not part of the main molecular filament, we notice a different morphology.  Warm molecular gas appears to connect together a compact elongated CO emitting clump which points towards the center of a ring of CO emission separated by 1.5 arcsec ($\sim$ 700 pc).  The  two CO structures have very different radial velocities, with the compact core having strongly blueshifted emission with respect to the ring. Both the ring and the core show broad ($\sim$ 100 \kms FWHM) CO line-widths, suggesting that they are highly turbulent. One possibility is that a dense molecular cloud (perhaps from the outer parts of the intruder galaxy) has impacted a low-density gas filament in the outer parts on NGC 7319, entraining gas and dissipating kinetic energy which heats the bridge between them. Such events are probably rare. Alternatively, the connection between the two CO clouds by the warm gas may be entirely coincidental.}

\item{The distribution of warm, cold and ionized gas in the SQ-A star forming region (ALMA Field 4) is more typical of the gas distribution of a small dwarf galaxy.  The warm and cold molecular gas are contained within a roughly elongated stellar distribution (as observed by NIRCam). The lack of clear rotation in the CO-emitting gas may argue against this object being created in a tidal stream. Instead, we suggest it may be more consistent with the collapse of a rare pre-existing dense molecular structure which has become gravitationally unstable after it was compressed by the passage of the large-scale shock, leading to a starburst.} 

We present a conceptual picture for the transfer of kinetic energy from the intruder galaxy's large scale shock into a pre-existing multi-phase tidal filament via the formation and subsequent molecular cloud shattering of cold molecular clouds in a post-shock layer. After the main shock passes over a multi-phase tidal filament, both X-ray emitting hot gas, and dense molecular clouds can form together if the molecular formation time on dust grains is shorter than the cloud-crushing time of the clouds. However, as the molecular clouds grow, the velocity of the denser forming clouds relative to the diffuse hot gas causes them to experience ram-pressure stripping from the hot component. Models show that in this situation, the cold molecular clouds can shatter into a fog of tiny molecular clouds, heated by an exchange in energy with the hot medium. The fog clouds are expected to have low internal velocity dispersion which we associate with the clouds that radiate mid-IR H$_2$ emission (low shock velocities in the molecular gas).  The warm and cold gas associated with ALMA Field 6 is likely one example of many such cold clouds being turned into a warm H$_2$ fog all along the main SQ large-scale shock front. Regions outside the main molecular shock front also seem to show a head-tail morphology (for example one extremely long, 20kpc tail to the SW of the SQ-A star forming region).  These may be examples of a similar process,  where the flow of hot gas may originate in the relative motion of the intruder galaxy's hot halo relative to possible pre-existing molecular clouds in the system. Future mid-IR spectroscopy will allow us to test this picture, as the kinematics of the warm H$_2$ emitting fog is likely to be very different from that of the cold clouds, as they rapidly accelerate up to the flow velocity of the hot gas,  before eventually being destroyed.

\end{itemize}
\begin{acknowledgements}
We acknowledge the remarkable dedication of the scientists and engineers who made the James Web Space Telescope possible, and especially the Early Release Observation (ERO) science team upon which some of our paper is based. 
This work is based, in part, on observations made with the NASA/ESA/CSA James Webb Space Telescope. The data were obtained from the Mikulski Archive for Space Telescopes at the Space Telescope Science Institute, which is operated by the Association of Universities for Research in Astronomy, Inc., under NASA contract NAS 5-03127 for JWST. These observations are associated with program \# 2732. 
This paper makes use ALMA data from program ID: 2015.1.000241 (P.I. P. Guillard). ALMA is a partnership of ESO (representing its member states), NSF (USA) and NINS (Japan), together with NRC (Canada), MOST and ASIAA (Taiwan), and KASI (Republic of Korea), in cooperation with the Republic of Chile. The Joint ALMA Observatory is operated by ESO, AUI/NRAO and NAOJ.
The National Radio Astronomy Observatory is a facility of the National Science Foundation operated under cooperative agreement by Associated Universities, Inc.
E.O'S. acknowledges support from NASA through \textit{Chandra} Award Number GO8-19112A. PG would like to thank the Sorbonne University, the Institut Universitaire de France, the Centre National d'Etudes Spatiales (CNES), the "Programme National de Cosmologie and Galaxies" (PNCG) and the "Physique Chimie du Milieu Interstellaire" (PCMI) programs of CNRS/INSU, with INC/INP co-funded by CEA and CNES,  for there financial supports. We wish to thank an anonymous referee for very helpful comments on the manuscript. 
\end{acknowledgements}
\newpage

\appendix
\restartappendixnumbering
\section{ALMA and CARMA CO Kinematics of the ALMA Field Regions 4,5 and 6} 

\begin{figure*}[ht]
\includegraphics[width=0.99\textwidth]{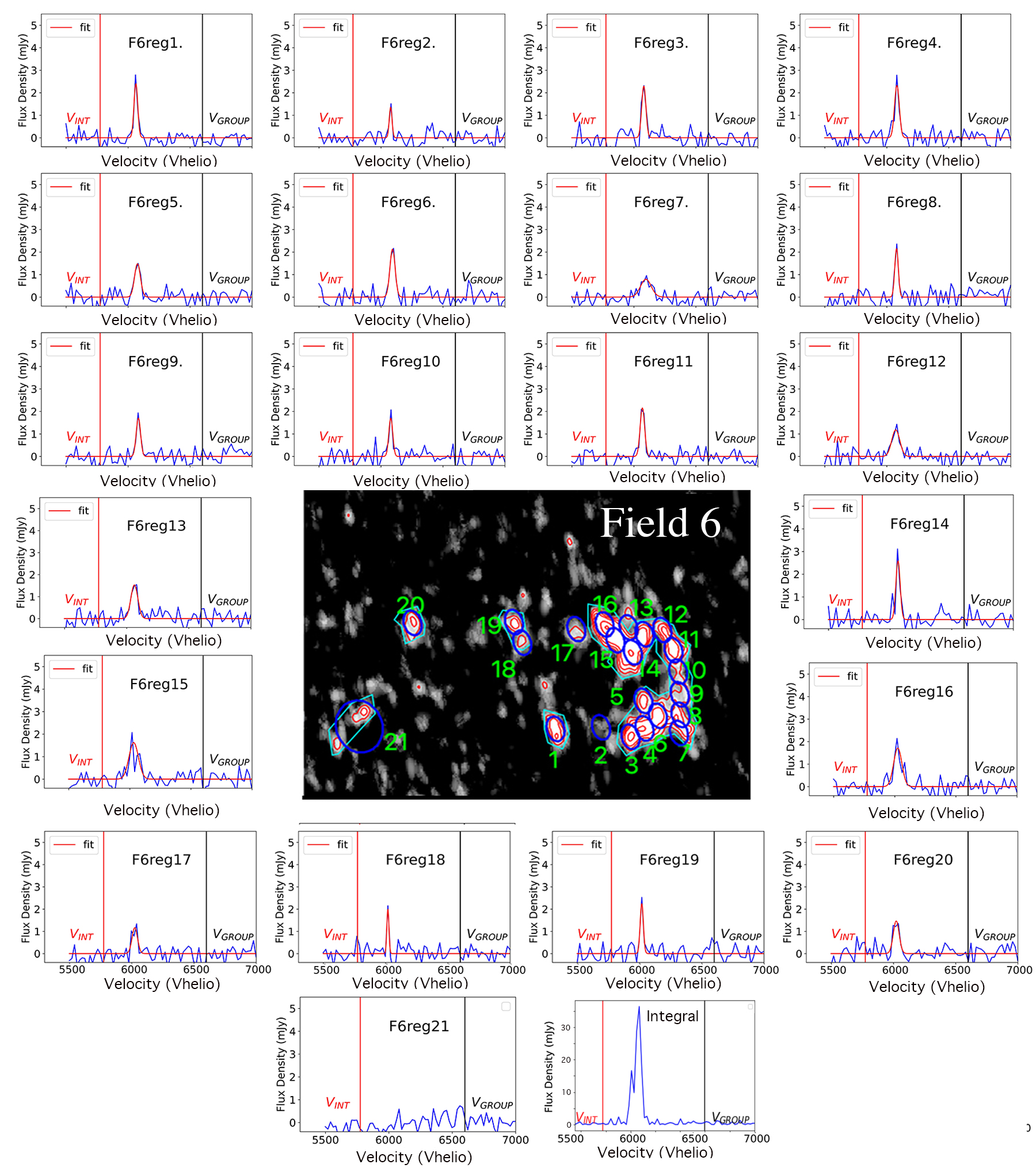}
\caption{Field 6 extracted spectra for CO (2-1) spectra covering, the CO beam-size extraction regions (blue ellipses on the inset) labeled in in inset at the center of the figure. The inset shows a grey-scale representation of the CO integrated emission.  The cyan polygons around the emission show the extracted areas which were summed for an integrated spectrum over the whole region. Gaussian fits to the CO line profiles are shown superimposed on the spectra with a red line.  Region 21 shows faint emission and was not fit. The red and blue vertical lines indicate the average velocity of the intruder galaxy NGC 7318b, and the average group velocity of the main SQ members respectively. The integrated spectrum obtained over the cyan polygons is also presented  as the last spectrum. The integrated properties of all the spectra are given in Table~\ref{tab:COobs}.} \label{fig:FigureA1}
\end{figure*}

\begin{figure*}
\includegraphics[width=0.99\textwidth]{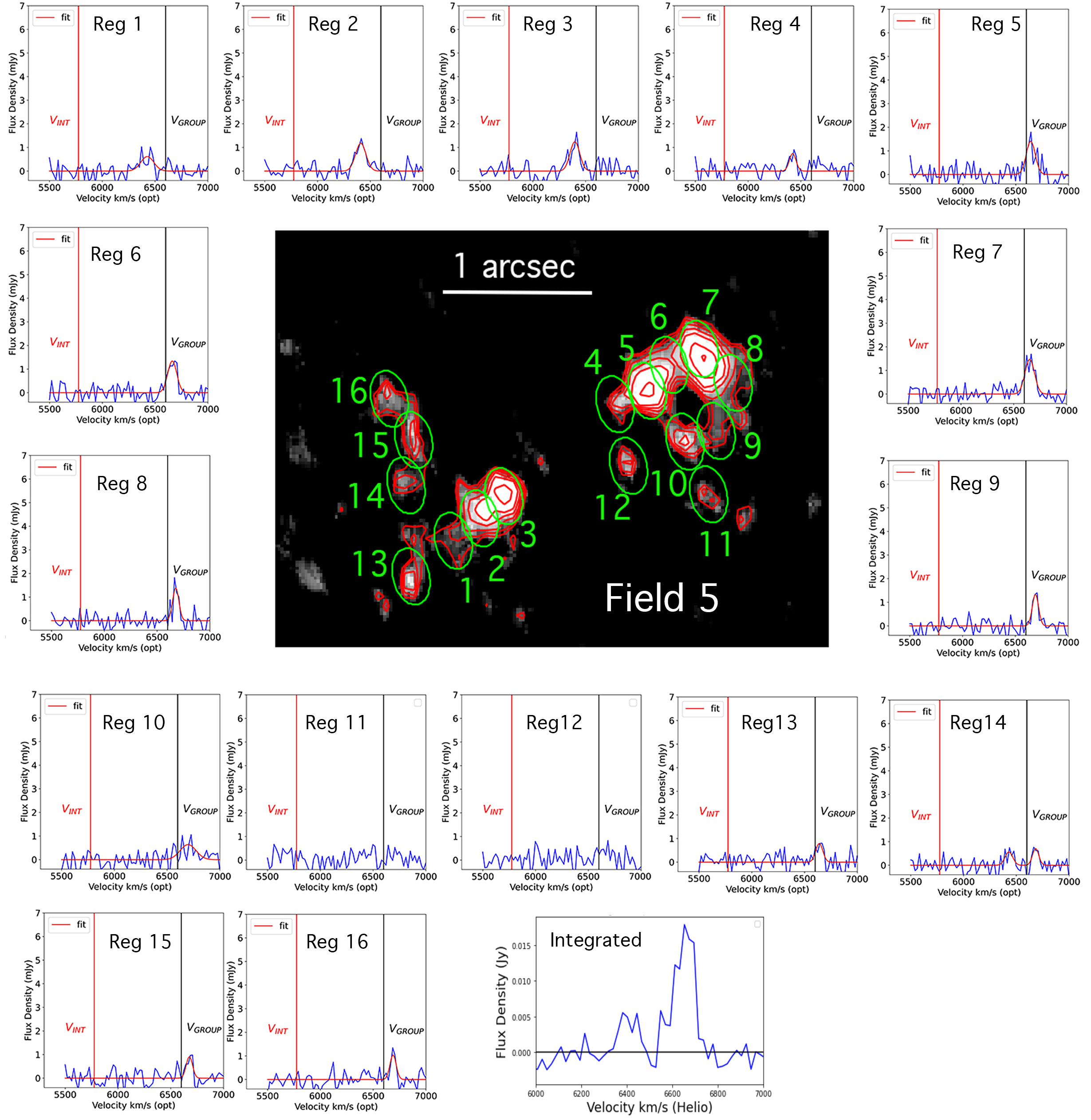}
\caption{Field 5 extracted spectra for CO (2-1) spectra covering the CO beam-size extraction regions (green ellipses) labeled on the inset at the center of the figure. The inset shows a grey-scale representation of the CO integrated emission, with contours of the same emission superimposed in red. Gaussian fits to the CO line profiles are shown overlayed on each of the spectra with a red line.  Regions 11 and 12 are too faint to fit. The red and blue vertical lines indicate the average velocity of the intruder galaxy NGC 7318b, and the average group velocity of the main SQ members respectively. The integrated spectrum obtained over the cyan polygons is also presented  as the last spectrum. The integrated properties of all the spectra are given in Table~\ref{tab:COobs2}. Note that Regions 1, 2, 3, and 4 have significantly different lower velocities than the rest of the emission. Region 14 also shows a double profile (labeled A and B), with faint emission from the low velocity structure as well as the higher velocity structure (13, 14, 15 and 16). } \label{fig:FigureA2}
\end{figure*}

\begin{figure*}
\includegraphics[width=0.99\textwidth]{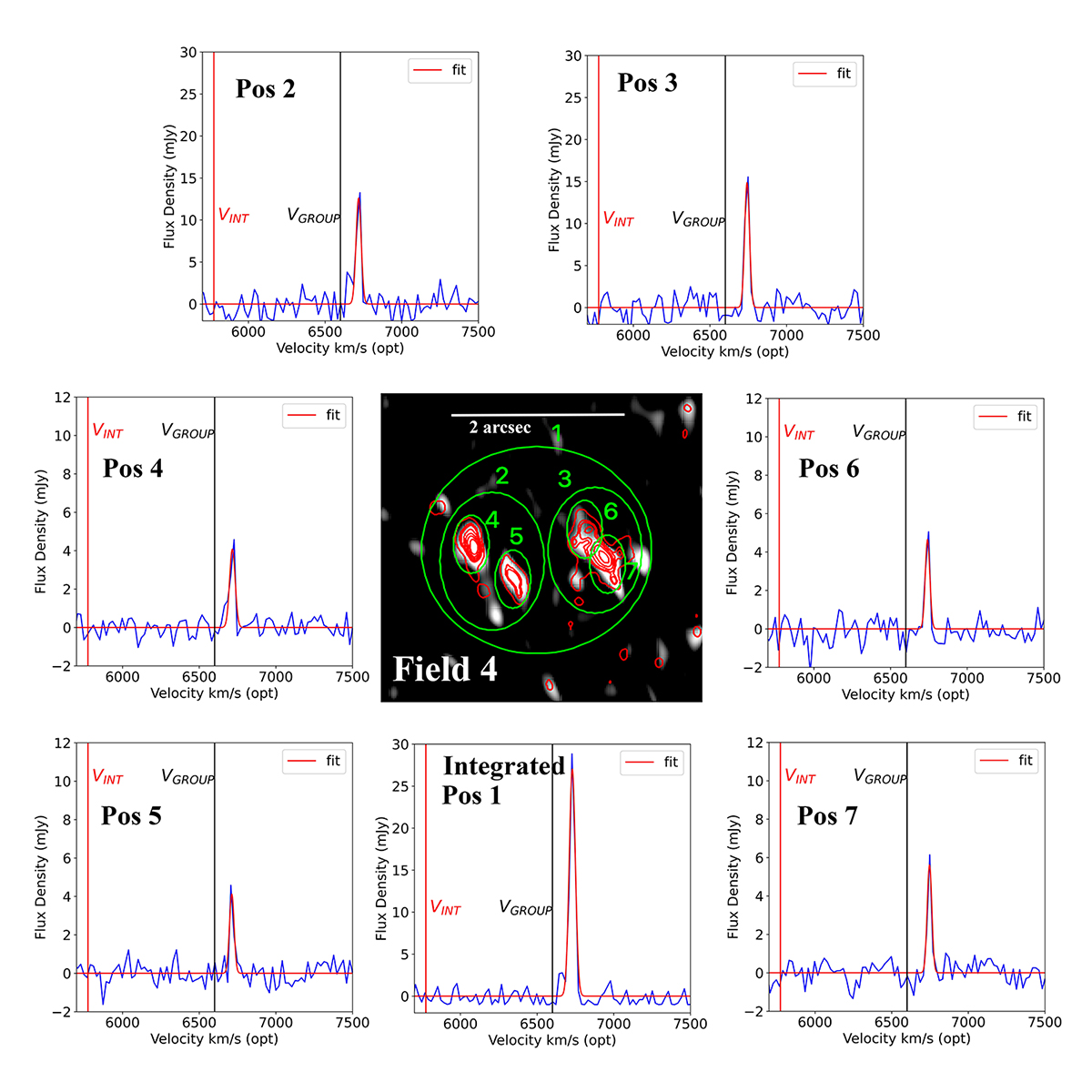}
\caption{Field 4 extracted spectra for CO (2-1) spectra covering the CO extraction regions (green ellipses) labeled on the inset at the center of the figure. The inset shows a grey-scale representation of the CO integrated emission, with contours of the same emission superimposed in red. Gaussian fits to the CO line profiles are shown overlayed on each of the spectra with a red line. The red and black vertical lines indicate the average velocity of the intruder galaxy NGC 7318b, and the average group velocity of the main SQ members respectively.  } \label{fig:FigureA3}
\end{figure*}

In this Appendix we present Table~\ref{tab:COobs}, Table~\ref{tab:COobs2} and Table~\ref{tab:COobs3}, which provide observed and derived properties of individually extracted spectra for Field6, Field 5 and Field4 respectively based on the CO (2-1) ALMA data. The regions (e. g. F6reg1) refer to the defined region extracted from each of the ALMA fields shown in Figure~\ref{fig:FigureA1}, Figure~\ref{fig:FigureA2}, and Figure~\ref{fig:FigureA3} respectively. Those figures show a grey-scale representation of the surface density map of the CO integrated emission with elliptical (blue) beam-size extraction areas marked. The spectra and fitted Gaussian profiles are also shown.  At the bottom of Figure~\ref{fig:FigureA1} and Figure~\ref{fig:FigureA2} we present the integrated emission profiles. For Figure~\ref{fig:FigureA3}, the integrated spectrum is F4p1.  

In each table we provide the following information. Column 1 is the region name, Column 2 is the heliocentric velocity of the emission, with uncertainty, Column 3 is the FWHM (and uncertainty) of the spectrum in \kms, and Columns 4 and 5 provide the CO line integral in Jy-\kms~ for the observed CO (2-1) emission profile and the derived equivalent CO (1-0) line integral respectively. The assumption for the conversion to a CO (1-0) line integral is given in the footnotes. Column 6 give the total (cold) H$_2$  mass in each extracted region assuming a Galactic value for the conversion from CO line intensity to molecular column density (see footnotes). Column 7 gives the total gas mass assuming 36$\%$ Helium content. 

In the shock dominated regions Field 5 and 6, the typical mass within the ALMA beam within the brighter regions is a few $\times 10^6 M_{\odot}$, whereas the brightest regions in Field 4 (SQ-A) are a factor of 8 to ten times higher, indicating significantly higher hydrogen columns in SQ-A. This difference may be even higher if the value for X$_{co}$ is larger in the shock-dominated regions.

\begin{deluxetable*}{lllllll}
\tablecolumns{7}
\tablewidth{0pc}
\tablecaption{Properties of CO emission for beam-sized extractions in Field 6 (see Figure~\ref{fig:FigureA1}) 
\label{tab:COobs}}
\tablehead{
\colhead{Region} &
\colhead{V$_{helio}$} & 
\colhead{FWHM} &
\colhead{$\Sigma(S_v\Delta$$V$)} &
\colhead{$\Sigma(S_v\Delta$$V$)\tablenotemark{a}} &
\colhead{M(H$_2$)\tablenotemark{a}} &
\colhead{M$_{gas}$\tablenotemark{b}} 
\\
\colhead{} & 
\colhead{} & 
\colhead{} &
\colhead{CO(2-1)} &
\colhead{CO(1-0)} &
\colhead{$\times10^6$} &
\colhead{$\times10^6$}
\\
\colhead{} & 
\colhead{(\kms)} & 
\colhead{(\kms)} &
\colhead{(mJy-km s$^{-1}$)} &
\colhead{(mJy-km s$^{-1}$)} &
\colhead{(M$_{\odot}$)} &
\colhead{(M$_{\odot}$)}
}
\startdata
F6reg1 &  6052 ($\pm$1) &    36 ($\pm$3) &       92 ($\pm$ 8) &       29 ($\pm$2) &    1.9 ($\pm$0.2) &    2.6 ($\pm$0.2) \\
F6reg2 &  6067 ($\pm$2) &    27 ($\pm$5) &       38 ($\pm$ 7) &       12 ($\pm$2) &    0.8 ($\pm$0.1) &    1.1 ($\pm$0.2) \\
F6reg3 &  6069 ($\pm$2) &    40 ($\pm$4) &       98 ($\pm$10) &       31 ($\pm$3) &    2.0 ($\pm$0.2) &    2.8 ($\pm$0.3) \\
F6reg4 &  6071 ($\pm$1) &    46 ($\pm$3) &      113 ($\pm$ 9) &       35 ($\pm$3) &    2.4 ($\pm$0.2) &    3.2 ($\pm$0.3) \\
F6reg5 &  6065 ($\pm$2) &    56 ($\pm$6) &       88 ($\pm$10) &       28 ($\pm$3) &    1.8 ($\pm$0.2) &    2.5 ($\pm$0.3) \\
F6reg6 &  6082 ($\pm$2) &    50 ($\pm$4) &      113 ($\pm$10) &       35 ($\pm$3) &    2.4 ($\pm$0.2) &    3.2 ($\pm$0.3) \\
F6reg7 &  6086 ($\pm$6) &   102 ($\pm$13) &       85 ($\pm$12) &       27 ($\pm$4) &    1.8 ($\pm$0.2) &    2.4 ($\pm$0.3) \\
F6reg8 &  6068 ($\pm$2) &    34 ($\pm$4) &       78 ($\pm$9) &       24 ($\pm$3) &    1.6 ($\pm$0.2) &    2.2 ($\pm$0.3) \\
F6reg9 &  6072 ($\pm$2) &    42 ($\pm$4) &       75 ($\pm$8) &       23 ($\pm$3) &    1.6 ($\pm$0.2) &    2.1 ($\pm$0.2) \\
F6reg10 &  6068 ($\pm$2) &    37 ($\pm$4) &       68 ($\pm$8) &       21 ($\pm$3) &    1.4 ($\pm$0.2) &    1.9 ($\pm$0.2) \\
F6reg11 &  6058 ($\pm$1) &    43 ($\pm$3) &       99 ($\pm$8) &       31 ($\pm$3) &    2.1 ($\pm$0.2) &    2.8 ($\pm$0.2) \\
F6reg12 &  6061 ($\pm$3) &    77 ($\pm$7) &       96 ($\pm$10) &       30 ($\pm$3) &    2.0 ($\pm$0.2) &    2.7 ($\pm$0.3) \\
F6reg13 &  6051 ($\pm$3) &    66 ($\pm$6) &      107 ($\pm$11) &       34 ($\pm$4) &    2.2 ($\pm$0.2) &    3.1 ($\pm$0.3) \\
F6reg14 &  6053 ($\pm$1) &    41 ($\pm$3) &      113 ($\pm$9) &       35 ($\pm$3) &    2.4 ($\pm$0.2) &    3.2 ($\pm$0.3) \\
F6reg15 &  6021 ($\pm$3) &    91 ($\pm$7) &      156 ($\pm$14) &       49 ($\pm$4) &    3.3 ($\pm$0.3) &    4.4 ($\pm$0.4) \\
F6reg16 &  6013 ($\pm$2) &    79 ($\pm$6) &      144 ($\pm$11) &       45 ($\pm$4) &    3.0 ($\pm$0.2) &    4.1 ($\pm$0.3) \\
F6reg17 &  6014 ($\pm$3) &    52 ($\pm$7) &       64 ($\pm$10) &       20 ($\pm$3) &    1.3 ($\pm$0.2) &    1.8 ($\pm$0.3) \\
F6reg18 &  6007 ($\pm$1) &    20 ($\pm$3) &       42 ($\pm$6) &       13 ($\pm$2) &    0.9 ($\pm$0.1) &    1.2 ($\pm$0.2) \\
F6reg19 &  6007 ($\pm$1) &    30 ($\pm$3) &       72 ($\pm$8) &       23 ($\pm$3) &    1.5 ($\pm$0.2) &    2.1 ($\pm$0.2) \\
F6reg20 &  6012 ($\pm$3) &    56 ($\pm$6) &       88 ($\pm$11) &       27 ($\pm$3) &    1.8 ($\pm$0.2) &    2.5 ($\pm$0.3) \\
Integrated\tablenotemark{c} & -- & -- & 2380 ($\pm$230) & 743 ($\pm$70) & 49.69 ($\pm$5.0) & 67.6 ($\pm$7.0)  \\
\enddata
\tablenotetext{a}{We assume a conversion between a line flux at CO~(1-0) to that of CO~(2-1) of
S$_{CO(1-0)}$ /S$_{CO(2-1)}$ = $(\nu_{1-0}/\nu_{2-1})^2$ $(r_{21})^{-1}$, where $\nu_{1-0}$ and $\nu_{2-1}$, where $\nu_{1-0}$ and $\nu_{2-1}$ are the rest frequencies of the transitions, and $r_{21}$ is assumed to be 0.8, similar to that measured in the Taffy galaxy bridge \citep{Zhu2007}, which shares many similarities with the gas in Stephan's Quintet (see \citealt{Appleton2022}). The H$_2$ masses presented here are for an X$_{CO}$ value =  X$_{CO,20}$ = N(H$_2$)/I$_{CO}$ = 2 $\times$ $10^{20}$ cm$^{-2}$ ($K-$\kms)$^{-1}$, the standard value assumed for our Galaxy.  In the text we discuss how this may not be applicable in such a turbulent region. $M_{H2}$ = $7.72 \times 10^3 D^2 \Sigma(S_v$ $\Delta$$V$)(1+z)$^{-1}$, where D = 94 Mpc, and $ \Sigma(S_v$$\Delta$$V$) is the CO$_{(1-0)}$ line integral with $S_{v}$ in Jy and the velocity $V$ of the gas in \kms. We assume z = 0.02.}
\tablenotetext{b}{Total gas mass M$_{gas}$ includes a 36$\%$ correction for Helium \citep{Bolatto2013}.}
\tablenotetext{c}{Integral properties of the whole structure are based on the summed extracted spectrum obtained from the cyan polygons shown in the inset to Figure~\ref{fig:FigureA1} and the last spectrum in the same figure.}
\end{deluxetable*}

\begin{deluxetable*}{lllllll}
\tablecolumns{7}
\tablewidth{0pc}
\tablecaption{Properties of CO emission for beam-sized extractions in Field 5 (see Figure~\ref{fig:FigureA2})  
\label{tab:COobs2}
}
\tablehead{
\colhead{Region} &
\colhead{V$_{helio}$} & 
\colhead{FWHM} &
\colhead{$\Sigma($S$_v\Delta$$V$)} &
\colhead{$\Sigma($S$_v\Delta$$V$)\tablenotemark{a}} &
\colhead{M(H$_2$)\tablenotemark{a}} &
\colhead{M$_{gas}$\tablenotemark{b}} 
\\
\colhead{} & 
\colhead{} & 
\colhead{} &
\colhead{CO(2-1)} &
\colhead{CO(1-0)} &
\colhead{$\times10^6$} &
\colhead{$\times10^6$}
\\
\colhead{} & 
\colhead{(\kms)} & 
\colhead{(\kms)} &
\colhead{(mJy-km s$^{-1}$)} &
\colhead{(mJy-km s$^{-1}$)} &
\colhead{(M$_{\odot}$)} &
\colhead{(M$_{\odot}$)}
}
\startdata
F5reg1  &  6416 ($\pm$10) &   145 ($\pm$23) &    0.10 ($\pm$0.02) &    0.030 ($\pm$0.005) &    2.0 ($\pm$0.3) &    2.7 ($\pm$0.5) \\
F5reg2  &  6399 ($\pm$4) &   117 ($\pm$10) &    0.15 ($\pm$0.01) &    0.045 ($\pm$0.004) &    3.0 ($\pm$0.3) &    4.1 ($\pm$0.4) \\
F5reg3  &  6388 ($\pm$5) &   108 ($\pm$12) &    0.14 ($\pm$0.02) &    0.045 ($\pm$0.005) &    3.0 ($\pm$0.4) &    4.1 ($\pm$0.5) \\
F5reg4  &  6412 ($\pm$6) &    75 ($\pm$14) &    0.06 ($\pm$0.01) &    0.018 ($\pm$0.004) &    1.2 ($\pm$0.2) &    1.7 ($\pm$0.3) \\
F5reg5  &  6631 ($\pm$4) &   102 ($\pm$10) &    0.15 ($\pm$0.02) &    0.047 ($\pm$0.005) &    3.2 ($\pm$0.3) &    4.3 ($\pm$0.4) \\
F5reg6  &  6651 ($\pm$4) &   101 ($\pm$8) &    0.14 ($\pm$0.01) &    0.045 ($\pm$0.004) &    3.0 ($\pm$0.3) &    4.1 ($\pm$0.4) \\
F5reg7  &  6640 ($\pm$3) &   107 ($\pm$8) &    0.17 ($\pm$0.01) &    0.052 ($\pm$0.004) &    3.5 ($\pm$0.3) &    4.7 ($\pm$0.4) \\
F5reg8  &  6666 ($\pm$2) &    60 ($\pm$6) &    0.10 ($\pm$0.01) &    0.030 ($\pm$0.003) &    2.0 ($\pm$0.2) &    2.7 ($\pm$0.3) \\
F5reg9  &  6679 ($\pm$2) &    61 ($\pm$6) &    0.09 ($\pm$0.01) &    0.029 ($\pm$0.003) &    1.9 ($\pm$0.2) &    2.6 ($\pm$0.3) \\
F5reg10 &  6688 ($\pm$12) &   177 ($\pm$29) &    0.12 ($\pm$0.04) &    0.037 ($\pm$0.014) &    2.5 ($\pm$0.8) &    3.4 ($\pm$1.2) \\
F5reg11\tablenotemark{d} &  --  &    --  &    --  &    --  &    --  &    --  \\
F5reg12\tablenotemark{d} &  --  &    --  &    --  &    --  &    --  &    --   \\
F5reg13 &  6628 ($\pm$6) &    79 ($\pm$15) &    0.07 ($\pm$0.01) &    0.021 ($\pm$0.004) &    1.4 ($\pm$0.3) &    1.9 ($\pm$0.4) \\
F5reg14A\tablenotemark{e} &  6419 ($\pm$9) &    87 ($\pm$21) &    0.052 ($\pm$0.02) &    0.016 ($\pm$0.008) &    1.1 ($\pm$0.6) &    1.5 ($\pm$0.8) \\
F5reg14B\tablenotemark{e} &  6671 ($\pm$5) &    59 ($\pm$12) &    0.05 ($\pm$0.01) &    0.015 ($\pm$0.003) &    1.0 ($\pm$0.2) &    1.4 ($\pm$0.3) \\
F5reg15 &  6671 ($\pm$5) &    59 ($\pm$12) &    0.05 ($\pm$0.01) &    0.015 ($\pm$0.003) &    1.0 ($\pm$0.2) &    1.4 ($\pm$0.3) \\
F5reg16 &  6680 ($\pm$3) &    48 ($\pm$6) &    0.07 ($\pm$0.01) &    0.021 ($\pm$0.003) &    1.4 ($\pm$0.2) &    1.9 ($\pm$0.3) \\
Integrated & -- & -- & 2300 ($\pm$230) & 718 ($\pm$70) & 48.1 ($\pm$4) & 6.54 ($\pm$7) 
\\
\enddata
\tablenotetext{a}{We assume a conversion between a line flux at CO~(1-0) to that of CO~(2-1) of
S$_{CO(1-0)}$ /S$_{CO(2-1)}$ = $(\nu_{1-0}/\nu_{2-1})^2$ $(r_{21})^{-1}$, where $\nu_{1-0}$ and $\nu_{2-1}$ are the rest frequencies of the transitions, and $r_{21}$ is assumed to be 0.8, similar to that measured in the Taffy galaxy bridge \citep{Zhu2007}, which shares many similarities with the gas in Stephan's Quintet (see \citealt{Appleton2022}). The H$_2$ masses presented here are for an X$_{CO}$ value =  X$_{CO,20}$ = N(H$_2$)/I$_{CO}$ = 2 $\times$ $10^{20}$ cm$^{-2}$ ($K-$\kms)$^{-1}$, the standard value assumed for our Galaxy.  In the text we discuss how this may not be applicable in such a turbulent region. $M_{H2}$ = $7.72 \times 10^3 D^2 \Sigma(S_v$ $\Delta$$V$)(1+z)$^{-1}$, where D = 94 Mpc, and $ \Sigma(S_v$$\Delta$$V$) is the CO$_{(1-0)}$ line integral with $S_{v}$ in Jy and the velocity $V$ of the gas in \kms. We assume z = 0.02.}
\tablenotetext{b}{Total gas mass includes a 36$\%$ correction for Helium \citep{Bolatto2013}.}
\tablenotetext{c}{Integral properties of the whole structure are based on the summed extracted spectrum obtained from the cyan polygons shown in the inset to Figure~\ref{fig:FigureA2} and the last spectrum in the same figure.}
\tablenotetext{d}{low signal to noise region}
\tablenotetext{e}{Two features present. 14A = faint broad emission, 14B = brighter higher velocity emission.}
\end{deluxetable*}

\begin{deluxetable*}{lllllll}
\tablecolumns{7}
\tablewidth{0pc}
\tablecaption{Properties of CO emission for beam-sized extractions in Field 4 (see Figure~\ref{fig:FigureA3})} \label{tab:COobs3}
\tablehead{
\colhead{Region} &
\colhead{V$_{helio}$} & 
\colhead{FWHM} &
\colhead{$\Sigma($S$_v\Delta V$)} &
\colhead{$\Sigma($S$_v\Delta V$)\tablenotemark{a}} &
\colhead{M(H$_2$)\tablenotemark{a}} &
\colhead{M$_{gas}$\tablenotemark{b}} 
\\
\colhead{} & 
\colhead{} & 
\colhead{} &
\colhead{CO(2-1)} &
\colhead{CO(1-0)} &
\colhead{$\times10^6$} &
\colhead{$\times10^6$}
\\
\colhead{} & 
\colhead{(\kms)} & 
\colhead{(\kms)} &
\colhead{(mJy-km s$^{-1}$)} &
\colhead{(mJy-km s$^{-1}$)} &
\colhead{(M$_{\odot}$)} &
\colhead{(M$_{\odot}$)}
}
\startdata
F4reg2 &  6708 ($\pm$1) &    40 ($\pm$3) &    0.53 ($\pm$0.04) &    0.166 ($\pm$0.013) &   11.1 ($\pm$0.9) &   15.1 ($\pm$1.2) \\
F4reg3 &  6731 ($\pm$1) &    37 ($\pm$2) &    0.59 ($\pm$0.04) &    0.183 ($\pm$0.012) &   12.3 ($\pm$0.8) &   16.7 ($\pm$1.1) \\
F4reg4 &  6706 ($\pm$1) &    43 ($\pm$3) &    0.19 ($\pm$0.02) &    0.058 ($\pm$0.005) &    3.9 ($\pm$0.3) &    5.3 ($\pm$0.5) \\
F4reg5 &  6702 ($\pm$1) &    35 ($\pm$3) &    0.15 ($\pm$0.02) &    0.048 ($\pm$0.005) &    3.2 ($\pm$0.3) &    4.4 ($\pm$0.5) \\
F4reg6 &  6731 ($\pm$1) &    35 ($\pm$3) &    0.17 ($\pm$0.02) &    0.054 ($\pm$0.006) &    3.6 ($\pm$0.4) &    5.0 ($\pm$0.5) \\
F4reg7 &  6736 ($\pm$1) &    35 ($\pm$2) &    0.21 ($\pm$0.02) &    0.066 ($\pm$0.005) &    4.4 ($\pm$0.3) &    6.0 ($\pm$0.4) \\
F4reg1(int)\tablenotemark{c} &  6719 ($\pm$1) &    47 ($\pm$1) &     1.4 ($\pm$0.04) &    0.422 ($\pm$0.012) &   28.2 ($\pm$0.8) &   38.4 ($\pm$1.1) \\
\enddata
\tablenotetext{a}{We assume a conversion between a line flux at CO~(1-0) to that of CO~(2-1) of
S$_{CO(1-0)}$ /S$_{CO(2-1)}$ = $(\nu_{1-0}/\nu_{2-1})^2$ $(r_{21})^{-1}$, where $\nu_{1-0}$ and $\nu_{2-1}$ are the ratios of the rest frequencies of the transitions, and $r_{21}$ is assumed to be 0.8, similar to that measured in the Taffy galaxy bridge \citep{Zhu2007}, which shares many similarities with the gas in Stephan's Quintet (see \citealt{Appleton2022}). The H$_2$ masses presented here are for an X$_{CO}$ value =  X$_{CO,20}$ = N(H$_2$)/I$_{CO}$ = 2 $\times$ $10^{20}$ cm$^{-2}$ ($K-$\kms)$^{-1}$, the standard value assumed for our Galaxy.  In the text we discuss how this may not be applicable in such a turbulent region. $M_{H2}$ = $7.72 \times 10^3 D^2 \Sigma(S_v$ $\Delta$$V$)(1+z)$^{-1}$, where D = 94 Mpc, and $ \Sigma(S_v$$\Delta$$V$) is the CO$_{(1-0)}$ line integral with $S_{v}$ in Jy and the velocity $V$ of the gas in \kms. We assume z = 0.02.}
\tablenotetext{b}{Total gas mass M$_{gas}$ = M$_{H2}$ $\times$ 1.36, includes a 36$\%$ correction for Helium \citep{Bolatto2013}.}
\tablenotetext{c}{Integral properties of the whole structure are based on the summed extracted spectrum obtained from position 1 (see Figure~\ref{fig:FigureA3}.) }
\end{deluxetable*}

\section{Photometry and Age Determination from the Fedotov (2014)  Catalog}

Photometry of the Stephan's Quintet system was performed by \citet{Fedotov2014} using UBV$_m$VI photometry based on observations made using HST and the F336W (U), F438W (B), F547W (V$_m$), F606W (V) and F814W (I)
with WFC3/UV.  The magnitudes where converted to the Vega system assuming zeropoints of 23.484 (U), 24.974 (B), 24.748 (V$_m$), 25.987 (V) and 24.680 (I). Corrections for foreground extinction assumed
A$_{336}$ = 0.353 mag, A$_{438}$ = 0.288 mag, A$_{606}$ = 0.197 mag, and A$_{814}$ = 0.122 mag. Super Star Clusters(SSCs) were extracted from the images using DAOfind. At the distance of Stephan's Quintet, the WFC3 resolution means that objects as large as 12 pc would be classified as SSCs, but may include more than one cluster. SSCs were selected with a color cut of B-V $<$  1.5  or V-I $<$ 1.0 in order to minimize detection of foreground objects. In addition, to be considered a SSC, the object had to have a photometric error of $<$ 0.3 mag, fullfill a sharpness criterion, and exhibit a goodness-of-fit criterion based on the shape of the object compared with the point-spread-function. 

Table~\ref{tab:tablephot} presents B, V$_m$, V, I photometry for objects we identify in the U-band images which correspond to the youngest objects associated with the three studied fields. For Field 4 (SQ-A) most of the objects with U-band detections were of moderate age (F4A-F) ranging from 72-360 Myrs. NIRCam detects a large number of potential star clusters in the bidy of this likely forming dwarf galaxy for which we defer discussion to another paper. In Field 5, only one object (F5D) is clearly detected in U-band with an age of 8 Myr. This SSC is one of the objects that lies in the string of potentially background galaxies to the East and North of the CO ring. In Field 6, F6A and B are the bright young stellar objects (3-5 Myr) with young ages seen associated with the PAH region in the NW CO clump structure seen in Figure\ref{fig:Field6}c. The other clumps (F6C, F6D and F6E) as quite close to the triangular head of the warm H$_2$ emission, and also exhibit young ages.  The RA and Dec positions are from the original catalog, and were found to be shifted slightly (0.25 arcsec) relative to the GAIA coordinate system used in the paper.  

The method used by \citep{Fedotov2014} to estimate the intrinsic  A$_v$ and age of the clusters was by comparison of the SED with population synthesis models. The use of the U-band and the V$_{574}$ filter, helps to break the age-reddening degeneracy, as well as avoiding significant contamination with H$\alpha$ emission in the F606W filter.  The age and extinction of each cluster was found by least-squares fitting the SEDs to predictions from (unpublished) updates to the \citet{Bruzual2003} models (known as CB07), and made available on request by the authors (S. Charlot and G. Bruzual). These models incorporate the work of \citep{Marigo2007} into the evolutionary code, and are described in more detail by \citet{Bruzual2011}. They assume a \citet{Chabrier2003} initial mass function. 
The best fits values for E(B-V), A$_v$ and age forthe clusters are given in the Table.

\begin{deluxetable*}{llllllllll}
\tablecolumns{10}
\tablewidth{0pc}
\tablecaption{UBVmI Photometric Properties of clumps in Field 4, 5 and 6 from catalog by Fedotov (2011) 
\label{tab:tablephot}} 
\tablehead{
\colhead{Name} &
\colhead{RA} &
\colhead{Dec} &
\colhead{U$_{F336W}$} &
\colhead{B$_{F435W}$} &
\colhead{Vm$_{F547M}$} &
\colhead{I$_{F814W}$} &
\colhead{ebv} &
\colhead{A$_v$} &
\colhead{Age}        
\\
\colhead{} &
\colhead{(J2000)} & 
\colhead{(J2000)} & 
\colhead{(mag (unc))} &
\colhead{(mag (unc))} &
\colhead{(mag (unc))} &
\colhead{(mag (unc))}&
\colhead{(mag)})&
\colhead{(mag)} &
\colhead{$\times10^7$~yr} 
\\
\colhead{} &
\colhead{[1]} & 
\colhead{[2]} & 
\colhead{[3]} &
\colhead{[4]} &
\colhead{[5]} &
\colhead{[6]} &
\colhead{[7]} &
\colhead{[8]} &
\colhead{[9]} 
}
\startdata
F4A & 338.99469 & 33.98040   & 24.371 (0.035) & 24.758 (0.05) & 24.253 (0.042) & 23.1780 (0.032) & 0.42 & 1.3 & 6.5 \\ 
F4B & 338.99490  & 33.980267 & 24.543 (0.085) & 24.732 (0.08) & 23.961 (0.075) & 22.864 (0.091) & 0.50 & 1.55 & 11.0 \\  
F4C & 338.99496 & 33.980377 & 22.887 (0.066) & 23.258 (0.07) & 22.743 (0.076) & 21.7290 (0.111) &  0.40 &  1.24 & 7.2 \\ 
F4D & 338.99539 & 33.980259 & 25.552 (0.046) & 25.234 (0.05) & 24.322 (0.041) & 22.805 (0.031) &  0.50 & 1.55 &  36.1 \\
F4E & 338.99536 & 33.98069  & 24.911 (0.041) & 24.902 (0.05) & 24.119 (0.032) & 22.686 (0.069) &  0.50 & 1.55 & 16.1 \\
F4F & 338.99564 & 33.980526 & 22.891 (0.14)  & 23.523 (0.12) & 22.675 (0.116) & 22.588 (0.149) &  -- &  -- & 36.2 \\
F4G & 338.99606 & 33.980625 & 22.373 (0.03)  & 22.982 (0.05) & 22.378 (0.055) & 21.534 (0.039) &  0.34 & 1.05 &  7.20 \\
F5D & 339.00900 &   33.974075 & 27.17 (0.12) &  28.58 (0.278) &  28.32 (0.27) &  26.70 (0.23) & 0.28 & 0.86 & 0.79  \\
F6A & 338.99918 &  33.972874 & 24.70 (0.08)  &   26.04 (0.09) &  26.10 (0.08) &  25.91 (0.09) &  0.08 &  0.24 & 0.53  \\
F6B &  338.99915 & 33.972858 & 24.41 (0.10)  &   25.62 (0.08) &  25.74 (0.10) &  25.91 (0.13) &  0.10 & 0.31 & 0.35 \\
F6C & 338.99896 & 33.972748  & 26.86 (0.14)  &   28.05 (0.20)  &  27.62 (0.11) &  27.065 (0.16) &  0.5  & 1.55 & 0.3 \\
F6D & 338.99896 & 33.972588  & 27.028 (0.15) &   28.31 (0.17) &   27.914 (0.17) & 27.711 (0.25) &  0.5  & 1.55 & 0.28\\
F6E & 338.99915 & 33.972572  & 27.225 (0.17) &   28.33 (0.19) &   28.594 (0.28) & 28.395 (0.39) &  0.06 & 0.18 & 0.52
\enddata
\end{deluxetable*}

\bibliographystyle{aasjournal}
\bibliography{paper.bib}

\begin{thebibliography}{}
\expandafter\ifx\csname natexlab\endcsname\relax\def\natexlab#1{#1}\fi
\providecommand{\url}[1]{\href{#1}{#1}}
\providecommand{\dodoi}[1]{doi:~\href{http://doi.org/#1}{\nolinkurl{#1}}}
\providecommand{\doeprint}[1]{\href{http://ascl.net/#1}{\nolinkurl{http://ascl.net/#1}}}
\providecommand{\doarXiv}[1]{\href{https://arxiv.org/abs/#1}{\nolinkurl{https://arxiv.org/abs/#1}}}

\bibitem[{{Alatalo} {et~al.}(2013){Alatalo}, {Davis}, {Bureau}, {Young},
  {Blitz}, {Crocker}, {Bayet}, {Bois}, {Bournaud}, {Cappellari}, {Davies}, {de
  Zeeuw}, {Duc}, {Emsellem}, {Khochfar}, {Krajnovi{\'c}}, {Kuntschner},
  {Lablanche}, {Morganti}, {McDermid}, {Naab}, {Oosterloo}, {Sarzi}, {Scott},
  {Serra}, \& {Weijmans}}]{Alatalo2013}
{Alatalo}, K., {Davis}, T.~A., {Bureau}, M., {et~al.} 2013, \mnras, 432, 1796,
  \dodoi{10.1093/mnras/sts299}

\bibitem[{{Allen} \& {Hartsuiker}(1972)}]{Allen1972}
{Allen}, R.~J., \& {Hartsuiker}, J.~W. 1972, \nat, 239, 324,
  \dodoi{10.1038/239324a0}

\bibitem[{{Appleton} {et~al.}(2006){Appleton}, {Xu}, {Reach}, {Dopita}, {Gao},
  {Lu}, {Popescu}, {Sulentic}, {Tuffs}, \& {Yun}}]{Appleton2006}
{Appleton}, P.~N., {Xu}, K.~C., {Reach}, W., {et~al.} 2006, \apjl, 639, L51,
  \dodoi{10.1086/502646}

\bibitem[{{Appleton} {et~al.}(2013){Appleton}, {Guillard}, {Boulanger},
  {Cluver}, {Ogle}, {Falgarone}, {Pineau des For{\^e}ts}, {O'Sullivan}, {Duc},
  {Gallagher}, {Gao}, {Jarrett}, {Konstantopoulos}, {Lisenfeld}, {Lord}, {Lu},
  {Peterson}, {Struck}, {Sturm}, {Tuffs}, {Valchanov}, {van der Werf}, \&
  {Xu}}]{Appleton2013}
{Appleton}, P.~N., {Guillard}, P., {Boulanger}, F., {et~al.} 2013, \apj, 777,
  66, \dodoi{10.1088/0004-637X/777/1/66}

\bibitem[{{Appleton} {et~al.}(2017){Appleton}, {Guillard}, {Togi}, {Alatalo},
  {Boulanger}, {Cluver}, {Pineau des For{\^e}ts}, {Lisenfeld}, {Ogle}, \&
  {Xu}}]{Appleton2017}
{Appleton}, P.~N., {Guillard}, P., {Togi}, A., {et~al.} 2017, \apj, 836, 76,
  \dodoi{10.3847/1538-4357/836/1/76}

\bibitem[{{Appleton} {et~al.}(2022){Appleton}, {Emonts}, {Lisenfeld},
  {Falgarone}, {Guillard}, {Boulanger}, {Braine}, {Ogle}, {Struck}, {Vollmer},
  \& {Yeager}}]{Appleton2022}
{Appleton}, P.~N., {Emonts}, B., {Lisenfeld}, U., {et~al.} 2022, \apj, 931,
  121, \dodoi{10.3847/1538-4357/ac63b2}

\bibitem[{{Bolatto} {et~al.}(2013){Bolatto}, {Wolfire}, \&
  {Leroy}}]{Bolatto2013}
{Bolatto}, A.~D., {Wolfire}, M., \& {Leroy}, A.~K. 2013, \araa, 51, 207,
  \dodoi{10.1146/annurev-astro-082812-140944}

\bibitem[{{Braine} \& {Combes}(1992)}]{Braine1992}
{Braine}, J., \& {Combes}, F. 1992, \aap, 264, 433

\bibitem[{{Bruzual} \& {Charlot}(2003)}]{Bruzual2003}
{Bruzual}, G., \& {Charlot}, S. 2003, \mnras, 344, 1000,
  \dodoi{10.1046/j.1365-8711.2003.06897.x}

\bibitem[{{Bruzual A.}(2011)}]{Bruzual2011}
{Bruzual A.}, G. 2011, in Revista Mexicana de Astronomia y Astrofisica
  Conference Series, Vol.~40, Revista Mexicana de Astronomia y Astrofisica
  Conference Series, 36--41

\bibitem[{{CASA Team} {et~al.}(2022){CASA Team}, {Bean}, {Bhatnagar}, {Castro},
  {Donovan Meyer}, {Emonts}, {Garcia}, {Garwood}, {Golap}, {Gonzalez Villalba},
  {Harris}, {Hayashi}, {Hoskins}, {Hsieh}, {Jagannathan}, {Kawasaki},
  {Keimpema}, {Kettenis}, {Lopez}, {Marvil}, {Masters}, {McNichols},
  {Mehringer}, {Miel}, {Moellenbrock}, {Montesino}, {Nakazato}, {Ott}, {Petry},
  {Pokorny}, {Raba}, {Rau}, {Schiebel}, {Schweighart}, {Sekhar}, {Shimada},
  {Small}, {Steeb}, {Sugimoto}, {Suoranta}, {Tsutsumi}, {van Bemmel},
  {Verkouter}, {Wells}, {Xiong}, {Szomoru}, {Griffith}, {Glendenning}, \&
  {Kern}}]{Casa2022}
{CASA Team}, {Bean}, B., {Bhatnagar}, S., {et~al.} 2022, \pasp, 134, 114501,
  \dodoi{10.1088/1538-3873/ac9642}

\bibitem[{{Chabrier}(2003)}]{Chabrier2003}
{Chabrier}, G. 2003, \pasp, 115, 763, \dodoi{10.1086/376392}

\bibitem[{{Clark} {et~al.}(2019){Clark}, {De Vis}, {Baes}, {Bianchi},
  {Casasola}, {Cassar{\`a}}, {Davies}, {Dobbels}, {Lianou}, {De Looze},
  {Evans}, {Galametz}, {Galliano}, {Jones}, {Madden}, {Mosenkov}, {Verstocken},
  {Viaene}, {Xilouris}, \& {Ysard}}]{Clark2019}
{Clark}, C.~J.~R., {De Vis}, P., {Baes}, M., {et~al.} 2019, \mnras, 489, 5256,
  \dodoi{10.1093/mnras/stz2257}

\bibitem[{{Cluver} {et~al.}(2010){Cluver}, {Appleton}, {Boulanger}, {Guillard},
  {Ogle}, {Duc}, {Lu}, {Rasmussen}, {Reach}, {Smith}, {Tuffs}, {Xu}, \&
  {Yun}}]{Cluver2010}
{Cluver}, M.~E., {Appleton}, P.~N., {Boulanger}, F., {et~al.} 2010, \apj, 710,
  248, \dodoi{10.1088/0004-637X/710/1/248}

\bibitem[{{Daddi} {et~al.}(2015){Daddi}, {Dannerbauer}, {Liu}, {Aravena},
  {Bournaud}, {Walter}, {Riechers}, {Magdis}, {Sargent}, {B{\'e}thermin},
  {Carilli}, {Cibinel}, {Dickinson}, {Elbaz}, {Gao}, {Gobat}, {Hodge}, \&
  {Krips}}]{Daddi2015}
{Daddi}, E., {Dannerbauer}, H., {Liu}, D., {et~al.} 2015, \aap, 577, A46,
  \dodoi{10.1051/0004-6361/201425043}

\bibitem[{{Duarte Puertas} {et~al.}(2019){Duarte Puertas},
  {Iglesias-P{\'a}ramo}, {Vilchez}, {Drissen}, {Kehrig}, \&
  {Martin}}]{DuartePuertas2019}
{Duarte Puertas}, S., {Iglesias-P{\'a}ramo}, J., {Vilchez}, J.~M., {et~al.}
  2019, \aap, 629, A102, \dodoi{10.1051/0004-6361/201935686}

\bibitem[{{Duarte Puertas} {et~al.}(2021){Duarte Puertas}, {Vilchez},
  {Iglesias-P{\'a}ramo}, {Drissen}, {Kehrig}, {Martin}, {P{\'e}rez-Montero}, \&
  {Arroyo-Polonio}}]{DuartePuertas2021}
{Duarte Puertas}, S., {Vilchez}, J.~M., {Iglesias-P{\'a}ramo}, J., {et~al.}
  2021, \aap, 645, A57, \dodoi{10.1051/0004-6361/202038734}

\bibitem[{{Farber} \& {Gronke}(2022{\natexlab{a}})}]{Farber2022a}
{Farber}, R.~J., \& {Gronke}, M. 2022{\natexlab{a}}, arXiv e-prints,
  arXiv:2209.13622.
\newblock \doarXiv{2209.13622}

\bibitem[{{Farber} \& {Gronke}(2022{\natexlab{b}})}]{Farber2022b}
---. 2022{\natexlab{b}}, \mnras, 510, 551, \dodoi{10.1093/mnras/stab3412}

\bibitem[{{Fedotov}(2014)}]{Fedotov2014}
{Fedotov}, K. 2014, Ph. D. Thesis, University of Western Ontario, Canda

\bibitem[{{Fedotov} {et~al.}(2011){Fedotov}, {Gallagher}, {Konstantopoulos},
  {Chandar}, {Bastian}, {Charlton}, {Whitmore}, \& {Trancho}}]{Fedotov2011}
{Fedotov}, K., {Gallagher}, S.~C., {Konstantopoulos}, I.~S., {et~al.} 2011,
  \aj, 142, 42, \dodoi{10.1088/0004-6256/142/2/42}

\bibitem[{{Gallagher} {et~al.}(2001){Gallagher}, {Charlton}, {Hunsberger},
  {Zaritsky}, \& {Whitmore}}]{Gallagher2001}
{Gallagher}, S.~C., {Charlton}, J.~C., {Hunsberger}, S.~D., {Zaritsky}, D., \&
  {Whitmore}, B.~C. 2001, \aj, 122, 163, \dodoi{10.1086/321111}

\bibitem[{{Gronke} \& {Oh}(2020)}]{Gronke2020}
{Gronke}, M., \& {Oh}, S.~P. 2020, \mnras, 492, 1970,
  \dodoi{10.1093/mnras/stz3332}

\bibitem[{{Gronke} \& {Oh}(2022)}]{Gronke2022}
---. 2022, arXiv e-prints, arXiv:2209.00732.
\newblock \doarXiv{2209.00732}

\bibitem[{{Guillard} {et~al.}(2010){Guillard}, {Boulanger}, {Cluver},
  {Appleton}, {Pineau Des For{\^e}ts}, \& {Ogle}}]{Guillard2010}
{Guillard}, P., {Boulanger}, F., {Cluver}, M.~E., {et~al.} 2010, \aap, 518,
  A59, \dodoi{10.1051/0004-6361/200913430}

\bibitem[{{Guillard} {et~al.}(2009){Guillard}, {Boulanger}, {Pineau Des
  For{\^e}ts}, \& {Appleton}}]{Guillard2009}
{Guillard}, P., {Boulanger}, F., {Pineau Des For{\^e}ts}, G., \& {Appleton},
  P.~N. 2009, \aap, 502, 515, \dodoi{10.1051/0004-6361/200811263}

\bibitem[{{Guillard} {et~al.}(2012){Guillard}, {Boulanger}, {Pineau des
  For{\^e}ts}, {Falgarone}, {Gusdorf}, {Cluver}, {Appleton}, {Lisenfeld},
  {Duc}, {Ogle}, \& {Xu}}]{Guillard2012}
{Guillard}, P., {Boulanger}, F., {Pineau des For{\^e}ts}, G., {et~al.} 2012,
  \apj, 749, 158, \dodoi{10.1088/0004-637X/749/2/158}

\bibitem[{{Guillard} {et~al.}(2022){Guillard}, {Appleton}, {Boulanger},
  {Shull}, {Lehnert}, {Pineau des Forets}, {Falgarone}, {Cluver}, {Xu},
  {Gallagher}, \& {Duc}}]{Guillard2022}
{Guillard}, P., {Appleton}, P.~N., {Boulanger}, F., {et~al.} 2022, \apj, 925,
  63, \dodoi{10.3847/1538-4357/ac313f}

\bibitem[{{H{\"o}gbom}(1974)}]{Hogbom1974}
{H{\"o}gbom}, J.~A. 1974, \aaps, 15, 417

\bibitem[{{Houck} {et~al.}(2004){Houck}, {Roellig}, {van Cleve}, {Forrest},
  {Herter}, {Lawrence}, {Matthews}, {Reitsema}, {Soifer}, {Watson}, {Weedman},
  {Huisjen}, {Troeltzsch}, {Barry}, {Bernard-Salas}, {Blacken}, {Brandl},
  {Charmandaris}, {Devost}, {Gull}, {Hall}, {Henderson}, {Higdon}, {Pirger},
  {Schoenwald}, {Sloan}, {Uchida}, {Appleton}, {Armus}, {Burgdorf},
  {Fajardo-Acosta}, {Grillmair}, {Ingalls}, {Morris}, \& {Teplitz}}]{Houck2004}
{Houck}, J.~R., {Roellig}, T.~L., {van Cleve}, J., {et~al.} 2004, \apjs, 154,
  18, \dodoi{10.1086/423134}

\bibitem[{{Iglesias-P{\'a}ramo} {et~al.}(2012){Iglesias-P{\'a}ramo},
  {L{\'o}pez-Mart{\'\i}n}, {V{\'\i}lchez}, {Petropoulou}, \&
  {Sulentic}}]{Iglesias-Paramo2012}
{Iglesias-P{\'a}ramo}, J., {L{\'o}pez-Mart{\'\i}n}, L., {V{\'\i}lchez}, J.~M.,
  {Petropoulou}, V., \& {Sulentic}, J.~W. 2012, \aap, 539, A127,
  \dodoi{10.1051/0004-6361/201118055}

\bibitem[{{Jennings} {et~al.}(2022){Jennings}, {Beckmann}, {Sijacki}, \&
  {Dubois}}]{Jennings2022}
{Jennings}, F., {Beckmann}, R., {Sijacki}, D., \& {Dubois}, Y. 2022, arXiv
  e-prints, arXiv:2211.09183.
\newblock \doarXiv{2211.09183}

\bibitem[{{Jones} \& {Nuth}(2011)}]{Jones2011}
{Jones}, A.~P., \& {Nuth}, J.~A. 2011, \aap, 530, A44,
  \dodoi{10.1051/0004-6361/201014440}

\bibitem[{{Konstantopoulos} {et~al.}(2014){Konstantopoulos}, {Appleton},
  {Guillard}, {Trancho}, {Cluver}, {Bastian}, {Charlton}, {Fedotov},
  {Gallagher}, {Smith}, \& {Struck}}]{Konstantopoulos2014}
{Konstantopoulos}, I.~S., {Appleton}, P.~N., {Guillard}, P., {et~al.} 2014,
  \apj, 784, 1, \dodoi{10.1088/0004-637X/784/1/1}

\bibitem[{{Lesaffre} {et~al.}(2013){Lesaffre}, {Pineau des For{\^e}ts},
  {Godard}, {Guillard}, {Boulanger}, \& {Falgarone}}]{Lesaffre2013}
{Lesaffre}, P., {Pineau des For{\^e}ts}, G., {Godard}, B., {et~al.} 2013, \aap,
  550, A106, \dodoi{10.1051/0004-6361/201219928}

\bibitem[{{Marigo} \& {Girardi}(2007)}]{Marigo2007}
{Marigo}, P., \& {Girardi}, L. 2007, in Astronomical Society of the Pacific
  Conference Series, Vol. 374, From Stars to Galaxies: Building the Pieces to
  Build Up the Universe, ed. A.~{Vallenari}, R.~{Tantalo}, L.~{Portinari}, \&
  A.~{Moretti}, 33

\bibitem[{{Mendes de Oliveira} {et~al.}(2001){Mendes de Oliveira}, {Plana},
  {Amram}, {Balkowski}, \& {Bolte}}]{mendes2001}
{Mendes de Oliveira}, C., {Plana}, H., {Amram}, P., {Balkowski}, C., \&
  {Bolte}, M. 2001, \aj, 121, 2524, \dodoi{10.1086/320390}

\bibitem[{{Moles} {et~al.}(1997){Moles}, {Sulentic}, \&
  {M{\'a}rquez}}]{Moles1997}
{Moles}, M., {Sulentic}, J.~W., \& {M{\'a}rquez}, I. 1997, \apjl, 485, L69,
  \dodoi{10.1086/310817}

\bibitem[{{Natale} {et~al.}(2010){Natale}, {Tuffs}, {Xu}, {Popescu},
  {Fischera}, {Lisenfeld}, {Lu}, {Appleton}, {Dopita}, {Duc}, {Gao}, {Reach},
  {Sulentic}, \& {Yun}}]{Natale2010}
{Natale}, G., {Tuffs}, R.~J., {Xu}, C.~K., {et~al.} 2010, \apj, 725, 955,
  \dodoi{10.1088/0004-637X/725/1/955}

\bibitem[{{O'Sullivan} {et~al.}(2009){O'Sullivan}, {Giacintucci}, {Vrtilek},
  {Raychaudhury}, \& {David}}]{OSullivan2009}
{O'Sullivan}, E., {Giacintucci}, S., {Vrtilek}, J.~M., {Raychaudhury}, S., \&
  {David}, L.~P. 2009, \apj, 701, 1560, \dodoi{10.1088/0004-637X/701/2/1560}

\bibitem[{{Papadopoulos} {et~al.}(2008){Papadopoulos}, {Feain}, {Wagg}, \&
  {Wilner}}]{Papadopoulos2008}
{Papadopoulos}, P.~P., {Feain}, I.~J., {Wagg}, J., \& {Wilner}, D.~J. 2008,
  \apj, 684, 845, \dodoi{10.1086/590233}

\bibitem[{{Pereira-Santaella} {et~al.}(2022){Pereira-Santaella},
  {{\'A}lvarez-M{\'a}rquez}, {Garc{\'\i}a-Bernete}, {Labiano}, {Colina},
  {Alonso-Herrero}, {Bellocchi}, {Garc{\'\i}a-Burillo}, {H{\"o}nig}, {Ramos
  Almeida}, \& {Rosario}}]{Pereira-Santaella2022}
{Pereira-Santaella}, M., {{\'A}lvarez-M{\'a}rquez}, J., {Garc{\'\i}a-Bernete},
  I., {et~al.} 2022, \aap, 665, L11, \dodoi{10.1051/0004-6361/202244725}

\bibitem[{{Pontoppidan} {et~al.}(2022){Pontoppidan}, {Blome}, {Braun}, {Brown},
  {Carruthers}, {Coe}, {DePasquale}, {Espinoza}, {Garcia Marin}, {Gordon},
  {Henry}, {Hustak}, {James}, {Koekemoer}, {LaMassa}, {Law}, {Lockwood},
  {Moro-Martin}, {Mullally}, {Pagan}, {Player}, {Proffitt}, {Pulliam},
  {Ramsay}, {Ravindranath}, {Reid}, {Robberto}, {Sabbi}, \&
  {Ubeda}}]{Pontoppidan2022}
{Pontoppidan}, K., {Blome}, C., {Braun}, H., {et~al.} 2022, arXiv e-prints,
  arXiv:2207.13067.
\newblock \doarXiv{2207.13067}

\bibitem[{{Sandstrom} {et~al.}(2013){Sandstrom}, {Leroy}, {Walter}, {Bolatto},
  {Croxall}, {Draine}, {Wilson}, {Wolfire}, {Calzetti}, {Kennicutt}, {Aniano},
  {Donovan Meyer}, {Usero}, {Bigiel}, {Brinks}, {de Blok}, {Crocker}, {Dale},
  {Engelbracht}, {Galametz}, {Groves}, {Hunt}, {Koda}, {Kreckel}, {Linz},
  {Meidt}, {Pellegrini}, {Rix}, {Roussel}, {Schinnerer}, {Schruba}, {Schuster},
  {Skibba}, {van der Laan}, {Appleton}, {Armus}, {Brandl}, {Gordon}, {Hinz},
  {Krause}, {Montiel}, {Sauvage}, {Schmiedeke}, {Smith}, \&
  {Vigroux}}]{Sandstrom2013}
{Sandstrom}, K.~M., {Leroy}, A.~K., {Walter}, F., {et~al.} 2013, \apj, 777, 5,
  \dodoi{10.1088/0004-637X/777/1/5}

\bibitem[{{Sault} {et~al.}(1995){Sault}, {Teuben}, \& {Wright}}]{Sault1995}
{Sault}, R.~J., {Teuben}, P.~J., \& {Wright}, M.~C.~H. 1995, in Astronomical
  Society of the Pacific Conference Series, Vol.~77, Astronomical Data Analysis
  Software and Systems IV, ed. R.~A. {Shaw}, H.~E. {Payne}, \& J.~J.~E.
  {Hayes}, 433.
\newblock \doarXiv{astro-ph/0612759}

\bibitem[{{Smith} {et~al.}(2007){Smith}, {Draine}, {Dale}, {Moustakas},
  {Kennicutt}, {Helou}, {Armus}, {Roussel}, {Sheth}, {Bendo}, {Buckalew},
  {Calzetti}, {Engelbracht}, {Gordon}, {Hollenbach}, {Li}, {Malhotra},
  {Murphy}, \& {Walter}}]{Smith2007}
{Smith}, J.~D.~T., {Draine}, B.~T., {Dale}, D.~A., {et~al.} 2007, \apj, 656,
  770, \dodoi{10.1086/510549}

\bibitem[{{Sulentic} {et~al.}(2001){Sulentic}, {Rosado}, {Dultzin-Hacyan},
  {Verdes-Montenegro}, {Trinchieri}, {Xu}, \& {Pietsch}}]{Sulentic2001}
{Sulentic}, J.~W., {Rosado}, M., {Dultzin-Hacyan}, D., {et~al.} 2001, \aj, 122,
  2993, \dodoi{10.1086/324455}

\bibitem[{{Togi} \& {Smith}(2016)}]{Togi2016}
{Togi}, A., \& {Smith}, J.~D.~T. 2016, \apj, 830, 18,
  \dodoi{10.3847/0004-637X/830/1/18}

\bibitem[{{Trinchieri} {et~al.}(2005){Trinchieri}, {Sulentic}, {Pietsch}, \&
  {Breitschwerdt}}]{Trinchieri2005}
{Trinchieri}, G., {Sulentic}, J., {Pietsch}, W., \& {Breitschwerdt}, D. 2005,
  \aap, 444, 697, \dodoi{10.1051/0004-6361:20052910}

\bibitem[{{Xu} {et~al.}(1999){Xu}, {Sulentic}, \& {Tuffs}}]{Xu1999}
{Xu}, C., {Sulentic}, J.~W., \& {Tuffs}, R. 1999, \apj, 512, 178,
  \dodoi{10.1086/306771}

\bibitem[{{Xu} {et~al.}(2003){Xu}, {Lu}, {Condon}, {Dopita}, \&
  {Tuffs}}]{Xu2003}
{Xu}, C.~K., {Lu}, N., {Condon}, J.~J., {Dopita}, M., \& {Tuffs}, R.~J. 2003,
  \apj, 595, 665, \dodoi{10.1086/377445}

\bibitem[{{Xu} {et~al.}(2005){Xu}, {Iglesias-P{\'a}ramo}, {Burgarella}, {Rich},
  {Neff}, {Lauger}, {Barlow}, {Bianchi}, {Byun}, {Forster}, {Friedman},
  {Heckman}, {Jelinsky}, {Lee}, {Madore}, {Malina}, {Martin}, {Milliard},
  {Morrissey}, {Schiminovich}, {Siegmund}, {Small}, {Szalay}, {Welsh}, \&
  {Wyder}}]{Xu2005}
{Xu}, C.~K., {Iglesias-P{\'a}ramo}, J., {Burgarella}, D., {et~al.} 2005, \apjl,
  619, L95, \dodoi{10.1086/425130}

\bibitem[{{Zhu} {et~al.}(2007){Zhu}, {Gao}, {Seaquist}, \& {Dunne}}]{Zhu2007}
{Zhu}, M., {Gao}, Y., {Seaquist}, E.~R., \& {Dunne}, L. 2007, \aj, 134, 118,
  \dodoi{10.1086/517996}

\end{thebibliography}

\end{document}